\documentclass[%
 reprint,
 amsmath,amssymb,
 aps,
prb,
]{revtex4-2}

\usepackage{graphicx}% Include figure files
\usepackage{dcolumn}% Align table columns on decimal point
\usepackage{bm}% bold math
\usepackage{xcolor}
\usepackage{hyperref}
\hypersetup{
	colorlinks,
	linkcolor={blue!90!black},
	citecolor={blue!90!black},
	urlcolor=	{blue!90!black}}
\usepackage[utf8]{inputenc}
\usepackage[T1]{fontenc}
\usepackage{mathtools}

\DeclareMathOperator{\re}{Re}
\DeclareMathOperator{\im}{Im}
\DeclareMathOperator{\tr}{Tr}

\newcommand{\DPS}{\Delta_\mathrm{PS}}
\newcommand{\omc}{\omega_\mathrm{c}}
\newcommand{\oml}{\omega_\mathrm{L}}

\newcommand{\GG}{\arrowvert \textit{G} \rangle \langle \textit{G} \arrowvert}
\newcommand{\GX}{\arrowvert \textit{G} \rangle \langle \textit{X} \arrowvert}
\newcommand{\XG}{\arrowvert \textit{X} \rangle \langle \textit{G} \arrowvert}
\newcommand{\XX}{\arrowvert \textit{X} \rangle \langle \textit{X} \arrowvert}

\newcommand{\rhoS}{\rho^{\mathrm{S}}}
\newcommand{\tf}{\textit{t}_\mathrm{f}}
\newcommand{\ti}{\textit{t}_\mathrm{i}}
\newcommand{\supL}{\mathcal{L}}
\newcommand{\tildesupL}{\tilde{\mathcal{L}}}
\newcommand{\TP}{\textit{T}_\mathrm{P}}
\newcommand{\TPconj}{\textit{T}^\dag_\mathrm{P}}
\makeatletter
\def\subsubsection{%
  \@startsection{subsubsection}{3}{\z@}%
  {-1.5ex plus -0.5ex minus -0.2ex}%
  {0.5ex plus 0.2ex}%
  {\normalfont\small\bfseries}% <--- Tu zmieniliśmy \itshape na \bfseries
}
\makeatother

\begin{document}

%\preprint{APS/123-QED}

\title{Fano profile in the resonance fluorescence spectrum of a solid-state quantum emitter coupled to phonons}% Force line breaks with \\
\author{Rafał A. Bogaczewicz}
\author{Paweł Machnikowski}%
\affiliation{Institute of Theoretical Physics, Wrocław University of Science and Technology, Wybrzeże Wyspiańskiego 27, 50-370 Wrocław, Poland}

\date{\today}% It is always \today, today,

\begin{abstract}
We present a theory of resonance fluorescence (RF) of a solid-state quantum emitter in the regime of weak optical excitation. The emitter is coupled to the phonon modes of the surrounding bulk semiconductor, described by a super-Ohmic spectral density. We show that the RF spectrum of this system consists of a central elastic line, a broad phonon sideband known from other linear and non-linear spectra of such systems, and a narrow inelastic contribution characteristic of scattering spectra, stemming from noise-induced transient dynamics. At moderate phonon couplings or low temperatures, the interplay between the broad sideband and the inelastic feature leads to a Fano-like profile near the resonant energy with the Fano parameter determined by laser detuning. In the weak-coupling limit (where only single-phonon processes are included), the spectrum becomes an exact Fano shape, with the inelastic component vanishing completely at the exact laser frequency. The amplitude of this spectral feature grows linearly with temperature, while its width depends solely on the spontaneous emission rate of the emitter. Because the spectral line shape is extremely sensitive to dephasing, resonance fluorescence can serve as a direct, non-interferometric probe of coherence at dephasing rates critical for generating highly indistinguishable single photons.
We relate the quantum character of the reservoir to the non-commutativity of noise observables and show that the Fano resonance persists in the classical (commutative) limit. We also discuss how the redistribution of optical coupling efficiency between the central line and the sidebands affects the total scattering rate under various excitation conditions. 
\end{abstract}

\maketitle

%\tableofcontents

\section{\label{sec:Introduction} Introduction} 

Resonance fluorescence (RF) of a single quantum emitter \cite{ScullyZubairy1997,Meystre2007} has found numerous applications in quantum optics and quantum information processing due to the nontrivial properties of scattered light. Within the RF scheme, it is possible, for instance, to generate non-classical states of photons with unique coherence properties \cite{Dalibard1983,Nienhuis1993} and create indistinguishable \cite{Scholl2019} or anti-bunched \cite{Wang2025} photons. The last decades have witnessed a rapid development of RF as a tool to characterize the quantum properties not only of atomic and molecular systems \cite{wrigge_efficient_2008}, but also of various kinds of solid-state ``artificial atoms'' \cite{Muller2007,Astaview2010,Toyli2016}.

Semiconductor self-assembled quantum dots (QDs) have been widely used as quantum emitters in various optical schemes. Combining atomic-like QD properties with the RF technique has allowed one to observe spin dynamics \cite{Vamivakas2010,Delteil2014}, interface \cite{Ylmaz2010}, and entangle \cite{Delley2017} QD spins with single photons or read out spin states from the optical signal \cite{Lu2010,Vamivakas2009}. Recently, QD emitters have been used in acousto-optic quantum hybrid systems as transducers between acoustical and optical signals \cite{Weiss2018,Delsing2019, Weiss2021,Wigger2021,Metcalfe2010,krenner2026roadmap}. QDs generate antibunched light \cite{Weiss2021}, making them perfect candidates for a single-photon source \cite{Senellart2017}, which opens the door to advanced applications in quantum information processing, such as quantum multiplexing \cite{Piparo2019}, time and frequency bin encoding \cite{Lu2023,Pan2012}, or quantum acousto-optic transduction \cite{Stannigel2010}. Periodic acoustic modulation enables Floquet engineering of the spectral properties of these systems \cite{Groll_2026}. A QD can also be used as a quantum sensor for an acoustic cavity, the fluorescence signal carrying sufficient information to retrieve the phonon number statistics \cite{Groll2025}. Another important family of solid-state quantum emitters in the optical domain, which overcomes the low-temperature limitations of QDs, is that of defect centers in various material systems. These have been shown to exhibit photon antibunching up to room temperature \cite{Tran2016,Martinez2016,Jungwirth2016} or far above \cite{Kianinia2017}, making them a valuable complement of QDs as emitters used in quantum information processing.

Solid-state quantum technologies critically depend on the coherence of quantum emitters. This requirement has become particularly stringent in the context of measurement-based quantum computing \cite{raussendorf2001}, where the scalable generation of multi-dimensional cluster states is driven by optical fusion operations \cite{browne2005}. Because these fusion gates demand nearly perfect interference between sequentially emitted photons, maintaining an ultra-high level of indistinguishability across long photon trains is strictly required. Consequently, demonstrating negligibly small pure dephasing to reach the strict life time limit represents an essential step in the development of such highly coherent sources.
In fact, photon indistinguishability measurements using resonant and deterministic \cite{ding2016, wang2016} excitation schemes, employing low-noise QD structures \cite{kuhlmann2015, somaschi2016}, or benefiting from a strong Purcell effect \cite{ding2016, tomm2021} vastly suppress such dephasing. In such devices, Hong-Ou-Mandel visibilities reach 0.995 \cite{wei2014, somaschi2016}, implying a ratio of pure dephasing to recombination rates on the order of $10^{-3}$. The measured values are most likely limited by experimental imperfections of the interferometric setup \cite{wei2014a, somaschi2016, wang2019}, and there seems to be no fundamental reason for the unavoidable presence of pure dephasing in every system. 

In a solid-state matrix, phonons are one of the main sources of environmental noise that perturbs the transition energy of the emitter, even though they cannot induce transitions because of substantial energy mismatch. 
A common signature of carrier-phonon interactions in the spectra of solid-state quantum emitters are the broad phonon sidebands surrounding the central (``zero-phonon'') line: the exchange of acoustic phonons accompanying an optical process modifies the energy of the absorbed or emitted photon, leading to spectral features on the high or low energy side of the central line. Such features are observed in QD luminescence \cite{Besombes2001,Favero2003} and four-wave-mixing spectroscopy \cite{Vagov2004} and have been predicted in absorption \cite{Krummheuer2002}. As the range of efficiently coupled phonon wavelengths in QDs is limited from below by the QD size, the spectral width of the resulting sidebands reaches a few meV. 
The coupling strength between phonons and exciton is represented by a Huang-Rhys factor \cite{Huang1950}, which reflects the spectral weight accumulated in the phonon sidebands at null temperature. 
This experimentally accessible parameter may have a wide range of values depending on the shape and size of the emitter, the properties of the surrounding material, or the coupling mechanism. For InAs/GaAs self-assembled QDs, the Huang-Rhys factor is on the order of 0.1 \cite{Krummheuer2002,Favero2003,Vagov2004}. Defect centers show similar phonon sidebands, but they are much wider due to the presumably point-like nature of the defect, with the Huang-Rhys parameter one order of magnitude larger in the case of defects in hBN \cite{Jungwirth2016,Wigger2019,Khatri2019,Preuss2022}. The spectral weight of this phonon feature increases with temperature. Apart from this primary phonon coupling, which cannot broaden the transition line due to the super-Ohmic spectral density of the phonon reservoir, phonons contribute to higher-order processes which effectively lead to pure-dephasing-induced line broadening. Known mechanisms include coupling to higher confined states \cite{muljarov2004} and phonon anharmonicity \cite{machnikowski2006prl,machnikowski2006pss}.

Absorption or emission of an acoustic phonon can also take place during scattering of a photon, leading to various effects under strong excitation or induced by optical phonons \cite{Kabuss2010,Kabuss2011b,McCutcheon2013}. In the case of a weakly excited system coupled to acoustic phonons, as studied here, correlation expansion modeling under pulsed excitation has shown that sidebands indeed appear in the resonance fluorescence spectra \cite{Ahn2005}, alongside the narrow central line corresponding to elastic scattering, which is centered at the laser frequency, as predicted by the standard, unperturbed model \cite{ScullyZubairy1997,Meystre2007}. On the other hand, classical noise under cw excitation gives rise to inelastic scattering, manifested by a line at the QD transition frequency, with a width dependent on the noise strength in the noise models studied so far \cite{Bogaczewicz2023,Bogaczewicz2025}. 

In this work, we study theoretically the RF of a single weakly excited emitter coupled to phonons in a bulk material. We develop a general description of the weak-excitation RF that includes phonon effects exactly to all orders. We apply this method to a two-level system coupled to bulk acoustic phonons, analyzing both the scattering spectra at resonant or nearly resonant excitation, and the total scattering intensity as a function of excitation frequency. For definiteness, we focus on a self-assembled QD, but we cover both weak and strong phonon coupling regimes to ensure a general scope of the theory. We show that, in addition to a broad phonon sideband, the inelastic scattering spectrum shows a narrow Fano feature with temperature-dependent intensity but fixed width. Remarkably, in the independent-boson model studied here, this feature appears in scattering although it is known to be absent in absorption \cite{Krummheuer2002}, unlike in the more widely studied Fano-Anderson models \cite{Fano1961,Riffe2011}. In contrast to some previous works concerning Fano resonances involving phonons, where a discrete phonon mode couples to electronic continuum \cite{leeUltrafastFanoResonance2006, misochkoFanoInterferenceExcitation2015, yoshinoDynamicalFanoLikeInterference2015, watanabePolaronicQuasiparticlePicture2017, vinodFanoResonanceCoherent2018, watanabeUltrafastAsymmetricRosenZenerlike2019, vasileiadisFrequencydomainStudyNonthermal2020} or to phonon bath \cite{kitzmanQuantumAcousticFano2023}, in our model the sharp resonance pertains to electronic excitation, while the wide background stems from mechanical degrees of freedom. We discuss the impact of dephasing on the predicted feature and show that a clear trace of the Fano profile persists in the presence of inhomogeneous line broadening, while additional homogeneous broadening beyond the radiatively limited linewidth is detrimental to this spectral feature. In this way, we demonstrate that the well-characterized phonon effects are not only decoupled from other sources of pure dephasing but also contribute to a mechanism by which these more enigmatic and presumably extremely weak sources of dephasing can be characterized. We thus establish the RF spectrum as a unique, highly sensitive probe of coherence.  

The paper is structured as follows. In Sec.~\ref{sec:System_and_model} we define the model of the system. Next, we present the theory describing the evolution of the system (Sec.~\ref{sec:evol-polaron-pict}), the steady state of the system (Sec.~\ref{sec:steady-state}) the quantum autocorrelation function (Sec.~\ref{sec:G}) and the RF spectrum (Sec.~\ref{sec:RF_spectrum}). Sec.~\ref{sec:Results_discussions} discusses the RF spectra and scattering intensities under various conditions. The paper is concluded in Sec.~\ref{sec:conclusions}.

\section{\label{sec:System_and_model}System and model}

We consider an atomic-like solid-state quantum emitter, weakly excited by a continuous-wave (cw) laser, tuned resonantly or near resonantly to the emitter's fundamental transition frequency. To facilitate presentation, we will think of a self-assembled semiconductor QD, in which the optically induced excitation is a bound exciton (electron-hole pair). However, the presented approach does not depend on this physical picture and is also valid for strong phonon coupling, so it can be used, e.g., for defect centers. The  lifetime of the fundamental excitation of the system is limited by spontaneous emission (radiative recombination). In addition, we take into account pure dephasing of the optical coherence (homogeneous dephasing) as well as the effect of low-frequency fluctuations of the transition energy due, e.g., to charge noise in the environment (inhomogeneous broadening). We assume that the frequency and polarization selectivity of the exciting beam allow us to focus on a single transition. The scattered light is spectrally resolved and integrated in time, leading to the RF spectrum.

The exciton in the system interacts with the lattice degrees of freedom, that is, with the phonon modes of the bulk host material,
which are assumed to be in a thermal state. 
We assume that phonons have a typical super-Ohmic spectral density with dependence $\propto \omega^3$ at low frequencies, which is typical for bulk solid-state environments \cite{Weiss1998,Reiter2019}. The high-frequency cut-off for phonons, which may be related to the Debye frequency or to the size of the mesoscopic system \cite{Lueker2017,Khatri2019}, is the largest frequency scale of the problem; hence, the time span of the memory of the phonon reservoir is short compared to the typical time scales of the problem. For a QD, this cut-off is at several ps$^{-1}$, which is indeed much larger than the exciton decay rate or optical Rabi frequency in the weak excitation regime ($\sim$ ns$^{-1}$). 

The emitter is modeled as a two-level system driven by a classical laser beam with the bosonic reservoir coupled via an independent-boson Hamiltonian \cite{Mahan2000,Huang1950}. The ground and excited system states are denoted $|G\rangle$ and $|X\rangle$, respectively, and the bare transition energy between them is $\hbar\omega_0$. Phonon modes are labeled by their spectral branch $\lambda$ and wave vector $\bm{k}$ and described by the creation and annihilation operators $b^{\dagger}_{\lambda,\bm{k}}$, $b_{\lambda,\bm{k}}$, with the corresponding frequencies $\omega_{\lambda,\bm{k}}$.
The Hamiltonian of the system is 
\begin{equation*}
H = H_0 + H_\mathrm{int},
\end{equation*}
where $H_0$ describes the system, the free phonon subsystem, and the system-phonon interaction, while $H_\mathrm{int}$ accounts for the optical driving.

In the frame rotating with the laser frequency $\oml$, the first part is
\begin{align}
H_0 & =  \hbar\left[\omega_0-\oml\right]\XX + H_\mathrm{ph} \label{eq:H_0} \\
& \quad+ \hbar \XX \sum_{\lambda,\bm{k}}\left(g_{\lambda,\bm{k}} b^\dagger_{\lambda,\bm{k}} + g_{\lambda,\bm{k}}^* b_{\lambda,\bm{k}}\right) ,\nonumber
\end{align}
where $g_{\lambda,\bm{k}}$ denotes the coupling constant between the emitter and a phonon mode, and
\begin{equation*}
H_\mathrm{ph} 
= \sum_{\lambda,\bm{k}}\hbar \omega_{\lambda,\bm{k}} b^{\dagger}_{\lambda,\bm{k}}b_{\lambda,\bm{k}} %\label{eq:H_ph}
\end{equation*}
is the free-phonon Hamiltonian. 
The coupling to the phonon reservoir is characterized by the spectral density, 
\begin{equation*}
J(\omega) = \sum_{\lambda,\bm{k}} \left| g_{\lambda,\bm{k}} \right|^2 \delta\left(\omega - \omega_{\lambda,\bm{k}}\right), %\label{eq:J_w}
\end{equation*}
or by the memory function
\begin{align}
\phi(t) & = \int_{0}^{\infty}d\omega \frac{J(\omega)}{\omega^2}
\left(e^{-i\omega t}\left[n(\omega)+1\right] + e^{i\omega t}n(\omega)\right)  \label{eq:Phi_x_quantum} \\
& = \sum_{\lambda,\bm{k}} \left| \frac{g_{\lambda,\bm{k}}}{\omega_{\lambda,\bm{k}}} \right|^2
\left(e^{-i\omega_{\lambda,\bm{k}} t}\left[n_{\lambda,\bm{k}}+1\right] + e^{i\omega_{\lambda,\bm{k}} t}n_{\lambda,\bm{k}}\right), \nonumber
\end{align}
where $n(\omega)$ is the Bose distribution and $n_{\lambda,\bm{k}} = n(\omega_{\lambda,\bm{k}})$ is the number of phonons in the mode $(\lambda,\bm{k})$. 

The overall strength of the carrier-phonon coupling is quantified by the Huang-Rhys parameter \cite{Huang1950}
\begin{equation*}
F_\mathrm{HR} = \int_0^{\infty} d\omega \frac{J(\omega)}{\omega^2} 
= \sum_{\lambda,\bm{k}} \left| \frac{g_{\lambda,\bm{k}}}{\omega_{\lambda,\bm{k}}} \right|^2.
\end{equation*}

The second contribution to the Hamiltonian describes the interaction between laser light and the emitter, where the optical driving strength is characterized by the Rabi frequency $\Omega$. In the rotating frame and rotating wave approximation, this reads
\begin{equation*}
H_\mathrm{int} = - \frac{\hbar\Omega}{2}\left(\GX + \XG\right). %\label{eq:H_int} 
\end{equation*}
In addition, spontaneous emission is described by the standard Lindblad dissipator,
\begin{equation*}
L^{(\mathrm{se})}\left[O\right] = 
\gamma_{1}\left(\GX O \XG -\frac{1}{2}\left\{\XX , O\right\}\right),
\end{equation*}
where $\gamma_{1}$ is the emission rate, $\left\{A,B\right\} = AB + BA$, and $O$ denotes an arbitrary operator. Pure dephasing with rate $\gamma_{2}$ is included via
\begin{equation*}
L^{(\mathrm{pd})}\left[O\right] = 
\gamma_{2}\left(\XX O \XX -\frac{1}{2}\left\{\XX , O\right\}\right),
\end{equation*}
which includes, in particular, phonon processes beyond the independent boson model \cite{muljarov2004,machnikowski2006prl,machnikowski2006pss}.

The system evolves according to the master equation
\begin{equation} 
\frac{d\rho(t)}{dt} = L_0\left[\rho(t)\right] + L_1\left[\rho(t)\right], \label{eq:master_euqation}   
\end{equation}
where
\begin{equation}
L_0\left[O\right] = -\frac{i}{\hbar}\left[H_0,O\right] + L^{(\mathrm{se})}\left[O\right] + L^{(\mathrm{pd})}\left[O\right] \label{eq:L0}
\end{equation}
represents the unperturbed (free) evolution of the system
and
\begin{equation*}
L_1\left[O\right] = -\frac{i}{\hbar}\left[H_\mathrm{int}, O\right] %\label{eq:L_1_O}
\end{equation*}
accounts for the optical excitation and will be treated as a perturbation in the weak excitation regime of interest.

The optical spectrum is given by
\begin{equation}    \label{eq:RF_spectrum1}
S(\omega) = \re  \int_0^{\infty}d\tau e^{i(\omega-\omega_L)\tau} G(\tau)
\end{equation}
with the steady-state first-order autocorrelation function
\begin{equation}
G(\tau) = \langle \sigma_+(0)\sigma_-(\tau)\rangle ,  \label{eq:autocorrelation_0}
\end{equation}
where we use the fact that the system is stationary and introduce the standard transition operators $\sigma_+ = \sigma_-^\dagger = \XG $ which are written here in the Heisenberg picture and in the rotating frame.  
The total scattering intensity is 
\begin{equation} \label{eq:Intensity}
I_\mathrm{tot} = \int_{-\infty}^{\infty} d\omega S(\omega) = \pi G(0). 
\end{equation}
Finally, inhomogeneous broadening of the transition energy can be included by averaging the spectrum over a Gaussian distribution of transition energies with the standard deviation $\sigma$.

\section{\label{sec:autocorrelation_function} Theory of the RF response with phonons}

In this section, we present a method for calculating the RF spectrum of a quantum emitter coupled to phonons to the second order in the laser driving and including the phonon effects exactly within the independent boson model. 
Prior to this, we introduce some formal definitions.

We denote the formal solution of Eq.~\eqref{eq:master_euqation} as
\begin{equation*}
\rho(t+\tau) = \supL_{\tau}\left[\rho(t)\right],
\end{equation*}
which, together with Eq.~\eqref{eq:master_euqation}, defines the propagator $\supL_{\tau}$ generated by $L_0 + L_1$. The propagator depends only on the time interval because the system is stationary. Although primarily defined for the density matrix, the equation of motion, as well as the propagator, can be extended to the entire Liouville space of operators on the system (emitter and phonons) Hilbert space, which we will exploit in the following. As is customary, we refer to propagators and generators, that is, operators acting on the Liouville space, as superoperators.
We will also use the decomposition of operators in the emitter basis,
\begin{equation*}
O = \sum_{i,j=G,X} \arrowvert i \rangle \langle j \arrowvert O_{ij}, %\label{eq:rhoS_parts} 
\end{equation*}
where $O_{ij}$ are operators on the Hilbert space of the phonon subsystem.

Since the effect of inhomogeneous dephasing amounts to averaging over a distribution of transition energies, i.e., detuning values, we perform the derivations at a fixed detuning $\Delta$ and average the result in the final phase.

\subsection{System evolution and autocorrelation function in the polaron picture}\label{sec:evol-polaron-pict}

By applying the quantum regression theorem, Eq.~(\ref{eq:autocorrelation_0}) can then be written in the form 
\begin{align}
G(\tau) & = \tr(\GX \supL_{\tau}\left[\rhoS\XG\right]) \nonumber \\
& = \tr_\mathrm{ph} \langle X \arrowvert \supL_{\tau}\left[\rhoS\XG\right] \arrowvert G \rangle,
\label{eq:autocorrelation_1}
\end{align}
where $\rhoS$ is the steady state of the system. In the final step, we have split the total trace into partial traces over the emitter and phonon degrees of freedom, ($\tr_{\mathrm{ph}}$), and explicitly taken the former. It should be noted that the regression theorem is applied only with respect to the optical reservoir (spontaneous emission process) for which the Markovian approximation works perfectly, while the phonon reservoir is treated exactly.

To evaluate the autocorrelation function to order $\Omega^2$, the propagator is expressed as a second-order Born expansion
\begin{align}
\supL_{\tf-\ti} &  = \supL_{\tf-\ti}^{(0)} + \int_{\ti}^{\tf} dt_1 \supL_{\tf-t_1}^{(0)} L_1 \supL_{t_1-\ti}^{(0)} \label{eq:L_total} \\
&\quad + \int_{\ti}^{\tf} dt_1 \int_{\ti}^{t_1} dt_2 \supL_{\tf-t_1}^{(0)} L_1 \supL_{t_1-t_2}^{(0)} L_1 \supL_{t_2-\ti}^{(0)}  +\ldots \nonumber 
%\\
%&\quad + O\left(\Omega^3\right), \nonumber
\end{align}
where the terms correspond to successive orders of $\Omega$. 
Here, $\supL_{\tau}^{(0)}$ denotes the propagator of the unperturbed evolution, generated by $L_0$, that is,
\begin{equation*}%\label{eq:gen-L0}
\frac{d}{dt}\supL^{(0)}_t O = L_0 \supL^{(0)}_t O.
\end{equation*}
Here and in the following, we adopt the convention that all superoperators act on all terms to their right, so we can omit brackets.

The emitter-phonon coupling in the Hamiltonian of Eq.~\eqref{eq:H_0} is removed by the unitary transformation \cite{Mahan2000}
\begin{equation}\label{eq:TP}
\TP =  \GG + \XX D\left\{g_{\lambda,\bm{k}}\right\}, 
\end{equation}
which defines the \textit{polaron frame of reference} in which the excited state of the emitter is dressed with a lattice deformation restoring the lowest-energy configuration in the presence of interaction. Here $D$ is the displacement (Weyl) operator, with $\{g_{\lambda,\bm{k}}\}$ representing the full set of coupling constants,
\begin{equation}
D\left\{ g_{\lambda,\bm{k}} \right\} =
%& V(t) D\left\{ g(0) \right\} V^\dagger(t) =  \label{eq:D_time} \\
%= &
\exp\sum_{\lambda,\bm{k}} \left(\frac{g_{\lambda,\bm{k}}}{\omega_{\lambda,\bm{k}}} b_{\lambda,\bm{k}}^\dagger - \frac{g^*_{\lambda,\bm{k}}}{\omega_{\lambda,\bm{k}}} b_{\lambda,\bm{k}} \right). \nonumber 
\end{equation}
In addition, it is convenient to perform the calculations in the interaction picture with respect to free phonon Hamiltonian, defined by the usual transformation
\begin{equation}
V(t) = \exp\left(i H_\mathrm{ph} t / \hbar\right). \label{eq:V_t} 
\end{equation}
For an operator $O(t)$, the transformed operator is
\begin{equation*}
\tilde{O}(t) = V(t)T_\mathrm{P} O(t) T^\dagger_\mathrm{P} V^\dagger (t). %\label{eq:transformations} 
\end{equation*}
The corresponding transformed superoperator $\tildesupL_{\tf,\ti}$ that propagates the state from $\ti$ to $\tf$ is, consistently, defined by
%\begin{align}
%& \tildesupL_{\tf,\ti} \tilde{O} = \label{eq:tilde_superoperator} \\
%& \hspace{0.7 cm} = V(\tf) \TP \left(\supL_{\tf,\ti} V(\ti) \TP O \TPconj %V^\dagger(\ti)\right) \TPconj V^\dagger(\tf) . \nonumber
%\end{align}
\begin{align} \label{eq:tilde_superoperator}
\MoveEqLeft{\tildesupL_{\tf,\ti} \tilde{O}} 
= \\
&  V(\tf) \TP \left(\supL_{\tf-\ti} \TPconj V^\dagger(\ti) \tilde{O} V(\ti) \TP  \right) \TPconj V^\dagger(\tf) . \nonumber
\end{align}
The transformed free propagator is defined in the same way,
\begin{equation*} %\label{eq:tilde_superoperator_0}
\tildesupL^{(0)}_{\tf,\ti} \tilde{O} 
= V(\tf) \TP \left(\supL^{(0)}_{\tf-\ti} \TPconj V^\dagger(\ti) \tilde{O} V(\ti) \TP  \right) \TPconj V^\dagger(\tf) .
\end{equation*}
By direct differentiation and using Eq.~\eqref{eq:TP}, Eq.~\eqref{eq:V_t}, and Eq.~\eqref{eq:L0} one finds
\begin{equation}\label{eq:gen-L0-tilde}
\frac{d}{dt}\tildesupL^{(0)}_{t,t'} \tilde{O} = \tilde{L}_0 \tildesupL^{(0)}_{t,t'} \tilde{O}
\end{equation}
with the generator
\begin{equation*}
\tilde{L}_0 \tilde{O} = -\frac{i}{\hbar}\left[H'_0,\tilde{O}\right] + \tilde{L}^{(\mathrm{se})}\tilde{O} + \tilde{L}^{(\mathrm{pd})}\tilde{O},
%\label{eq:L0-tilde}
\end{equation*}
where  
%\begin{equation*}
$H'_0 = - \hbar\Delta \XX$,
%\end{equation*}

\begin{align} 
\tilde{L}^{(\mathrm{se})}\tilde{O} & = 
- \frac{\gamma_{1}}{2} \{ \XX , \tilde{O}\} \label{eq:Lse-tilde} \\
& \quad + \gamma_{1} \GG D^\dagger\{g_{\lambda,\bm{k}}(t)\}\tilde{O}_{XX} D\{g_{\lambda,\bm{k}}(t)\}, \nonumber 
\end{align}
and
\begin{equation}
\tilde{L}^{(\mathrm{pd})}\tilde{O} = - \frac{\gamma_{2}}{2} \{ \XX , \tilde{O}\} + \gamma_{2} \XX \tilde{O}_{XX} . \label{eq:Lpd-tilde}
\end{equation}
Here
\begin{equation} \label{eq:Dt}
D\{g_{\lambda,\bm{k}}(t)\} 
= V(t) D\{g_{\lambda,\bm{k}}\} V^{\dag}(t),
\end{equation}
with 
$g_{\lambda,\bm{k}}(t) = g_{\lambda,\bm{k}}e^{i\omega_{\lambda,\bm{k}}t}$, and 
%\begin{align}
$\Delta = \oml-\left(\omega_0-\DPS\right)$ %\label{eq:Delta} 
%\end{align}
is the detuning between the laser frequency and the polaron-shifted transition frequency $\tilde{\omega}_0=\omega_0-\DPS$, with the polaron shift $\DPS=\sum_{\lambda,\bm{k}} \arrowvert g_{\lambda,\bm{k}} \arrowvert^2/\omega_{\lambda,\bm{k}}$.

Note that $\tilde{H}_0$ has become trivial, with the coupling and free-phonon contribution removed by the polaron transformation and interaction picture, respectively. The price one pays for this simplification is, on the one hand, a trivial time dependence in the displacement operator and, on the other hand, a much more involved modification of the dissipator in Eq.~\eqref{eq:Lse-tilde}. The latter is indeed reminiscent of the Franck-Condon principle: spontaneous emission leaves the lattice in a state adapted to the excited electronic state (polaronic dressing), which is a displaced state with respect to the lattice ground state in the absence of charge excitation, hence the displacement operators accompanying the ground-state component in Eq~\eqref{eq:Lse-tilde}.  

Eqs.~\eqref{eq:gen-L0-tilde}--\eqref{eq:Lpd-tilde} yield simple closed equations for three components of a general operator (here $\tilde{O}_\mathrm{ph}$ is a phonon operator),
\begin{subequations}
\begin{align}
\tildesupL^{(0)}_{t,t'}\XG \tilde{O}_\mathrm{ph} 
& = e^{\left(i\Delta -\frac{\gamma}{2} \right)(t-t')} \XG \tilde{O}_\mathrm{ph},  
\label{eq:L0-solution-XG}\\
\tildesupL^{(0)}_{t,t'}\GX \tilde{O}_\mathrm{ph} 
& = e^{\left(-i\Delta -\frac{\gamma}{2} \right)(t-t')} \GX \tilde{O}_\mathrm{ph}, 
\label{eq:L0-solution-GX} \\
\tildesupL^{(0)}_{t,t'}\GG \tilde{O}_\mathrm{ph} & = \GG \tilde{O}_\mathrm{ph}, \label{eq:L0-solution-GG} 
\end{align}
where $\gamma = \gamma_1 + \gamma_2$.
The fourth equation is more involved
\begin{align}
\MoveEqLeft \tildesupL^{(0)}_{t,t'}\XX \tilde{O}_\mathrm{ph} 
 = e^{-\gamma_{1}(t-t')} \XX \tilde{O}_\mathrm{ph} \label{eq:L0-solution-XX}  \\
& + \gamma_{1} \GG \int_{t'}^t d\tau D^\dagger\{g_{\lambda,\bm{k}}(\tau)\}\tilde{O}_\mathrm{ph} 
D\{g_{\lambda,\bm{k}}(\tau)\}.
\nonumber
\end{align}
However, the last term does not contribute to the autocorrelation function.
\end{subequations}

At the same time, the generator of the optical coupling is transformed as
\begin{align} 
\tilde{L}_1(t)\tilde{O}(t) 
& = V(\tf) \TP \left(L_1(t) \TPconj V^\dagger(\ti) \tilde{O} V(\ti) \TP  \right) \TPconj V^\dagger(\tf) 
\nonumber \\
&= -\frac{i}{\hbar}\left[\tilde{H}_\mathrm{int}(t),\tilde{O}(t)\right], \label{eq:tilde_L1}
\end{align}
with
\begin{equation*}
\tilde{H}_\mathrm{int}(t) = -\frac{\hbar\Omega}{2}\left(\GX D^\dagger \{g_{\lambda,\bm{k}}(t)\} + \mathrm{h.c.} \right) . %\label{eq:tilde_H_int} 
\end{equation*}
These transformations preserve the form of the Born expansion, Eq.~(\ref{eq:L_total}),
\begin{align}
\tildesupL_{\tf,\ti} \approx & \tildesupL_{\tf,\ti}^{(0)} + \int_{\ti}^{\tf} dt_1 \tildesupL_{\tf,t_1}^{(0)} \tilde{L}_1(t_1) \tildesupL_{t_1,\ti}^{(0)} \label{eq:tilde_L_total} \\
& + \int_{\ti}^{\tf} dt_1 \int_{\ti}^{t_1} dt_2 \tildesupL_{\tf,t_1}^{(0)} \tilde{L}_1(t_1) \tildesupL_{t_1,t_2}^{(0)} \tilde{L}_1(t_2) \tildesupL_{t_2,\ti}^{(0)} +\ldots \nonumber
\end{align}

Finally, inverting the definition in Eq.~\eqref{eq:tilde_superoperator}, we express $\supL_\tau$ by $\tildesupL_{\tau,0}$ in Eq.~\eqref{eq:autocorrelation_1} and write the autocorrelation function in terms of transformed quantities,
\begin{align}
G(\tau) & = \tr_\mathrm{ph} \left\langle X \left|\TPconj V^{\dag}(\tau) 
\right. \right. \label{eq:autocorrelation_3} \\
&\quad \times \left. \left. \left(\tilde{A}^\mathrm{(I)}(\tau)+\tilde{A}^\mathrm{(II)}(\tau)\right) 
V(\tau)\TP \right| G \right\rangle     \nonumber \\
& = \tr_\mathrm{ph} \left[D^\dagger\{g_{\bm{k}}(\tau)\} \left(\tilde{A}^{(\mathrm{I})}_{XG}(\tau)+\tilde{A}^{(\mathrm{II})}_{XG}(\tau)\right)\right], \nonumber
\end{align}
where we define two contributions for further reference,
\begin{subequations} \label{eq:A_I_II_tau}
\begin{align}
\tilde{A}^{(\mathrm{I})}(\tau) = & \tildesupL_{\tau.0}\XG \tilde{\rho}^{\mathrm{S}}_{XX}(0) D\{g_{\lambda,\bm{k}}\}, 
\label{eq:tilde_A_I} \\
\tilde{A}^{(\mathrm{II})}(\tau) = & \tildesupL_{\tau,0}\GG \tilde{\rho}^{\mathrm{S}}_{GX}(0) D\{g_{\lambda,\bm{k}}\}. \label{eq:tilde_A_II}
\end{align}    
\end{subequations}
In the last step of Eq.~\eqref{eq:autocorrelation_3} we use the explicit form of the unitary operators $\TP$ and $V$ given by Eqs.~\eqref{eq:TP} and ~\eqref{eq:V_t}, respectively, apply the identity $D^\dagger\{g_{\lambda,\bm{k}}\}V^\dagger (\tau) = V^\dagger (\tau) D^\dagger\{g_{\lambda,\bm{k}}(\tau)\}$, following directly from Eq.~\eqref{eq:Dt}, and then take advantage of the cyclic property of the trace. Note that, at this point, the autocorrelation function is expressed in terms of phonon operators only. 

\subsection{Steady state in the polaron picture}\label{sec:steady-state}

The next step is to find the relevant XX and GX components of the steady state density matrix $\tilde{\rho}^\mathrm{S}$, which appear in Eqs.~\eqref{eq:A_I_II_tau}. We find the steady state by formally propagating the initial state of the system from the far past. This is done up to the required second order in $\Omega$ using the Born expansion given by Eq.~\eqref{eq:tilde_L_total}. Note that terms off-diagonal in the emitter basis (emitter coherences) appear only in odd orders, while the diagonal ones (emitter occupations) contribute only in even orders. 

In zeroth order, i.e., without optical driving, the steady state of the system is the emitter ground state and the thermal equilibrium state of the phonon reservoir,
\begin{equation*}
\tilde{\rho}^{(\mathrm{S,0})}(t) = \GG \rho_{\mathrm{T}}.
\end{equation*}
This state is obviously invariant under the unperturbed evolution $\tildesupL^{(0)}$.

We write the first-order contribution, using the relevant term of Eq.~\eqref{eq:tilde_L_total}, in the form 
\begin{equation*}
\tilde{\rho}^{(\mathrm{S},1)}(t)=
\int_{-\infty}^{t} dt_1 \tildesupL_{t,t_1}^{(0)} \tilde{L}_1(t_1) \tilde{\rho}^{(\mathrm{S},0)}(t_1).
\end{equation*}
Then, we explicitly apply Eq.~\eqref{eq:tilde_L1}, Eq.~\eqref{eq:L0-solution-XG}, and Eq.~\eqref{eq:L0-solution-GX}, which yields
\begin{align*}
\MoveEqLeft{\tilde{\rho}_{GX}^{(\mathrm{S},1)}(t)
=  \tilde{\rho}_{XG}^{(\mathrm{S},1)*}(t)} \\ %\label{eq:tilde_rhoS_GX} \\
& = -\frac{i\Omega}{2}\int_{-\infty}^t dt_1 e^{-\left(\frac{\gamma}{2}+i\Delta\right)(t-t_1)} \rho_\mathrm{T}D^{\dagger}\left\{g_{\lambda,\bm{k}}(t_1)\right\}, %\nonumber
\end{align*}
while $\tilde{\rho}_{GG}^{(\mathrm{S},1)}(t) =  \tilde{\rho}_{XX}^{(\mathrm{S},1)}(t) = 0$.

The second-order term from Eq.~\eqref{eq:tilde_L_total}, is
\begin{equation*}
\tilde{\rho}^{(\mathrm{S},2)}(t)=
\int_{-\infty}^{t} dt_2 \tildesupL_{t,t_2}^{(0)} \tilde{L}_1(t_2) \tilde{\rho}^{(\mathrm{S},1)}(t_2).
\end{equation*}
Again, we apply Eq.~\eqref{eq:tilde_L1} explicitly, which leads to terms proportional to $\GG$ and $\XX$.  According to Eq.~\eqref{eq:tilde_A_I}, only $\tilde{\rho}^{\mathrm{S}}_{XX}$ is needed for the correlation function up to the second order. This is given by Eq.~\eqref{eq:L0-solution-XX}, where only the simple first term contributes. Altogether, we get
\begin{align*}
\tilde{\rho}_{XX}^{(\mathrm{S},2)}(t) & = \frac{\Omega^2}{4} 
\int_{-\infty}^{t}dt_1 e^{-\gamma_{1} (t-t_1)} \int_{-\infty}^{t_1}dt_2 
e^{-\left(\frac{\gamma}{2}+i\Delta\right)(t_1-t_2)} \\
& \quad \times D\left\{g_{\lambda,\bm{k}}(t_1)\right\} \rho_\mathrm{T} D^\dagger\left\{g_{\lambda,\bm{k}}(t_2)\right\} + \mathrm{h.c.},  
\end{align*}
which, after interchanging the variables in the ``h.c.'' term, can be reduced to 
\begin{align*}
\tilde{\rho}_{XX}^{(\mathrm{S},2)}(t) & = 
\frac{\Omega^2}{4} \int_{-\infty}^{t}dt_1 
e^{-\left(\frac{\gamma_{1}}{2}-i\Delta\right) (t-t_1)} \\
& \quad\times \int_{-\infty}^{t}dt_2 
e^{-\left(\frac{\gamma_{1}}{2}+i\Delta\right)(t-t_2)} \\ %\nonumber \\
& \quad \times e^{-\frac{\gamma_{2}}{2}|t_1-t_2|}D\left\{g_{\lambda,\bm{k}}(t_1)\right\} \rho_\mathrm{T} D^\dagger\left\{g_{\lambda,\bm{k}}(t_2)\right\}.  %\label{eq:tilde_rhoS_XX}
\end{align*}

\subsection{Autocorrelation function}
\label{sec:G}

The following evaluation of Eq.~\eqref{eq:autocorrelation_3} proceeds in two steps: First, we calculate the quantities $\tilde{A}_{XG}^{(\mathrm{I,II)}}$ to the second order in $\Omega$, and then we average over the phonon reservoir. 

In Eq.~\eqref{eq:tilde_A_I}, $\tilde{\rho}^{\mathrm{S}}_{XX}$ is already in the second order, so $\tildesupL_{\tau,0}$ must be taken in the zeroth order. This is given again by Eq.~\eqref{eq:L0-solution-XG}, which immediately yields
\begin{align}
\tilde{A}_{XG}^{(\mathrm{I})}(\tau) & = \frac{\Omega^2}{4} e^{-\frac{\gamma_{2}}{2}\tau} \int_{-\infty}^{0}dt_1 
e^{-\left(\frac{\gamma_{1}}{2}-i\Delta\right)(\tau - t_1)} \label{eq:tilde_A_XG_I} \\
& \quad \times \int_{-\infty}^{0} dt_2 e^{\left(\frac{\gamma_{1}}{2}+i\Delta\right)t_2} e^{-\frac{\gamma_{2}}{2}|t_1-t_2|}
\nonumber  \\
& \quad \times  D\left\{g_{\lambda,\bm{k}}(t_1)\right\} \rho_\mathrm{T} D^\dagger\left\{g_{\lambda,\bm{k}}(t_2)\right\} D\{g_{\lambda,\bm{k}}(0)\} . \nonumber
\end{align}

Eq.~\eqref{eq:tilde_A_II} contains $\tilde{\rho}^{\mathrm{S}}_{GX}$, which is of the first order in $\Omega$. Further evolution must, therefore, be evaluated to the first order. Since the operator proportional to $\GG$ is invariant under $\tildesupL^{(0)}$ [see Eq.~\eqref{eq:L0-solution-GG}], applying the first-order term from Eq.~\eqref{eq:tilde_L_total} again amounts to explicitly applying $L_1$ and propagating the resulting off-diagonal terms via 
Eq.~\eqref{eq:L0-solution-XG}. This yields
\begin{align}
\tilde{A}_{XG}^{\mathrm{(II)}}(\tau) & = \frac{\Omega^2}{4} e^{-\frac{\gamma_{2}}{2}\tau}
\int_0^\tau dt_1 
e^{-\left(\frac{\gamma_{1}}{2}-i\Delta\right)(\tau - t_1)} \label{eq:tilde_A_XG_II} \\
& \quad \times \int_{-\infty}^0 dt_2 e^{\left(\frac{\gamma_{1}}{2}+i\Delta\right)t_2}
e^{\frac{\gamma_{2}}{2}(t_1+t_2)}
\nonumber \\
& \quad \times  D\{g_{\lambda,\bm{k}}(t_1)\} \rho_\mathrm{T} D^\dagger \{g_{\lambda,\bm{k}}(t_2)\} D\{g_{\lambda,\bm{k}}(0)\}. \nonumber
\end{align}

When substituting Eqs.~(\ref{eq:tilde_A_XG_I}) and (\ref{eq:tilde_A_XG_II}) to Eq.~(\ref{eq:autocorrelation_3}), we extend the integration limits for $t_1$ in Eq.~(\ref{eq:tilde_A_XG_II}) to $(-\infty,\tau)$ and subtract the surplus from Eq.~(\ref{eq:tilde_A_XG_I}), which yields the correlation function composed of two parts,
\begin{align}
G_1(\tau) & =  \frac{\Omega^2}{4} \int_0^{\infty} du \int_0^{\infty} du' e^{-\frac{\gamma}{2}(u+u')+i\Delta(u-u')}    \label{eq:autocorrelation_5_1} \\
& \quad \times C(\tau,u,u'), \nonumber
\end{align}
and
\begin{align}
G_2(\tau) & =  \frac{\Omega^2}{4} 
e^{(-\gamma/2 + i\Delta)\tau}
\int_0^{\infty} du \int_0^{\infty} du' e^{-\frac{\gamma}{2}(u+u')+i\Delta(u-u')}    \label{eq:autocorrelation_5_2} \\
& \quad \times  \left[ e^{\gamma_2 \min(u,u')} -1\right]
C(\tau,\tau+u,u'), \nonumber
\end{align}
where we change the variables to $u=-t_1$, $u'=\tau-t_2$ and $u=-t_1$, $u'=-t_2$ in Eq.~\eqref{eq:autocorrelation_5_1} and Eq.~\eqref{eq:autocorrelation_5_2}, respectively, and we define the phonon correlation function 
\begin{align}
C(\tau,u,u') & =
\left\langle D^\dagger\{g_{\lambda,\bm{k}}(-u')\}D\{g_{\lambda,\bm{k}}(0)\} 
\right. \label{eq:C} \\
&\left.\quad \times D^\dagger\{g_{\lambda,\bm{k}}(\tau)\}D\{g_{\lambda,\bm{k}}(\tau-u)\} \right\rangle, \nonumber  
\end{align}
with angular brackets denoting the thermal average, 
$\langle O \rangle = \tr_{\mathrm{ph}}\rho_\mathrm{T} O$. 
Note that the contribution $G_2(\tau)$ in Eq.~\eqref{eq:autocorrelation_5_2} is nonzero only in the presence of additional pure dephasing. 

To calculate the correlation function $C(\tau,u,u')$ we first iterate twice the Baker-Campbell-Hausdorff formula 
\begin{equation*}
e^A e^B = e^{A+B}e^{\frac{1}{2}[A,B]}\;
\mathrm{for}\; [A,B]=c\mathbb{I},\; c\in \mathbb{C}
\end{equation*}
to pairs of displacement operators in Eq.~\eqref{eq:C}, which yields
\begin{align}
& C (\tau,u,u') =  \label{eq:phonon_correlation_function} \\
& = \left\langle D \{-g_{\lambda,\bm{k}}(-u')+g_{\lambda,\bm{k}}(0)-g_{\lambda,\bm{k}}(\tau)+g_{\lambda,\bm{k}}(\tau-u) \} \right\rangle \nonumber \\
& \quad \times \exp \left\{
i\im \sum_{\lambda,\bm{k}} \left|\frac{g_{\lambda,\bm{k}}}{\omega_{\lambda,\bm{k}}}\right|^2
\left[
e^{i\omega_{\lambda,\bm{k}}u'}+e^{-i\omega_{\lambda,\bm{k}}u} \right. \right.\nonumber \\
& \quad \left. \left. 
+ e^{i\omega_{\lambda,\bm{k}}\tau}\left(1-e^{i\omega_{\lambda,\bm{k}}u'} \right)
\left(1-e^{-i\omega_{\lambda,\bm{k}}u} \right)
\right] \right\}. \nonumber 
\end{align}
The average is then calculated according to the relation \cite{Mahan2000}
\begin{equation*}
\left\langle D\left(G_{\lambda,\bm{k}}\right) \right\rangle = 
e^{-\sum_{\lambda,\bm{k}}\left(n_{\lambda,\bm{k}}+\frac{1}{2} \right)
\left| \frac{G_{\lambda,\bm{k}}}{\omega_{\lambda,\bm{k}}} \right|^2},
\end{equation*}
which for $G_{\lambda,\bm{k}} 
= -g_{\lambda,\bm{k}}(-u')+g_{\lambda,\bm{k}}(0)-g_{\lambda,\bm{k}}(\tau)+g_{\lambda,\bm{k}}(\tau-u)$, as in Eq.~\eqref{eq:phonon_correlation_function}, yields
\begin{align*}
\MoveEqLeft \left\langle 
D \{-g_{\lambda,\bm{k}}(-u')+g_{\lambda,\bm{k}}(0)-g_{\lambda,\bm{k}}(\tau)+g_{\lambda,\bm{k}}(\tau-u) \} \right\rangle \nonumber \\
& = \exp \Bigg[-\sum_{\lambda,\bm{k}}\left(n_{\lambda,\bm{k}} + \frac{1}{2}\right) \left|\frac{g_{\lambda,\bm{k}}}{\omega_{\lambda,\bm{k}}}\right|^2 \\
% \label{eq:D_av}  \\
& \quad  \times
\left|1 - e^{-i\omega_{\lambda,\bm{k}}u'} - e^{i\omega_{\lambda,\bm{k}}\tau} + e^{i\omega_{\lambda,\bm{k}}(\tau - u)}\right|^2 
\vphantom{\left|\frac{g_{\lambda,\bm{k}}}{\omega_{\lambda,\bm{k}}}\right|^2}
\Bigg]. \\ \nonumber
\end{align*}
Rearranging the terms and invoking Eq.~\eqref{eq:Phi_x_quantum}, Eq.~\eqref{eq:C} then takes the form
\begin{align}
\MoveEqLeft C(\tau,u,u') = \nonumber \\
& \exp \left[-2\phi(0)+\phi(u)+\phi(-u')+\phi(-\tau)\right. \label{eq:C_3} \\
&\quad \left. +\phi(u-\tau-u')-\phi(u-\tau)-\phi(-u'-\tau) \right].  \nonumber
\end{align}
As expected, the optical properties of a system interacting with the phonon bath in a thermal state (which is Gaussian) are fully characterized by the bath spectral density or, equivalently, its memory function. 

It is interesting to note that the imaginary parts of the memory functions in Eq.~\eqref{eq:C_3} stem from the phase factor in Eq.~\eqref{eq:phonon_correlation_function}, which is a consequence of the non-commutativity of the phonon operators in the BCH formula. This means that we have captured the \textit{quantumness} of the noise originating from the bosonic reservoir in the most fundamental sense. The corresponding classical model would be defined by the Hamiltonian in Eq.~\eqref{eq:H_0} in the interaction picture with respect to phonons and with $b_{\lambda,\bm{k}}, b_{\lambda,\bm{k}}^\dag$ being a conjugate pair of complex random variables. The same can be written in a more intuitive manner by defining quadrature amplitudes 
%\begin{subequations} \label{eq:X}
\begin{align}
X_{1,\lambda,\bm{k}} & = \frac{1}{2} \left(b_{\lambda,\bm{k}} e^{-i\theta_{\lambda,\bm{k}}} + b_{\lambda,\bm{k}}^{\dag} e^{i\theta_{\lambda,\bm{k}}}\right) , \\ %\label{eq:X_1} \\
X_{2,\lambda,\bm{k}} & =  \frac{1}{2i} \left(b_{\lambda,\bm{k}} e^{-i\theta_{\lambda,\bm{k}}} - b_{\lambda,\bm{k}}^{\dag} e^{i\theta_{\lambda,\bm{k}}}\right), %\label{eq:X_2}
\end{align}
%\end{subequations}
where we have decomposed $g_{\lambda,\bm{k}} = | g_{\lambda,\bm{k}} | e^{i\theta_{\lambda,\bm{k}}}$. Then the Hamiltonian with classical noise is
\begin{equation} \label{eq:H_0_cl} 
H'_0 =  \hbar\left[\omega_0-\oml\right]\XX 
+ \hbar \Delta\omega(t) \XX,
\end{equation}
where 
\begin{equation*}
\Delta\omega(t) = \sum_{\lambda,\bm{k}} 2\left| g_{\lambda,\bm{k}}\right|
\left[X_{1,\lambda,\bm{k}} \cos(\omega_{\lambda,\bm{k}} t) + X_{2,\lambda,\bm{k}}\sin(\omega_{\lambda,\bm{k}} t) \right]
\end{equation*}
describes classical random fluctuations of the energy which are due to the coupling to the environment. For the classical model to be equivalent to the quantum one, we need to require that $X_{1,\lambda,\bm{k}},X_{2,\lambda,\bm{k}}$ are independent, equally distributed Gaussian random variables with zero mean and with variances
\begin{equation*}
\left\langle X_{1,\lambda,\bm{k}}^2 \right\rangle 
= \left\langle X_{2,\lambda,\bm{k}}^2 \right\rangle
=\frac{1}{4}\left( 2n_{\lambda,\bm{k}}+1 \right),
\end{equation*}
matching those of the quantum quadrature operators in the thermal state. These variances are somewhat artificial as they do not follow from the most obvious physical model of classical oscillators, which would result in classical statistics and energy equipartition. We opt for this definition because we want to separate the consequences of the fundamental non-commutativity of quantum noise operators from the differences between the quantum and classical statistics, which are, of course, also fundamental but much more obvious. With this choice, the classical memory function $\phi'(t)$ is consistent with the quantum one, with the non-commutativity-related imaginary part discarded, i.e.,
$\phi'(t) = \re \phi(t)$.

The result for the classical (commutative) case can alternatively be obtained by directly computing the autocorrelation function from the evolution driven by the parametrically time-dependent Hamiltonian in Eq.~\eqref{eq:H_0_cl} and then averaging over the random variables $X_{1,\lambda,\bm{k}},X_{2,\lambda,\bm{k}}$, which yields
\begin{equation*}
C(\tau,u,u') = \left\langle e^{i\Phi(0,-u')-i\Phi(\tau,\tau-u)} \right\rangle , 
\end{equation*}
where 
\begin{equation*}
\Phi(t_1,t_2) = \int_{t_1}^{t_2} dt \Delta\omega(t). 
\end{equation*}
This links the present result to our previous studies of systems subjected to classical noise \cite{Bogaczewicz2023,Bogaczewicz2023}, where the noise affects the optical response via the phase accumulated due to random fluctuations of the transition energy.

\subsection{\label{sec:RF_spectrum} RF spectrum}

%The analysis of the RF spectrum and its intensity will contain results relative to the characteristic magnitude of $S_0 = \Omega^2/(\gamma_{1}^2 \omc)$ and $I_0 = \Omega^2/\gamma_{1}^2$, respectively. We start from the case of no pure dephasing ($\gamma_{2}=0$) nor detuning blur ($\sigma_\Delta=0$). Then, 

The structure of the RF spectrum is revealed when we decompose the phonon correlation function in Eq.~(\ref{eq:C_3}) into contributions according to how they depend on $\tau$, which  determines the spectral properties according to the Fourier transform in Eq.~\eqref{eq:RF_spectrum1}.
For the one contributing via Eq.~\eqref{eq:autocorrelation_5_1} we use the decomposition
$C(\tau,u,u') = C_\mathrm{el}(u,u')+C_\mathrm{psb1}(\tau,u,u')+C_\mathrm{in1}(\tau,u,u')$,
where
\begin{subequations} %\label{eq:C1_split}
\begin{align}
C_\mathrm{el}(u,u') 
& = e^{-2\phi(0)+\phi(u)+\phi(-u')} , \label{eq:C_el} \\
C_\mathrm{psb1}(\tau,u,u') & = e^{-2\phi(0)+\phi(u)+\phi(-u')}\left[e^{\phi(-\tau)}-1\right] , \label{eq:C_ph1} \\
C_\mathrm{in1}(\tau,u,u') & = e^{-2\phi(0)+\phi(u)+\phi(-u')+\phi(-\tau)} \label{eq:C_in1} \\
& \quad \times \left[e^{\phi(u-\tau-u')-\phi(u-\tau)-\phi(-u'-\tau)} - 1\right]. \nonumber
\end{align}
\end{subequations}
Similarly, we decompose the function entering Eq.~\eqref{eq:autocorrelation_5_2} as $C(\tau,\tau+u,u') = C_\mathrm{in2}(u,u')+C_\mathrm{psb2}(\tau,u,u')+C_\mathrm{psb3}(\tau,u,u')$, with
\begin{subequations} %\label{eq:C2_split}
\begin{align}
C_\mathrm{in2}(u,u') 
& = e^{-2\phi(0)+\phi(-u')+\phi(u-u')-\phi(u)} , \label{eq:C_in2} \\
C_\mathrm{psb2}(\tau,u,u') & = e^{-2\phi(0)+\phi(-u')+\phi(u-u')-\phi(u)} \\
& \quad \times e^{\phi(\tau+u)-\phi(-\tau-u')}\left[e^{\phi(-\tau)}-1\right] , \label{eq:C_ph2} \\
C_\mathrm{psb3}(\tau,u,u') 
& = e^{-2\phi(0)+\phi(-u')+\phi(u-u')-\phi(u)} \label{eq:C_ph3} \\
& \quad \times \left[e^{\phi(\tau+u)-\phi(-u'-\tau)} - 1\right]. \nonumber
\end{align}
\end{subequations}
These contributions lead to the corresponding decomposition of the autocorrelation functions via Eq.~\eqref{eq:autocorrelation_5_1} and Eq.~\eqref{eq:autocorrelation_5_2}, and of the RF spectrum, following Eq.~\eqref{eq:RF_spectrum1}. 

The first component, $C_\mathrm{el}(u,u')$ in Eq.~(\ref{eq:C_el}), is $\tau$-independent and is responsible for an unbroadened central line in the spectrum, $S_\mathrm{el}(\omega)$, located at the laser frequency (which is zero frequency in our rotating frame). This line corresponds to elastic scattering of photons, reproducing a standard result for a weakly excited system \cite{ScullyZubairy1997}. The intensity (spectral weight) of this line is obtained by substituting $C_\mathrm{el}(u,u')$ into Eq.~\eqref{eq:autocorrelation_5_1} and then to Eq.~\eqref{eq:Intensity},
\begin{equation*}
I_\mathrm{el} = \frac{\pi\Omega^2}{4} | \xi(\gamma,\Delta) |^2 e^{-2\phi(0)}, %\label{eq:I_el}
\end{equation*}
where 
\begin{equation}\label{eq:xi}
\xi(\gamma,\Delta) = \int_0^\infty ds \zeta(\gamma,\Delta;s) 
\end{equation}
with 
\begin{equation*}
\zeta(\gamma,\Delta;s) = e^{-\left(\gamma/2-i\Delta \right) s + \phi(s)}
\end{equation*}

The contributions $C_\mathrm{psb1}(\tau,u,u')$ in Eq.~(\ref{eq:C_ph1}) and $C_\mathrm{psb2}(\tau,u,u')$ in Eq.~(\ref{eq:C_ph2}) are short-lived as a function of $\tau$, because $\phi(-\tau)\to 0$ as $\tau\gg 1/\omega_\mathrm{c}$, which is a picosecond time scale for a QD and even shorter for atomically localized defects. Therefore, they give rise to a broad spectral feature $S_\mathrm{psb}\left(\omega\right)$, which is the phonon sideband predicted for RF in the pulsed-excitation regime \cite{Ahn2005} and also known from other types of spectroscopy on zero-dimensional structures. Here, the low-frequency (red-shifted) part corresponds to energy loss at photon scattering, which means phonon emission, while the high-frequency (blue-shifted) side reflects energy gain or phonon absorption. 

Using $C_\mathrm{psb1}(\tau,u,u')$ in Eq.~\eqref{eq:autocorrelation_5_1} and substituting this into Eq.~\eqref{eq:RF_spectrum1}, we get
\begin{equation*} %\label{eq:S_ph} 
S_\mathrm{psb1}(\omega) = \frac{I_\mathrm{el}}{2\pi} \int_{-\infty}^{\infty} d\tau e^{-i\left(\omega-\oml\right)\tau} \left[e^{\phi(\tau)}-1\right],
\end{equation*}
where we used the symmetry $\phi^*(\tau)=\phi(-\tau)$.
The intensity is, from Eq.~\eqref{eq:Intensity},
\begin{equation*}
I_\mathrm{psb1} = I_\mathrm{el} \left(e^{\phi(0)}-1\right) . %\label{eq:I_ph1} 
\end{equation*}
Note that with our (somewhat arbitrary) decomposition, the shape of the phonon sideband given by this contribution is the same as in the absorption spectrum \cite{Krummheuer2002}. The total weight of the sideband is also in the same relation to the intensity of the central line. However, in the present case both these features attain a scaling factor that reflects the dependence of the total scattering intensity on the spectral position of the laser with respect to the fundamental transition and phonon sideband, contained in the factor $|\xi(\gamma,\Delta)|^2$.

The intensity of the component stemming from $C_\mathrm{psb2}(\tau,u,u')$ in Eq.~\eqref{eq:C_ph2} can be calculated in a similar way, yielding 
\begin{equation*}
I_\mathrm{psb2} = I_\mathrm{psb1} 
\frac{2\gamma_2\re \xi(\gamma,\Delta)}{\gamma^2|\xi(\gamma,\Delta)|^2} . %\label{eq:I_ph2} 
\end{equation*}
Since $\xi(\gamma,\Delta)$ is on the order of $1/\gamma$, this is reduced by a factor of $\gamma_2/\gamma$ compared to $I_\mathrm{psb1}$. Here we will assume $\gamma_2 \ll \gamma$, so that this term is a small correction to the previous one. 

The correlation function $C_\mathrm{psb3}(\tau,u,u')$ is not only short-lived in $\tau$ but also takes substantial values only when $u \lesssim \tau_{\mathrm{mem}}$ or $u'\lesssim \tau_{\mathrm{mem}}$. This restricts the integration to narrow ``tubes'' around the $u$ and $u'$ axes in the $(\tau,u,u')$ space, which together with the factor $[\exp(\gamma_2\min(u,u'))-1]$ in Eq.~\eqref{eq:autocorrelation_5_2} yields further scaling down by the very small factor $(\gamma\tau_{\mathrm{mem}})^2$, compared to $I_\mathrm{psb2}$. The spectrum in the broad frequency range (the phonon sidebands) is therefore determined mainly by the contribution from Eq.~\eqref{eq:C_ph1}.

The two remaining terms, given by Eq.~\eqref{eq:C_in1} and Eq.~\eqref{eq:C_in2} have very different character and lead to interesting physical features. 
The contribution of Eq.~\eqref{eq:C_in1} yields
\begin{align} \label{eq:S_in1} 
S_\mathrm{in1}(\omega) = & \frac{\Omega^2}{4} e^{-2\phi(0)} \re \int_{0}^{\infty} d\tau e^{i(\omega-\oml)\tau+\phi(-\tau)} \\
& \times \int_0^\infty du \int_0^\infty du' 
\zeta(\gamma,\Delta;u) \zeta(\gamma,\Delta;u') \nonumber \\
& \times \left[e^{\phi(u-\tau-u')-\phi(u-\tau)-\phi(-u'-\tau)}-1\right] \nonumber
\end{align}
To study its properties, we focus on the expression in the final square bracket for $\tau\gg 1/\omega_{\mathrm{c}}$ and $u,u'>0$ (cf. the limits of integration in Eq.~\eqref{eq:autocorrelation_5_1}). 
Then $\phi(-\tau-u')\ll 1$ but the first two memory functions in the exponent have non-zero values for arbitrary $\tau$, along the lines $u=\tau$ and $u=\tau+u'$. This term is therefore long-lived, up to the exponential cutoff at $\tau \sim 1/\gamma$. Thus, it will lead to a narrow spectral feature of width $\Delta\omega\sim \gamma$. 
Such a feature does not appear in absorption or emission and is therefore unique for light scattering spectroscopy. It has not been predicted in the previous works on resonance fluorescence of systems coupled to phonons \cite{Ahn2005}, which we attribute to the inherently low spectral resolution in that study, where pulsed excitation was assumed. 
The narrow feature is the quantum counterpart of the inelastic scattering line that emerges in the scattering spectrum for a system perturbed by classical noise \cite{Bogaczewicz2023,Bogaczewicz2025}.

The contribution from Eq.~\eqref{eq:C_in2} has a much simpler form since the correlation function $C_{\mathrm{in2}}$ does not depend on $\tau$. When substituted to Eq.~\eqref{eq:autocorrelation_5_2}, this simply yields a Lorentzian centered at $\omega = \oml - \Delta = \omega_0-\Delta_{\mathrm{PS}}$, which is the polaron-shifted transition frequency. This is a feature known from general noise studies \cite{Bogaczewicz2023,Bogaczewicz2025} that can be attributed to noise-induced transient dynamics of the interband dipole.

To gain more detailed insight into the spectrum in the narrow vicinity of the excitation frequency, we first note that the broad sideband does not vary much in the relevant very narrow range of frequencies and provides an essentially constant background (``pedestal''), which can be approximated by 
\begin{equation*}%\label{eq:Sw0}
S_{\mathrm{psb}}(\omega) \approx S_{\mathrm{psb1}}(\oml)
= \frac{I_{\mathrm{el}}}{2\pi}
\int_{-\infty}^{\infty} d\tau \left[e^{\phi(\tau)}-1\right]. 
\end{equation*}
On this background, the narrow spectral features described by Eq.~(\ref{eq:C_in1}) and Eq.~(\ref{eq:C_in2}) develop. 

We will start the discussion with the phonon-induced feature from Eq.~(\ref{eq:C_in1}), which does not depend on the presence of additional pure dephasing.
Here, the expression in square brackets is nonzero only in narrow ranges of $u$ around $\tau$ and $\tau+u'$ because the memory function is short-lived compared to the effective span of the integral over $u$, which is $1/\gamma$. Moreover, except for a negligibly narrow range of $u'$ near 0, these two parameter areas are disjoint so that they contribute additively, and one may write 
\begin{align*} %\label{eq:exp_approximation}
\MoveEqLeft e^{\phi(u-\tau-u')-\phi(u-\tau)-\phi(-\tau-u')} - 1 \\
& \approx (A+B) \delta(u-\tau-u') - (A-B) \delta(u-\tau),
\end{align*}
where
\begin{equation}\label{eq:AB}
A = \int_{-\infty}^{\infty}ds \sinh \phi(s),\;
B = \int_{-\infty}^{\infty}ds [\cosh \phi(s) -1].
\end{equation}
We neglect also the remaining memory functions in Eq.~(\ref{eq:C_in1}), which are non-zero only in a very narrow range around zero time and therefore yield only a minor correction to the overall value of the spectral function at low frequencies. The integrations in Eqs.~\eqref{eq:autocorrelation_5_1} and ~\eqref{eq:RF_spectrum1} then become trivial. Including the PSB ``pedestal'', the result is
\begin{align}\label{eq:Fano}
\MoveEqLeft S_{\mathrm{in1}}(\omega) + S_{\mathrm{psb}}(\oml) =  \\
&\frac{I_{\mathrm{el}}}{2\pi} 
\left\{A \frac{(\omega-\omega_{\mathrm{L}})^2}{(\gamma/2)^2 + (\omega-\tilde{\omega}_{0})^2} \right. \nonumber \\
& \quad + \left. B \left[ 2+
\frac{\gamma^2/2 -(\omega-\omega_{\mathrm{L}})^2+2\Delta^2}{(\gamma/2)^2 + (\omega-\tilde{\omega}_{0})^2}
\right] \right\}, \nonumber
\end{align}
where we have approximated $\zeta(\gamma_{1},\Delta;s)\approx \exp[-(\gamma_{1}/2-i\Delta)s]$ and neglected $\phi(-\tau)$ in Eq.~(\ref{eq:S_in1}), which is consistent with the previous approximations. Note that the width of these spectral features is always determined by the lifetime of the excited state and depends neither on the strength of the phonon couplings nor on the temperature.

In the leading order in the phonon coupling ($\sim |g_{\lambda,\bm{k}}|^2$ or linear in $\phi(t)$), one has $B=0$ and
\begin{equation}\label{eq:A}
A = \int_{-\infty}^{\infty}ds \phi(s) = \frac{2\pi k_{\mathrm{B}}T}{\hbar} 
\lim_{\omega\to 0} \frac{J(\omega)}{\omega^3},
\end{equation}
which is non-zero and finite for the actual super-Ohmic carrier-phonon spectral density. Simultaneously, the neglected memory functions in the exponents of Eq.~(\ref{eq:C_in1}) lead to higher-order corrections, hence Eq.~\eqref{eq:Fano} becomes the rigorous lowest-order result. The form of the spectrum in this limit therefore has a Fano shape,
\begin{equation*}%\label{eq:Fano-approx}
S_{\mathrm{in1}}(\omega) + S_{\mathrm{psb}}(\oml) \propto 
 \frac{(q\gamma/2 + \omega-\tilde{\omega}_{0})^2}{(\gamma/2)^2 + (\omega-\tilde{\omega}_{0})^2},
\end{equation*}
with the Fano factor $q = -2\Delta/\gamma$. It follows that for a weak phonon coupling, the inelastic scattering is completely suppressed at the laser frequency, while an approximately Lorentzian peak develops at the polaron-shifted transition frequency $\tilde{\omega}_0$ for $\Delta\gtrsim \gamma$. In the weak phonon coupling regime, the magnitude of the spectrum in the low-frequency range scales linearly with temperature. This limit corresponds to the single-phonon scattering regime.

For arbitrary phonon couplings, one finds the low-temperature asymptotics of A again given by Eq.~\eqref{eq:A}, up to corrections on the order of $T^3$ and higher, while 
\begin{equation*}
B = \pi\eta \left( \frac{k_{\mathrm{B}}T}{\hbar}\right)^3 
\left[\lim_{\omega\to 0} \frac{J(\omega)}{\omega^3}\right]^2,
\end{equation*}
up to corrections $\propto T^5$, where $\eta= 2\pi^2/3$,
%(see Appendix for the proof),
which means that at sufficiently low temperatures the spectrum is dominated by the Fano component. At high temperatures, the integrands in Eq.~\eqref{eq:AB} are dominated by the largest values of $\phi(t)\gg 1$, when 
$\sinh\phi(t)\approx\cosh\phi(t)\approx \exp[\phi(t)]/2$, so that $A\approx B$ and Eq.~\eqref{eq:Fano} yields a simple Lorentzian on a flat ``pedestal''. The coincidence of the low-temperature asymptotics with the single-phonon limit allows us to interpret the low-temperature Fano-like spectrum as resulting predominantly from single-phonon processes, while the transition to the Lorentzian shape is due to the growing contribution from multiple-phonon scattering. 

The intensity of the inelastic part of the RF spectrum stemming from the contribution in Eq.~(\ref{eq:C_in1}) is again obtained directly from Eq.~\eqref{eq:Intensity},
\begin{equation*}
I_\mathrm{in1} = I_\mathrm{el} e^{\phi(0)} \left(\frac{2\re\xi}{\gamma|\xi(\gamma,\Delta)}|^2-1\right) . %\label{eq:I_in}
\end{equation*}
In general, this includes not only the narrow feature discussed above but also a broad correction to the phonon sideband, which becomes important at strong phonon coupling. 

%Unlike the phonon sidebands, to which weak additional pure dephasing brings only a small quantitative correction, the narrow central feature is critically affected by this dephasing even when it is relatively small ($\gamma_2\ll\gamma$). Pure dephasing is equivalent to phase noise, which gives rise to an inelastic Lorentzian feature at the spectral position of the optical transition, reflecting the noise-induced transient dynamics of the system \cite{Bogaczewicz2023,Bogaczewicz2025}. 
The Lorentzian feature stemming from pure dephasing [Eq.~\eqref{eq:C_in2}] appears exactly at the same position as the Fano resonance and will dominate it for a sufficiently strong dephasing. To see when it happens, we find the contribution of Eq.~\eqref{eq:C_in2} in the zeroth order in phonon coupling, $C_{\mathrm{in2}}(\tau,u,u')=1$, and for $\gamma_2 \ll \gamma$, which yields
\begin{equation*}
S_{\mathrm{in2}}(\omega) \approx 
\frac{\gamma_2/\gamma}{(\gamma/2)^2 + \Delta^2}\frac{\Omega^2}{4}
\frac{\gamma/2}{(\gamma/2)^2 + (\omega-\oml+\Delta)^2}.
\end{equation*}
One can consider the pure dephasing contribution ``strong'', or dominating, when the value of the resulting Lorentzian at the laser frequency exceeds the level of the PSB ``pedestal''. This roughly corresponds to suppression of the Fano dip. Thus, the Fano feature is visible as long as
\begin{equation*}
S_{\mathrm{in2}}(\oml) \lesssim S_{\mathrm{psb1}}(\oml),
\end{equation*}
which explicitely gives
\begin{equation}\label{eq:gamma2-crit}
\frac{\gamma_2}{(\gamma/2)^2+\Delta^2} \lesssim 
\frac{2\pi k_{\mathrm{B}}T}{\hbar}\lim_{\omega\to 0} \frac{J(\omega)}{\omega^3}.
\end{equation}

When inhomogeneous broadening is present, the spectral features bound to the transition energies are broadened and ultimately can even be smeared out. This affects the peak of the Fano profile, the dephasing-induced Lorentzian feature and, to a much lesser degree, the phonon sidebands, which are typically much broader than the inhomogeneous effect. On the other hand, features bound to the laser frequency remain fixed, although their magnitude may change. Therefore, the central elastic line and the Fano dip around the laser frequency are not affected by inhomogeneous broadening.

Finally, we analyze the impact of phonons on the spectral response from another point of view: the total scattering intensity as a function of excitation frequency. 
The total intensity of the scattered light, $I_\mathrm{tot}$, neglecting the contributions induced by pure dephasing, is 
\begin{equation}
I_\mathrm{tot} = I_\mathrm{el} + I_\mathrm{psb} + I_\mathrm{in} =
 I_{\mathrm{el}} e^{-\phi(0)} \frac{\gamma}{2} \re \xi(\gamma,\Delta) . \label{eq:I_tot}
\end{equation} 
The parameter $\xi(\gamma,\Delta)$, defined in Eq.~\eqref{eq:xi}, governs not only the total scattering intensity, but also the intensities of the three components that we have separated in our analysis, via $I_{\mathrm{el}}$. This crucial parameter can be decomposed into two parts, $\xi(\gamma_{1},\Delta) = \xi'(\gamma_{1},\Delta) + \xi''(\gamma_{1},\Delta)$, with
\begin{equation}\label{eq:ksi1}
\xi'(\gamma,\Delta) = \int_0^\infty ds e^{-\left(\gamma/2-i\Delta \right) s}
=\frac{1}{\gamma/2-i\Delta}
\end{equation}
and
\begin{equation}\label{eq:ksi2}
\xi''(\gamma,\Delta) = \int_0^\infty ds e^{-\left(\gamma/2-i\Delta \right) s} \left(e^{\phi(s)}-1 \right).
\end{equation}
The first component yields
\begin{equation*}
\re \xi'(\gamma,\Delta) = \frac{\gamma/2}{(\gamma/2)^2 + \Delta^2} 
= \frac{\gamma}{2}|\xi'(\gamma,\Delta)|^2.
\end{equation*}
This Lorentzian dependence on detuning is known from the standard theory of resonance fluorescence and accounts for the excitation via the central line (fundamental transition). The second contribution gives rise to
\begin{equation*}
\re \xi''(\gamma,\Delta) \approx \frac{1}{2} \int_{-\infty}^\infty ds e^{i\Delta s} \left(e^{\phi(s)}-1 \right) \propto S_{\mathrm{psb1}}(\oml-\Delta),
\end{equation*}
where we omitted the damping with the rate $\gamma/2$ in view of the very strong localization of the integrand in the range $s\ll 1/\gamma$. 
Thus, this part reproduces the phonon-sideband, but inverted with respect to the laser frequency, i.e., absorption-like. It accounts for the photon scattering via phonon-assisted excitation.

\section{\label{sec:Results_discussions} Numerical results and discussion}

In this section, we present the results of numerical calculations of the RF spectrum and discuss its properties. 
For simplicity, we assume the phonon spectral density in the form
\begin{equation}
J(\omega) = \frac{F_\mathrm{HR}}{\omc^2}\omega^3e^{-\omega^2/(2\omc^2)}, \label{eq:J_spherical}
\end{equation}
which accounts for the dependence $\propto \omega^3$ at low frequencies and a high-frequency cutoff at $\omega = \omc$, while disregarding structural details irrelevant to the discussed phenomena. In the simulations, we set $\gamma_{1} = 1$~ns$^{-1}$ and $\omc = 1$~ps$^{-1}$.

\subsection{\label{sec:Results_moderate_coupling} Moderate phonon coupling}

\begin{figure}[tb]
\includegraphics[width=\linewidth]{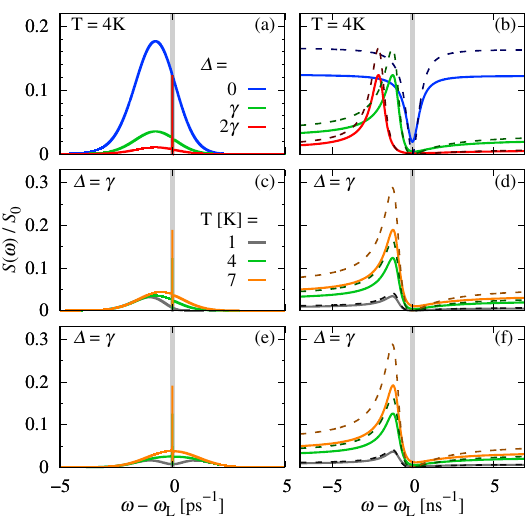}
\caption{\label{fig:RF_spectrum_moderate_coupling} RF spectrum at weak phonon coupling ($F_\mathrm{HR}=0.1$), in a broad (left) and narrow (right) spectral ranges around the resonance (notice the different units on the frequency axes). (a,b) Spectra at various detunings, as shown, at a fixed temperature $T=4$~K. (c,d) Spectra at various temperatures, as shown in (c), for a fixed laser detuning $\Delta=\gamma$. Vertical gray belts schematically show the position of the central elastic-scattering line. Dashed lines with colors corresponding to the respective solid lines present the RF spectra in the weak coupling limit for the same parameters of temperature and detuning. (e,f) As in (c,d) but for classical noise.}
\end{figure}

Figure~\ref{fig:RF_spectrum_moderate_coupling} shows the RF spectra for a moderate phonon coupling, where we set $F_\mathrm{HR}=0.1$. 
Figures~\ref{fig:RF_spectrum_moderate_coupling}(a) and \ref{fig:RF_spectrum_moderate_coupling}(b) present the results at a fixed temperature $T=4K$. In Fig.~\ref{fig:RF_spectrum_moderate_coupling}(a) we see the interplay between the broad PSB, the narrow inelastic feature, and the central elastic-scattering line (marked here and elsewhere as a vertical grey bar). Fig.~\ref{fig:RF_spectrum_moderate_coupling}(b) zooms in by a factor of 1000, focusing on the narrow spectral range around the resonance, where the PSB is essentially flat and together with the narrow feature creates a Fano-like feature as predicted in Sec.~\ref{sec:RF_spectrum}: As the detuning decreases, the RF spectrum evolves from a mostly absorptive shape (red line), through dispersive (green line), to a spectral dip (inverted Lorentzian) around the laser frequency (blue line) for resonant excitation. At the same time, the overall intensity decreases as the excitation becomes increasingly detuned, which is simply due to reduced excitation efficiency, governed by the prefactor in Eq.~\eqref{eq:Fano}. In the weak coupling limit (dashed lines), the RF spectrum shows exact Fano form with entirely suppressed scattering at the laser frequency ($\omega=\oml$).

Figs.~\ref{fig:RF_spectrum_moderate_coupling}c,d present the temperature dependence of the RF spectra at slightly detuned excitation ($\Delta=\gamma_{1}$), in a broad and narrow spectral range, respectively. As the temperature increases, the PSB grows and becomes more symmetric due to enhanced phonon absorption, as is known e.g. in absorption spectroscopy \cite{Krummheuer2002}. Already in this relatively low temperature range, the growth of scattering in the low-frequency sector [Fig.~\ref{fig:RF_spectrum_moderate_coupling}(d)] starts to deviate from the linear scaling with temperature predicted by Eq.~\eqref{eq:Fano} and Eq.~\eqref{eq:A}. In accordance with our general theory, the width of the inelastic profile remains constant. 

Figs.~\ref{fig:RF_spectrum_moderate_coupling}e,f present the RF spectrum for classical Gaussian noise, for the same parameters as in Figs.~\ref{fig:RF_spectrum_moderate_coupling}c,d. Classical noise generates a symmetric broad sideband, since the difference between phonon absorption and emission is not applicable in this case. Alternatively, one can say that classical noise is a (renormalized) high-temperature limit of the quantum case, where the sideband becomes symmetric. Comparing Fig.~\ref{fig:RF_spectrum_moderate_coupling}(f) with Fig.~\ref{fig:RF_spectrum_moderate_coupling}(d), one can see that the inelastic feature is the same in the quantum and classical regimes, showing Fano behavior in both these cases. Fano resonances are known to be generic for systems in which a discrete transition is located on a broad background and hence also to appear in purely classical systems \cite{Riffe2011,Jayich2012,Timberlake2019,Tribelsky2008,Liu2009,Hao2008,Verellen2009,Li2011,Lee2004,Iizawa2021} that may be as simple as coupled damped oscillators \cite{GarridoAlzar2002,Joe2006,Satpathy2012,Stassi2017,Iizawa2021}. The present result demonstrates this feature for an inherently quantum two-level system in which the spectral background originates from environmental noise, the classical and quantum regimes of which can be treated on equal footing, showing no difference in the spectral properties between these two cases. 

\begin{figure}[tb]
\includegraphics[width=\linewidth]{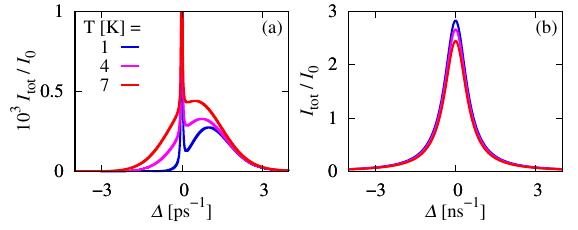}
\caption{\label{fig:Intensity_moderate_coupling} Total scattering intensity as a function of the excitation detuning from the fundamental transition for broad (a) and narrow (b) detuning ranges, for a weak phonon coupling ($F_\mathrm{HR}=0.1$) at different temperatures.}
\end{figure}

Figure~\ref{fig:Intensity_moderate_coupling} presents the total scattering intensity as a function of the detuning of the laser frequency from the polaron-shifted fundamental transition. At small detunings, Fig.~\ref{fig:Intensity_moderate_coupling}(b), a Lorentzian profile appears, in accordance with 
Eq.~\eqref{eq:I_tot} and Eq.~\eqref{eq:ksi1}. In a wider range of detunings [Fig.~\ref{fig:Intensity_moderate_coupling}(a)], the inverted broad phonon sideband is reproduced, as predicted by Eq.~\eqref{eq:ksi2}. Note that the intensity of this phonon-assisted scattering is three orders of magnitude weaker than that at direct excitation. 
The scattering intensities shown in Fig.~\ref{fig:Intensity_moderate_coupling} show temperature dependence that reflects the transfer of spectral weight from the central line to phonon sidebands: the scattering intensity at nearly-resonant excitation decreases, while at phonon-assisted excitation it increases as the temperature grows. In addition, the increasing intensity at red-detuned excitation reflects the growing contribution from phonon absorption processes.  

\subsection{\label{sec:results_strong_coupling} Strong phonon coupling}

\begin{figure}[tb]
\includegraphics[width=\linewidth]{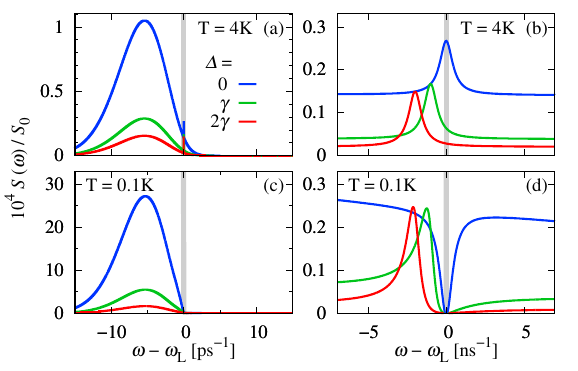}
\caption{\label{fig:RF_spectrum_strong_coupling}  RF spectrum for a strong phonon coupling ($F_\mathrm{HR}=5$), in a broad (left) and narrow (right) spectral range around the resonance. Results are shown at fixed temperatures $T$. Vertical gray belts denote again the elastic-scattering, central line.}
\end{figure}

\begin{figure}[tb]
\includegraphics[width=\linewidth]{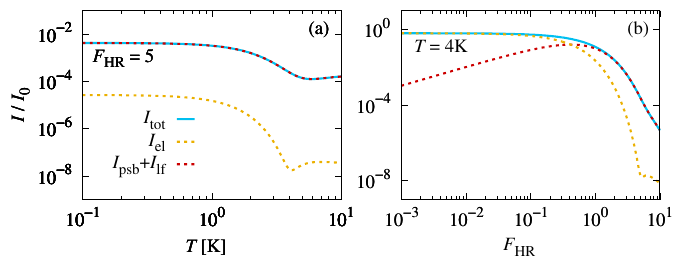}
\caption{\label{fig:Intensity_temperature_FHR} Total intensity of the RF spectrum ($I_\mathrm{tot}$), as well as intensity of the elastic ($I_\mathrm{el}$) and inelastic ($I_\mathrm{psb}+I_\mathrm{in}$) parts at nearly-resonant excitation ($\Delta=\gamma_{1}$). (a) Dependence on temperature for a fixed $F_\mathrm{HR}=5$. (b) Dependence on the Huang-Rhys factor for a fixed temperature $T=4$~K.}
\end{figure}

Figure~\ref{fig:RF_spectrum_strong_coupling} shows the RF spectrum for a strong phonon coupling ($F_\mathrm{HR}=5$) at two different temperatures. In the broad spectral range [Fig.~\ref{fig:RF_spectrum_strong_coupling}(a) and Fig.~\ref{fig:RF_spectrum_strong_coupling}(c)] the RF spectrum again consists of a central, unbroadened line, a broad PSB, and a narrow inelastic scattering feature, the structure of which is better visible in the narrow spectral range in Figs.~\ref{fig:RF_spectrum_strong_coupling}(b) and \ref{fig:RF_spectrum_strong_coupling}(d). Compared to the weak coupling case (Fig. ~\ref{fig:RF_spectrum_moderate_coupling}), the overall amplitude of the RF spectrum is  reduced by several orders of magnitude (see also Fig.~\ref{fig:Intensity_temperature_FHR}, discussed below). This is a consequence of the strongly reduced efficiency of nearly-resonant excitation due to the transfer of spectral weight from the fundamental to phonon-assisted transitions. For strong coupling, this exponential effect dominates over the relative enhancement of the phonon sidebands. Moreover, there is a greater contribution of multiphonon processes, which results in a broadening of the PSB. The most striking difference between the weak and strong-coupling cases is the disappearance of the Fano profile, which is replaced by a Lorentzian at $T=4$~K [Fig.~\ref{fig:RF_spectrum_strong_coupling}(b)]. Indeed, at this temperature and for the selected value of the Huang-Rhys factor, the parameters $A$ and $B$ in Eq.~\eqref{eq:AB} become nearly equal, hence Eq.~\eqref{eq:Fano} yields a Lorentzian, as discussed in Sec.~\ref{sec:RF_spectrum}. As follows from that discussion, at low temperatures single-phonon processes dominate even for strong coupling and the Fano line profile should be recovered, which is indeed the case, as shown in Fig.~\ref{fig:RF_spectrum_strong_coupling}(d). One feature of the spectra in Figs.~\ref{fig:RF_spectrum_strong_coupling}(a) and \ref{fig:RF_spectrum_strong_coupling}(c) that may seem surprising is the increase of the phonon sidebands with decreasing temperature. This is due to the temperature-induced redistribution of the spectral weight between the central line and the sidebands: On the one hand, the sidebands are relatively weaker at lower temperatures, whereas on the other hand, exactly for this reason, the excitation via the central line becomes more efficient. The latter turns out to dominate at strong phonon couplings. As we shall see, the relative intensity of phonon-assisted scattering decreases at lower temperatures, as expected. 
Finally, we note that for a strong phonon coupling, the width of the inelastic feature is again temperature-independent and equal to $\gamma/2$.

\begin{figure}[tb]
\includegraphics[width=\linewidth]{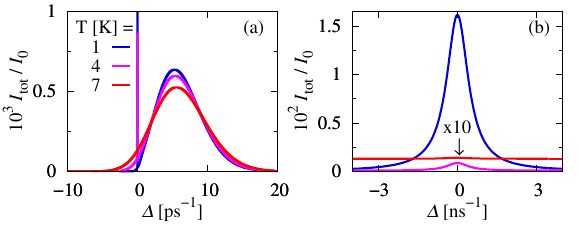}
\caption{\label{fig:Intensity_strong_coupling} Total scattering intensity for a strong coupling ($F_\mathrm{HR}=5$) as a function of the detuning $\Delta$ in broad (a) and narrow (b) spectral ranges.}
\end{figure}

The total scattering intensity at $F_\mathrm{HR}=5$ is presented in Fig.~\ref{fig:Intensity_strong_coupling}. The intensity at nearly resonant excitation, corresponding to excitation via central line, is much lower than for a weak coupling, as most of the spectral weight is transferred to sidebands. Strong temperature dependence leads to suppression of scattering under such excitation conditions already at $T=7$~K. Excitation via phonon sidebands leads to similar peak scattering intensity as for weak coupling, although the maximum is shifted to higher detunings and the range of detunings leading to efficient scattering is considerably extended due to multi-phonon processes. The slight decrease of the scattering intensity at growing temperatures may be attributed to redistribution of the spectral weight over a growing range of frequencies enabled by multiphonon processes at higher temperatures.

Figure~\ref{fig:Intensity_temperature_FHR}(a) shows the decomposition of the scattering intensity into elastic and inelastic (phonon-induced) components, the latter comprising both the low-frequency feature and the broad phonon sideband, as a function of temperature at a nearly resonant excitation for strong phonon-coupling. These results confirm that, at such strong phonon couplings, the scattering is dominated by the inelastic component and that the total intensity of the RF spectrum drops with growing temperature in the sub-Kelvin to few Kelvin range. The  growth of the scattering intensity at higher temperatures is due to the increasing efficiency of phonon-assisted excitation, which become the dominating excitation channel even in this nearly-resonant regime. The cross-over between the regimes of dominant elastic and inelastic scattering around $F_{\mathrm{HR}}=1$ is visible in Fig.~\ref{fig:Intensity_temperature_FHR}(b), where we show the same decomposition at a fixed temperature as a function of the phonon coupling strength. 

\begin{figure}[tb]
\includegraphics[width=0.7\linewidth]{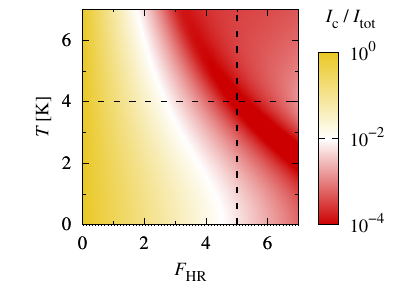}
\caption{\label{fig:Intensity_temperature_FHR_map} Map of the relative contribution of $I_\mathrm{el}$ to $I_\mathrm{tot}$ for different Huang-Rhys factors and temperatures. Results for $\Delta=\gamma_{1}$. Dashed vertical and horizontal black lines at correspond to the situations in Fig.~\ref{fig:Intensity_temperature_FHR}(a) and Fig.~\ref{fig:Intensity_temperature_FHR}(b), respectively.}
\end{figure}

Figure~\ref{fig:Intensity_temperature_FHR_map} presents the intensity of the elastic line relative to the total scattering intensity as a function of temperature and Huang-Rhys factor at nearly resonant excitation. As expected, the transition to the regime where the scattering to phonon sidebands strongly dominates over the elastic contribution ($I_\mathrm{el} \lesssim 0.01 I_\mathrm{tot}$, red area) shifts to lower phonon couplings as the temperature grows, although this change is not very strong. 

\subsection{\label{sec:results_dephasing} Impact of dephasing and inhomogeneous broadening}

\begin{figure}[tb]
\includegraphics[width=\linewidth]{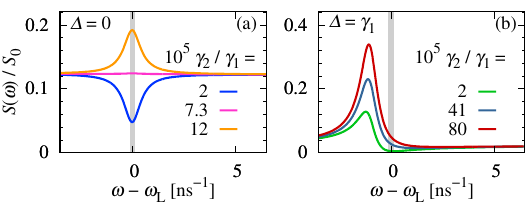}
\caption{\label{fig:RF_spectrum_pure_dephasing}RF spectra in the presence of additional Markovian dephasing at $T = 4$~K: (a) for resonant excitation ($\Delta=0$); (b) for $\Delta=\gamma_1$. Different lines correspond to growing dephasing, as shown in the plots.}
\end{figure}

Figure~\ref{fig:RF_spectrum_pure_dephasing} shows how the spectrum evolves in the presence of pure dephasing at $T = 4$~K, for resonant and detuned excitations. For our model of the spectral density, Eq.~\eqref{eq:J_spherical}, the critical value of $\gamma_2$ at which the Markovian dephasing effect dominates over phonons is, from Eq.~\eqref{eq:gamma2-crit},
\begin{equation*}
\gamma_{2\mathrm{c}} \approx \left[\frac{1}{4}+\frac{\Delta^2}{\gamma^2}\right]
\frac{2\pi k_{\mathrm{B}}T}{\hbar\omc}\frac{\gamma}{\omc}\gamma,
\end{equation*}
which is $0.8\times 10^{-4}\gamma$ and $4.1\times 10^{-4}\gamma$ for the set of parameters used here and for $\Delta=0$ and $\Delta=\gamma$, respectively. The numerical results in Fig.~\ref{fig:RF_spectrum_pure_dephasing} indeed show how the Fano dip inverts into a Lorentzian peak [Fig.~\ref{fig:RF_spectrum_pure_dephasing}(a)] and how the dip at the laser frequency disappears [Fig.~\ref{fig:RF_spectrum_pure_dephasing}(b)] with growing dephasing rate, with the former taking place earlier, in accordance with our estimate. This qualitative change is a witness to the presence of pure dephasing (other than the primary phonon-induced effects described by the independent boson model). By detuning the laser, the transition range can be shifted from $\gamma_2/\gamma_1\sim 10^{-4}$ to $\gamma_2/\gamma_1\sim 10^{-3}$, which covers the range of values beyond the sensitivity of interferometric measurements and may serve as a tool for further narrowing down the uncertainty range for dephasing rates in low-noise structures. 

\begin{figure}[tb]
\includegraphics[width=\linewidth]{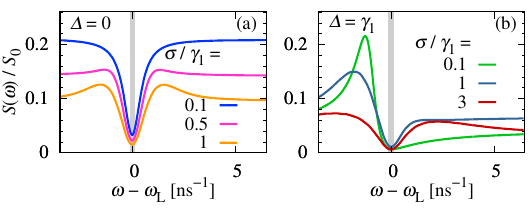}
\caption{\label{fig:RF_spectrum_inhom}
RF spectrum averaged over an inhomogeneous distribution of transition energies for a growing degree of inhomogeneity, as indicated by the Gaussian standard deviation. (a) Resonant excitation ($\Delta = 0$); (b) $\Delta=\gamma_1$. The results in (a) were computed using 15, 25, and 41 Gauss-Hermite nodes for $\sigma/\gamma_1 = 0.1, 0.5$, and 1, respectively, whereas the data in (b) required 15, 41, and 301 nodes for $\sigma/\gamma_1 = 0.1, 1$, and 3, respectively.}
\end{figure}

The effect of the inhomogeneous distribution of the transition energies is shown in Fig.~\ref{fig:RF_spectrum_inhom}, where we model the inhomogeneity by a Gaussian and compute the inhomogeneously broadened spectrum by direct averaging multiple spectra over the Gaussian distribution of detunings using the Gauss-Hermite rule. As expected, averaging over the variable detuning modifies the spectral profile and smears out the Fano feature at detuned excitation but preserves the dip around the resonant frequency. The sensitivity to inhomogeneous broadening is three to four orders of magnitude lower than in the homogeneous dephasing case, with the spectral line in the detuned case [Fig.~\ref{fig:RF_spectrum_inhom}(b)] qualitatively modified only when the broadening is a few times larger than the lifetime-limited width.

\section{\label{sec:conclusions}Conclusions}

We have shown that the low-excitation RF spectrum of a two-level system coupled to a bosonic bath with a super-Ohmic spectral density (e.g. phonons) shows a Fano-like profile of inelastic scattering intensity near the resonant energy for moderately strong phonon couplings. 
%which is due to the interplay of a PSB and a narrow inelastic scattering line stemming from noise-induced transient dynamics. 
The Fano profile in the independent-boson model describing this system is characteristic of the resonance fluorescence and does not appear in other types of linear or nonlinear spectroscopy, where phonon effects are manifested only by broad phonon sidebands, which exist also in fluorescence.
The amplitude of the Fano feature grows linearly with temperature. However, because of the super-Ohmic character of the phonon reservoir, its width depends solely on the exciton life time.
In the weak-coupling limit (single-phonon processes), the spectral profile becomes an exact Fano shape, where resonant light scattering is totally suppressed. 
For strong phonon couplings, the Fano profile disappears due to the growing role of  multi-phonon processes, unless the temperature is very low.

Phonon effects are manifested also in the dependence of the total scattering intensity on the excitation frequency, where broad sidebands appear due to phonon-assisted excitation. The total scattering intensity, as well as its distribution into elastic and inelastic (phonon-assisted) components, reflects the impact of phonons on both excitation and scattering efficiency, leading to different behaviors in different spectral ranges and for different coupling strengths. However, in general, the contribution of phonon-assisted scattering grows with both the coupling strength and the temperature, as expected. 

We have also formally captured the quantumness of the phonon noise by tracing it back to the non-commutativity of noise observables. We have shown that the Fano feature persists in the classical (commutative) limit of classical Gaussian noise with an appropriate spectral density.

In the presence of slow environmental fluctuations (inhomogeneous line broadening), the Fano feature smears out, but the suppression of scattering around the laser frequency persists. On the contrary, this characteristic spectral feature evolves towards a simple Lorentzian above a certain threshold value of pure dephasing (homogeneous broadening), which is a small fraction of the lifetime broadening for typical QD parameters and experimental conditions. Whether such pure dephasing (excluding phonons, which contribute differently as we have shown here) is substantial depends on the properties of a given system. Our results show that homogeneous pure dephasing processes leave a clear trace in the scattering spectra, with a qualitative transition in the range of dephasing rates three to four orders of magnitude below the recombination rate, which is beyond the accuracy of the current interferometric measurements and relevant for the state-of-the-art indistinguishable photon generation.

On the fundamental level, our results indicate the presence of a characteristic feature that appears only in the particular type of spectroscopy of a solid-state quantum emitter. The formalism itself relies on the description of the reservoir via its spectral density and can easily be applied to any system described by the independent boson model. From the perspective of applications, in view of the growing role of resonance fluorescence in characterizing and exploiting the quantum properties of solid-state emitters of light, the presented study may be important for designing and optimizing quantum light sources. In addition, the presence of the Fano feature is an ultra-sensitive spectral probe for the absence of pure dephasing effects, complementary to interferometric measurements of photon indistinguishability and perhaps less affected by imperfections of the optical setup than the latter. This can push the quality assessment of indistinguishable photon sources beyond the current limitations, which is crucial for scalable generation of indistinguishable photon trains with applications, e.g., to cluster state generation.

\begin{acknowledgments}

The authors thank D. Groll for discussions during the early stages of this work. 
This project was supported by the Polish National Science Centre (Narodowe Centrum Nauki, NCN) under grant no. 2023/50/A/ST3/00511
and by the German Federal Ministry of Education and Research via the Research Group Linkage Program of the Alexander von Humboldt Foundation. 
The authors acknowledge the use of the Gemini language model to identify key experimental benchmarks regarding pure dephasing rates and multi-photon interference, and to determine relevant citations, as well as for minor assistance in coding. The authors independently verified all extracted data, literature references, and code, and assume full responsibility for the scientific accuracy of the manuscript.

\end{acknowledgments}

\appendix

\bibliography{publications}% Produces the bibliography via BibTeX.

\begin{thebibliography}{87}%
\makeatletter
\providecommand \@ifxundefined [1]{%
 \@ifx{#1\undefined}
}%
\providecommand \@ifnum [1]{%
 \ifnum #1\expandafter \@firstoftwo
 \else \expandafter \@secondoftwo
 \fi
}%
\providecommand \@ifx [1]{%
 \ifx #1\expandafter \@firstoftwo
 \else \expandafter \@secondoftwo
 \fi
}%
\providecommand \natexlab [1]{#1}%
\providecommand \enquote  [1]{``#1''}%
\providecommand \bibnamefont  [1]{#1}%
\providecommand \bibfnamefont [1]{#1}%
\providecommand \citenamefont [1]{#1}%
\providecommand \href@noop [0]{\@secondoftwo}%
\providecommand \href [0]{\begingroup \@sanitize@url \@href}%
\providecommand \@href[1]{\@@startlink{#1}\@@href}%
\providecommand \@@href[1]{\endgroup#1\@@endlink}%
\providecommand \@sanitize@url [0]{\catcode `\\12\catcode `\$12\catcode `\&12\catcode `\#12\catcode `\^12\catcode `\_12\catcode `\%12\relax}%
\providecommand \@@startlink[1]{}%
\providecommand \@@endlink[0]{}%
\providecommand \url  [0]{\begingroup\@sanitize@url \@url }%
\providecommand \@url [1]{\endgroup\@href {#1}{\urlprefix }}%
\providecommand \urlprefix  [0]{URL }%
\providecommand \Eprint [0]{\href }%
\providecommand \doibase [0]{https://doi.org/}%
\providecommand \selectlanguage [0]{\@gobble}%
\providecommand \bibinfo  [0]{\@secondoftwo}%
\providecommand \bibfield  [0]{\@secondoftwo}%
\providecommand \translation [1]{[#1]}%
\providecommand \BibitemOpen [0]{}%
\providecommand \bibitemStop [0]{}%
\providecommand \bibitemNoStop [0]{.\EOS\space}%
\providecommand \EOS [0]{\spacefactor3000\relax}%
\providecommand \BibitemShut  [1]{\csname bibitem#1\endcsname}%
\let\auto@bib@innerbib\@empty
%</preamble>
\bibitem [{\citenamefont {Scully}\ and\ \citenamefont {Zubairy}(1997)}]{ScullyZubairy1997}%
  \BibitemOpen
  \bibfield  {author} {\bibinfo {author} {\bibfnamefont {M.~O.}\ \bibnamefont {Scully}}\ and\ \bibinfo {author} {\bibfnamefont {M.~S.}\ \bibnamefont {Zubairy}},\ }\href {https://doi.org/10.1017/CBO9780511813993} {\emph {\bibinfo {title} {Quantum Optics}}}\ (\bibinfo  {publisher} {Cambridge University Press},\ \bibinfo {year} {1997})\BibitemShut {NoStop}%
\bibitem [{\citenamefont {Meystre}\ and\ \citenamefont {Sargent}(2007)}]{Meystre2007}%
  \BibitemOpen
  \bibfield  {author} {\bibinfo {author} {\bibfnamefont {P.}~\bibnamefont {Meystre}}\ and\ \bibinfo {author} {\bibfnamefont {M.}~\bibnamefont {Sargent}},\ }\href {https://doi.org/10.1007/978-3-540-74211-1} {\emph {\bibinfo {title} {Elements of quantum optics}}}\ (\bibinfo  {publisher} {Springer Berlin Heidelberg},\ \bibinfo {year} {2007})\BibitemShut {NoStop}%
\bibitem [{\citenamefont {Dalibard}\ and\ \citenamefont {Reynaud}(1983)}]{Dalibard1983}%
  \BibitemOpen
  \bibfield  {author} {\bibinfo {author} {\bibfnamefont {J.}~\bibnamefont {Dalibard}}\ and\ \bibinfo {author} {\bibfnamefont {S.}~\bibnamefont {Reynaud}},\ }\href {https://doi.org/10.1051/jphys:0198300440120133700} {\bibfield  {journal} {\bibinfo  {journal} {Journal de Physique}\ }\textbf {\bibinfo {volume} {44}},\ \bibinfo {pages} {1337–1343} (\bibinfo {year} {1983})}\BibitemShut {NoStop}%
\bibitem [{\citenamefont {Nienhuis}(1993)}]{Nienhuis1993}%
  \BibitemOpen
  \bibfield  {author} {\bibinfo {author} {\bibfnamefont {G.}~\bibnamefont {Nienhuis}},\ }\href {https://doi.org/10.1103/PhysRevA.47.510} {\bibfield  {journal} {\bibinfo  {journal} {Phys. Rev. A}\ }\textbf {\bibinfo {volume} {47}},\ \bibinfo {pages} {510} (\bibinfo {year} {1993})}\BibitemShut {NoStop}%
\bibitem [{\citenamefont {Sch{\"{o}}ll}\ \emph {et~al.}(2019)\citenamefont {Sch{\"{o}}ll}, \citenamefont {Hanschke}, \citenamefont {Schweickert}, \citenamefont {Zeuner}, \citenamefont {Reindl}, \citenamefont {Covre~da Silva}, \citenamefont {Lettner}, \citenamefont {Trotta}, \citenamefont {Finley}, \citenamefont {M{\"{u}}ller}, \citenamefont {Rastelli}, \citenamefont {Zwiller},\ and\ \citenamefont {J{\"{o}}ns}}]{Scholl2019}%
  \BibitemOpen
  \bibfield  {author} {\bibinfo {author} {\bibfnamefont {E.}~\bibnamefont {Sch{\"{o}}ll}}, \bibinfo {author} {\bibfnamefont {L.}~\bibnamefont {Hanschke}}, \bibinfo {author} {\bibfnamefont {L.}~\bibnamefont {Schweickert}}, \bibinfo {author} {\bibfnamefont {K.~D.}\ \bibnamefont {Zeuner}}, \bibinfo {author} {\bibfnamefont {M.}~\bibnamefont {Reindl}}, \bibinfo {author} {\bibfnamefont {S.~F.}\ \bibnamefont {Covre~da Silva}}, \bibinfo {author} {\bibfnamefont {T.}~\bibnamefont {Lettner}}, \bibinfo {author} {\bibfnamefont {R.}~\bibnamefont {Trotta}}, \bibinfo {author} {\bibfnamefont {J.~J.}\ \bibnamefont {Finley}}, \bibinfo {author} {\bibfnamefont {K.}~\bibnamefont {M{\"{u}}ller}}, \bibinfo {author} {\bibfnamefont {A.}~\bibnamefont {Rastelli}}, \bibinfo {author} {\bibfnamefont {V.}~\bibnamefont {Zwiller}},\ and\ \bibinfo {author} {\bibfnamefont {K.~D.}\ \bibnamefont {J{\"{o}}ns}},\ }\href {https://doi.org/10.1021/acs.nanolett.8b05132} {\bibfield  {journal} {\bibinfo  {journal} {Nano Lett.}\ }\textbf {\bibinfo
  {volume} {19}},\ \bibinfo {pages} {2404} (\bibinfo {year} {2019})}\BibitemShut {NoStop}%
\bibitem [{\citenamefont {Wang}\ \emph {et~al.}(2025)\citenamefont {Wang}, \citenamefont {Huang}, \citenamefont {Li}, \citenamefont {Wang}, \citenamefont {Liu}, \citenamefont {Wu}, \citenamefont {Liu}, \citenamefont {Ni}, \citenamefont {Niu}, \citenamefont {Ji}, \citenamefont {Jiao}, \citenamefont {Yin},\ and\ \citenamefont {Yuan}}]{Wang2025}%
  \BibitemOpen
  \bibfield  {author} {\bibinfo {author} {\bibfnamefont {X.-J.}\ \bibnamefont {Wang}}, \bibinfo {author} {\bibfnamefont {G.}~\bibnamefont {Huang}}, \bibinfo {author} {\bibfnamefont {M.-Y.}\ \bibnamefont {Li}}, \bibinfo {author} {\bibfnamefont {Y.-Z.}\ \bibnamefont {Wang}}, \bibinfo {author} {\bibfnamefont {L.}~\bibnamefont {Liu}}, \bibinfo {author} {\bibfnamefont {B.}~\bibnamefont {Wu}}, \bibinfo {author} {\bibfnamefont {H.}~\bibnamefont {Liu}}, \bibinfo {author} {\bibfnamefont {H.}~\bibnamefont {Ni}}, \bibinfo {author} {\bibfnamefont {Z.}~\bibnamefont {Niu}}, \bibinfo {author} {\bibfnamefont {W.}~\bibnamefont {Ji}}, \bibinfo {author} {\bibfnamefont {R.}~\bibnamefont {Jiao}}, \bibinfo {author} {\bibfnamefont {H.-L.}\ \bibnamefont {Yin}},\ and\ \bibinfo {author} {\bibfnamefont {Z.}~\bibnamefont {Yuan}},\ }\href {https://doi.org/10.1038/s41467-025-61884-x} {\bibfield  {journal} {\bibinfo  {journal} {Nature Communications}\ }\textbf {\bibinfo {volume} {16}},\ \bibinfo {pages} {6453} (\bibinfo {year}
  {2025})}\BibitemShut {NoStop}%
\bibitem [{\citenamefont {Wrigge}\ \emph {et~al.}(2008)\citenamefont {Wrigge}, \citenamefont {Gerhardt}, \citenamefont {Hwang}, \citenamefont {Zumofen},\ and\ \citenamefont {Sandoghdar}}]{wrigge_efficient_2008}%
  \BibitemOpen
  \bibfield  {author} {\bibinfo {author} {\bibfnamefont {G.}~\bibnamefont {Wrigge}}, \bibinfo {author} {\bibfnamefont {I.}~\bibnamefont {Gerhardt}}, \bibinfo {author} {\bibfnamefont {J.}~\bibnamefont {Hwang}}, \bibinfo {author} {\bibfnamefont {G.}~\bibnamefont {Zumofen}},\ and\ \bibinfo {author} {\bibfnamefont {V.}~\bibnamefont {Sandoghdar}},\ }\href {https://doi.org/10.1038/nphys812} {\bibfield  {journal} {\bibinfo  {journal} {Nature Phys.}\ }\textbf {\bibinfo {volume} {4}},\ \bibinfo {pages} {60} (\bibinfo {year} {2008})}\BibitemShut {NoStop}%
\bibitem [{\citenamefont {Muller}\ \emph {et~al.}(2007)\citenamefont {Muller}, \citenamefont {Flagg}, \citenamefont {Bianucci}, \citenamefont {Wang}, \citenamefont {Deppe}, \citenamefont {Ma}, \citenamefont {Zhang}, \citenamefont {Salamo}, \citenamefont {Xiao},\ and\ \citenamefont {Shih}}]{Muller2007}%
  \BibitemOpen
  \bibfield  {author} {\bibinfo {author} {\bibfnamefont {A.}~\bibnamefont {Muller}}, \bibinfo {author} {\bibfnamefont {E.~B.}\ \bibnamefont {Flagg}}, \bibinfo {author} {\bibfnamefont {P.}~\bibnamefont {Bianucci}}, \bibinfo {author} {\bibfnamefont {X.~Y.}\ \bibnamefont {Wang}}, \bibinfo {author} {\bibfnamefont {D.~G.}\ \bibnamefont {Deppe}}, \bibinfo {author} {\bibfnamefont {W.}~\bibnamefont {Ma}}, \bibinfo {author} {\bibfnamefont {J.}~\bibnamefont {Zhang}}, \bibinfo {author} {\bibfnamefont {G.~J.}\ \bibnamefont {Salamo}}, \bibinfo {author} {\bibfnamefont {M.}~\bibnamefont {Xiao}},\ and\ \bibinfo {author} {\bibfnamefont {C.~K.}\ \bibnamefont {Shih}},\ }\href {https://doi.org/10.1103/PhysRevLett.99.187402} {\bibfield  {journal} {\bibinfo  {journal} {Phys. Rev. Lett.}\ }\textbf {\bibinfo {volume} {99}},\ \bibinfo {pages} {187402} (\bibinfo {year} {2007})}\BibitemShut {NoStop}%
\bibitem [{\citenamefont {Astafiev}\ \emph {et~al.}(2010)\citenamefont {Astafiev}, \citenamefont {Zagoskin}, \citenamefont {Abdumalikov}, \citenamefont {Pashkin}, \citenamefont {Yamamoto}, \citenamefont {Inomata}, \citenamefont {Nakamura},\ and\ \citenamefont {Tsai}}]{Astaview2010}%
  \BibitemOpen
  \bibfield  {author} {\bibinfo {author} {\bibfnamefont {O.}~\bibnamefont {Astafiev}}, \bibinfo {author} {\bibfnamefont {A.~M.}\ \bibnamefont {Zagoskin}}, \bibinfo {author} {\bibfnamefont {A.~A.}\ \bibnamefont {Abdumalikov}}, \bibinfo {author} {\bibfnamefont {Y.~A.}\ \bibnamefont {Pashkin}}, \bibinfo {author} {\bibfnamefont {T.}~\bibnamefont {Yamamoto}}, \bibinfo {author} {\bibfnamefont {K.}~\bibnamefont {Inomata}}, \bibinfo {author} {\bibfnamefont {Y.}~\bibnamefont {Nakamura}},\ and\ \bibinfo {author} {\bibfnamefont {J.~S.}\ \bibnamefont {Tsai}},\ }\href {https://doi.org/10.1126/science.1181918} {\bibfield  {journal} {\bibinfo  {journal} {Science}\ }\textbf {\bibinfo {volume} {327}},\ \bibinfo {pages} {840} (\bibinfo {year} {2010})}\BibitemShut {NoStop}%
\bibitem [{\citenamefont {Toyli}\ \emph {et~al.}(2016)\citenamefont {Toyli}, \citenamefont {Eddins}, \citenamefont {Boutin}, \citenamefont {Puri}, \citenamefont {Hover}, \citenamefont {Bolkhovsky}, \citenamefont {Oliver}, \citenamefont {Blais},\ and\ \citenamefont {Siddiqi}}]{Toyli2016}%
  \BibitemOpen
  \bibfield  {author} {\bibinfo {author} {\bibfnamefont {D.~M.}\ \bibnamefont {Toyli}}, \bibinfo {author} {\bibfnamefont {A.~W.}\ \bibnamefont {Eddins}}, \bibinfo {author} {\bibfnamefont {S.}~\bibnamefont {Boutin}}, \bibinfo {author} {\bibfnamefont {S.}~\bibnamefont {Puri}}, \bibinfo {author} {\bibfnamefont {D.}~\bibnamefont {Hover}}, \bibinfo {author} {\bibfnamefont {V.}~\bibnamefont {Bolkhovsky}}, \bibinfo {author} {\bibfnamefont {W.~D.}\ \bibnamefont {Oliver}}, \bibinfo {author} {\bibfnamefont {A.}~\bibnamefont {Blais}},\ and\ \bibinfo {author} {\bibfnamefont {I.}~\bibnamefont {Siddiqi}},\ }\href {https://doi.org/10.1103/PhysRevX.6.031004} {\bibfield  {journal} {\bibinfo  {journal} {Phys. Rev. X}\ }\textbf {\bibinfo {volume} {6}},\ \bibinfo {pages} {031004} (\bibinfo {year} {2016})}\BibitemShut {NoStop}%
\bibitem [{\citenamefont {Vamivakas}\ \emph {et~al.}(2010)\citenamefont {Vamivakas}, \citenamefont {Lu}, \citenamefont {Matthiesen}, \citenamefont {Zhao}, \citenamefont {F{\"{a}}lt}, \citenamefont {Badolato},\ and\ \citenamefont {Atat{\"{u}}re}}]{Vamivakas2010}%
  \BibitemOpen
  \bibfield  {author} {\bibinfo {author} {\bibfnamefont {A.~N.}\ \bibnamefont {Vamivakas}}, \bibinfo {author} {\bibfnamefont {C.-Y.}\ \bibnamefont {Lu}}, \bibinfo {author} {\bibfnamefont {C.}~\bibnamefont {Matthiesen}}, \bibinfo {author} {\bibfnamefont {Y.}~\bibnamefont {Zhao}}, \bibinfo {author} {\bibfnamefont {S.}~\bibnamefont {F{\"{a}}lt}}, \bibinfo {author} {\bibfnamefont {A.}~\bibnamefont {Badolato}},\ and\ \bibinfo {author} {\bibfnamefont {M.}~\bibnamefont {Atat{\"{u}}re}},\ }\href {https://doi.org/10.1038/nature09359} {\bibfield  {journal} {\bibinfo  {journal} {Nature}\ }\textbf {\bibinfo {volume} {467}},\ \bibinfo {pages} {297} (\bibinfo {year} {2010})}\BibitemShut {NoStop}%
\bibitem [{\citenamefont {Delteil}\ \emph {et~al.}(2014)\citenamefont {Delteil}, \citenamefont {Gao}, \citenamefont {Fallahi}, \citenamefont {Miguel-Sanchez},\ and\ \citenamefont {Imamo{\u{g}}lu}}]{Delteil2014}%
  \BibitemOpen
  \bibfield  {author} {\bibinfo {author} {\bibfnamefont {A.}~\bibnamefont {Delteil}}, \bibinfo {author} {\bibfnamefont {W.-b.}\ \bibnamefont {Gao}}, \bibinfo {author} {\bibfnamefont {P.}~\bibnamefont {Fallahi}}, \bibinfo {author} {\bibfnamefont {J.}~\bibnamefont {Miguel-Sanchez}},\ and\ \bibinfo {author} {\bibfnamefont {A.}~\bibnamefont {Imamo{\u{g}}lu}},\ }\href {https://doi.org/10.1103/PhysRevLett.112.116802} {\bibfield  {journal} {\bibinfo  {journal} {Phys. Rev. Lett.}\ }\textbf {\bibinfo {volume} {112}},\ \bibinfo {pages} {116802} (\bibinfo {year} {2014})}\BibitemShut {NoStop}%
\bibitem [{\citenamefont {Yılmaz}\ \emph {et~al.}(2010)\citenamefont {Yılmaz}, \citenamefont {Fallahi},\ and\ \citenamefont {Imamo{\u{g}}lu}}]{Ylmaz2010}%
  \BibitemOpen
  \bibfield  {author} {\bibinfo {author} {\bibfnamefont {S.~T.}\ \bibnamefont {Yılmaz}}, \bibinfo {author} {\bibfnamefont {P.}~\bibnamefont {Fallahi}},\ and\ \bibinfo {author} {\bibfnamefont {A.}~\bibnamefont {Imamo{\u{g}}lu}},\ }\href {https://doi.org/10.1103/PhysRevLett.105.033601} {\bibfield  {journal} {\bibinfo  {journal} {Phys. Rev. Lett.}\ }\textbf {\bibinfo {volume} {105}},\ \bibinfo {pages} {33601} (\bibinfo {year} {2010})}\BibitemShut {NoStop}%
\bibitem [{\citenamefont {Delley}\ \emph {et~al.}(2017)\citenamefont {Delley}, \citenamefont {Kroner}, \citenamefont {Faelt}, \citenamefont {Wegscheider},\ and\ \citenamefont {İmamo{\u{g}}lu}}]{Delley2017}%
  \BibitemOpen
  \bibfield  {author} {\bibinfo {author} {\bibfnamefont {Y.~L.}\ \bibnamefont {Delley}}, \bibinfo {author} {\bibfnamefont {M.}~\bibnamefont {Kroner}}, \bibinfo {author} {\bibfnamefont {S.}~\bibnamefont {Faelt}}, \bibinfo {author} {\bibfnamefont {W.}~\bibnamefont {Wegscheider}},\ and\ \bibinfo {author} {\bibfnamefont {A.}~\bibnamefont {İmamo{\u{g}}lu}},\ }\href {https://doi.org/10.1103/PhysRevB.96.241410} {\bibfield  {journal} {\bibinfo  {journal} {Phys. Rev. B}\ }\textbf {\bibinfo {volume} {96}},\ \bibinfo {pages} {241410} (\bibinfo {year} {2017})}\BibitemShut {NoStop}%
\bibitem [{\citenamefont {Lu}\ \emph {et~al.}(2010)\citenamefont {Lu}, \citenamefont {Zhao}, \citenamefont {Vamivakas}, \citenamefont {Matthiesen}, \citenamefont {F{\"{a}}lt}, \citenamefont {Badolato},\ and\ \citenamefont {Atat{\"{u}}re}}]{Lu2010}%
  \BibitemOpen
  \bibfield  {author} {\bibinfo {author} {\bibfnamefont {C.-Y.}\ \bibnamefont {Lu}}, \bibinfo {author} {\bibfnamefont {Y.}~\bibnamefont {Zhao}}, \bibinfo {author} {\bibfnamefont {A.~N.}\ \bibnamefont {Vamivakas}}, \bibinfo {author} {\bibfnamefont {C.}~\bibnamefont {Matthiesen}}, \bibinfo {author} {\bibfnamefont {S.}~\bibnamefont {F{\"{a}}lt}}, \bibinfo {author} {\bibfnamefont {A.}~\bibnamefont {Badolato}},\ and\ \bibinfo {author} {\bibfnamefont {M.}~\bibnamefont {Atat{\"{u}}re}},\ }\href {https://doi.org/10.1103/PhysRevB.81.035332} {\bibfield  {journal} {\bibinfo  {journal} {Phys. Rev. B}\ }\textbf {\bibinfo {volume} {81}},\ \bibinfo {pages} {035332} (\bibinfo {year} {2010})}\BibitemShut {NoStop}%
\bibitem [{\citenamefont {Nick~Vamivakas}\ \emph {et~al.}(2009)\citenamefont {Nick~Vamivakas}, \citenamefont {Zhao}, \citenamefont {Lu},\ and\ \citenamefont {Atat{\"{u}}re}}]{Vamivakas2009}%
  \BibitemOpen
  \bibfield  {author} {\bibinfo {author} {\bibfnamefont {A.}~\bibnamefont {Nick~Vamivakas}}, \bibinfo {author} {\bibfnamefont {Y.}~\bibnamefont {Zhao}}, \bibinfo {author} {\bibfnamefont {C.-Y.}\ \bibnamefont {Lu}},\ and\ \bibinfo {author} {\bibfnamefont {M.}~\bibnamefont {Atat{\"{u}}re}},\ }\href {https://doi.org/10.1038/nphys1182} {\bibfield  {journal} {\bibinfo  {journal} {Nature Physics}\ }\textbf {\bibinfo {volume} {5}},\ \bibinfo {pages} {198} (\bibinfo {year} {2009})}\BibitemShut {NoStop}%
\bibitem [{\citenamefont {Weiß}\ and\ \citenamefont {Krenner}(2018)}]{Weiss2018}%
  \BibitemOpen
  \bibfield  {author} {\bibinfo {author} {\bibfnamefont {M.}~\bibnamefont {Weiß}}\ and\ \bibinfo {author} {\bibfnamefont {H.~J.}\ \bibnamefont {Krenner}},\ }\href {https://doi.org/10.1088/1361-6463/aace3c} {\bibfield  {journal} {\bibinfo  {journal} {Journal of Physics D: Applied Physics}\ }\textbf {\bibinfo {volume} {51}},\ \bibinfo {pages} {373001} (\bibinfo {year} {2018})}\BibitemShut {NoStop}%
\bibitem [{\citenamefont {Delsing}\ \emph {et~al.}(2019)\citenamefont {Delsing}, \citenamefont {Cleland}, \citenamefont {Schuetz}, \citenamefont {Knörzer}, \citenamefont {Giedke}, \citenamefont {Cirac}, \citenamefont {Srinivasan}, \citenamefont {Wu}, \citenamefont {Balram}, \citenamefont {Bäuerle}, \citenamefont {Meunier}, \citenamefont {Ford}, \citenamefont {Santos}, \citenamefont {Cerda-Méndez}, \citenamefont {Wang}, \citenamefont {Krenner}, \citenamefont {Nysten}, \citenamefont {Weiß}, \citenamefont {Nash}, \citenamefont {Thevenard}, \citenamefont {Gourdon}, \citenamefont {Rovillain}, \citenamefont {Marangolo}, \citenamefont {Duquesne}, \citenamefont {Fischerauer}, \citenamefont {Ruile}, \citenamefont {Reiner}, \citenamefont {Paschke}, \citenamefont {Denysenko}, \citenamefont {Volkmer}, \citenamefont {Wixforth}, \citenamefont {Bruus}, \citenamefont {Wiklund}, \citenamefont {Reboud}, \citenamefont {Cooper}, \citenamefont {Fu}, \citenamefont {Brugger}, \citenamefont {Rehfeldt},\ and\ \citenamefont
  {Westerhausen}}]{Delsing2019}%
  \BibitemOpen
  \bibfield  {author} {\bibinfo {author} {\bibfnamefont {P.}~\bibnamefont {Delsing}}, \bibinfo {author} {\bibfnamefont {A.~N.}\ \bibnamefont {Cleland}}, \bibinfo {author} {\bibfnamefont {M.~J.~A.}\ \bibnamefont {Schuetz}}, \bibinfo {author} {\bibfnamefont {J.}~\bibnamefont {Knörzer}}, \bibinfo {author} {\bibfnamefont {G.}~\bibnamefont {Giedke}}, \bibinfo {author} {\bibfnamefont {J.~I.}\ \bibnamefont {Cirac}}, \bibinfo {author} {\bibfnamefont {K.}~\bibnamefont {Srinivasan}}, \bibinfo {author} {\bibfnamefont {M.}~\bibnamefont {Wu}}, \bibinfo {author} {\bibfnamefont {K.~C.}\ \bibnamefont {Balram}}, \bibinfo {author} {\bibfnamefont {C.}~\bibnamefont {Bäuerle}}, \bibinfo {author} {\bibfnamefont {T.}~\bibnamefont {Meunier}}, \bibinfo {author} {\bibfnamefont {C.~J.~B.}\ \bibnamefont {Ford}}, \bibinfo {author} {\bibfnamefont {P.~V.}\ \bibnamefont {Santos}}, \bibinfo {author} {\bibfnamefont {E.}~\bibnamefont {Cerda-Méndez}}, \bibinfo {author} {\bibfnamefont {H.}~\bibnamefont {Wang}}, \bibinfo {author}
  {\bibfnamefont {H.~J.}\ \bibnamefont {Krenner}}, \bibinfo {author} {\bibfnamefont {E.~D.~S.}\ \bibnamefont {Nysten}}, \bibinfo {author} {\bibfnamefont {M.}~\bibnamefont {Weiß}}, \bibinfo {author} {\bibfnamefont {G.~R.}\ \bibnamefont {Nash}}, \bibinfo {author} {\bibfnamefont {L.}~\bibnamefont {Thevenard}}, \bibinfo {author} {\bibfnamefont {C.}~\bibnamefont {Gourdon}}, \bibinfo {author} {\bibfnamefont {P.}~\bibnamefont {Rovillain}}, \bibinfo {author} {\bibfnamefont {M.}~\bibnamefont {Marangolo}}, \bibinfo {author} {\bibfnamefont {J.-Y.}\ \bibnamefont {Duquesne}}, \bibinfo {author} {\bibfnamefont {G.}~\bibnamefont {Fischerauer}}, \bibinfo {author} {\bibfnamefont {W.}~\bibnamefont {Ruile}}, \bibinfo {author} {\bibfnamefont {A.}~\bibnamefont {Reiner}}, \bibinfo {author} {\bibfnamefont {B.}~\bibnamefont {Paschke}}, \bibinfo {author} {\bibfnamefont {D.}~\bibnamefont {Denysenko}}, \bibinfo {author} {\bibfnamefont {D.}~\bibnamefont {Volkmer}}, \bibinfo {author} {\bibfnamefont {A.}~\bibnamefont {Wixforth}}, \bibinfo
  {author} {\bibfnamefont {H.}~\bibnamefont {Bruus}}, \bibinfo {author} {\bibfnamefont {M.}~\bibnamefont {Wiklund}}, \bibinfo {author} {\bibfnamefont {J.}~\bibnamefont {Reboud}}, \bibinfo {author} {\bibfnamefont {J.~M.}\ \bibnamefont {Cooper}}, \bibinfo {author} {\bibfnamefont {Y.}~\bibnamefont {Fu}}, \bibinfo {author} {\bibfnamefont {M.~S.}\ \bibnamefont {Brugger}}, \bibinfo {author} {\bibfnamefont {F.}~\bibnamefont {Rehfeldt}},\ and\ \bibinfo {author} {\bibfnamefont {C.}~\bibnamefont {Westerhausen}},\ }\href {https://doi.org/10.1088/1361-6463/ab1b04} {\bibfield  {journal} {\bibinfo  {journal} {Journal of Physics D: Applied Physics}\ }\textbf {\bibinfo {volume} {52}},\ \bibinfo {pages} {353001} (\bibinfo {year} {2019})}\BibitemShut {NoStop}%
\bibitem [{\citenamefont {Weiß}\ \emph {et~al.}(2021)\citenamefont {Weiß}, \citenamefont {Wigger}, \citenamefont {Nägele}, \citenamefont {Müller}, \citenamefont {Finley}, \citenamefont {Kuhn}, \citenamefont {Machnikowski},\ and\ \citenamefont {Krenner}}]{Weiss2021}%
  \BibitemOpen
  \bibfield  {author} {\bibinfo {author} {\bibfnamefont {M.}~\bibnamefont {Weiß}}, \bibinfo {author} {\bibfnamefont {D.}~\bibnamefont {Wigger}}, \bibinfo {author} {\bibfnamefont {M.}~\bibnamefont {Nägele}}, \bibinfo {author} {\bibfnamefont {K.}~\bibnamefont {Müller}}, \bibinfo {author} {\bibfnamefont {J.~J.}\ \bibnamefont {Finley}}, \bibinfo {author} {\bibfnamefont {T.}~\bibnamefont {Kuhn}}, \bibinfo {author} {\bibfnamefont {P.}~\bibnamefont {Machnikowski}},\ and\ \bibinfo {author} {\bibfnamefont {H.~J.}\ \bibnamefont {Krenner}},\ }\href {https://doi.org/10.1364/OPTICA.412201} {\bibfield  {journal} {\bibinfo  {journal} {Optica}\ }\textbf {\bibinfo {volume} {8}},\ \bibinfo {pages} {291} (\bibinfo {year} {2021})}\BibitemShut {NoStop}%
\bibitem [{\citenamefont {Wigger}\ \emph {et~al.}(2021)\citenamefont {Wigger}, \citenamefont {Weiß}, \citenamefont {Lienhart}, \citenamefont {Müller}, \citenamefont {Finley}, \citenamefont {Kuhn}, \citenamefont {Krenner},\ and\ \citenamefont {Machnikowski}}]{Wigger2021}%
  \BibitemOpen
  \bibfield  {author} {\bibinfo {author} {\bibfnamefont {D.}~\bibnamefont {Wigger}}, \bibinfo {author} {\bibfnamefont {M.}~\bibnamefont {Weiß}}, \bibinfo {author} {\bibfnamefont {M.}~\bibnamefont {Lienhart}}, \bibinfo {author} {\bibfnamefont {K.}~\bibnamefont {Müller}}, \bibinfo {author} {\bibfnamefont {J.~J.}\ \bibnamefont {Finley}}, \bibinfo {author} {\bibfnamefont {T.}~\bibnamefont {Kuhn}}, \bibinfo {author} {\bibfnamefont {H.~J.}\ \bibnamefont {Krenner}},\ and\ \bibinfo {author} {\bibfnamefont {P.}~\bibnamefont {Machnikowski}},\ }\href {https://doi.org/10.1103/PhysRevResearch.3.033197} {\bibfield  {journal} {\bibinfo  {journal} {Phys. Rev. Res.}\ }\textbf {\bibinfo {volume} {3}},\ \bibinfo {pages} {033197} (\bibinfo {year} {2021})}\BibitemShut {NoStop}%
\bibitem [{\citenamefont {Metcalfe}\ \emph {et~al.}(2010)\citenamefont {Metcalfe}, \citenamefont {Carr}, \citenamefont {Muller}, \citenamefont {Solomon},\ and\ \citenamefont {Lawall}}]{Metcalfe2010}%
  \BibitemOpen
  \bibfield  {author} {\bibinfo {author} {\bibfnamefont {M.}~\bibnamefont {Metcalfe}}, \bibinfo {author} {\bibfnamefont {S.~M.}\ \bibnamefont {Carr}}, \bibinfo {author} {\bibfnamefont {A.}~\bibnamefont {Muller}}, \bibinfo {author} {\bibfnamefont {G.~S.}\ \bibnamefont {Solomon}},\ and\ \bibinfo {author} {\bibfnamefont {J.}~\bibnamefont {Lawall}},\ }\href {https://doi.org/10.1103/PhysRevLett.105.037401} {\bibfield  {journal} {\bibinfo  {journal} {Phys. Rev. Lett.}\ }\textbf {\bibinfo {volume} {105}},\ \bibinfo {pages} {037401} (\bibinfo {year} {2010})}\BibitemShut {NoStop}%
\bibitem [{\citenamefont {Krenner}\ \emph {et~al.}(2026)\citenamefont {Krenner}, \citenamefont {Santos}, \citenamefont {Westerhausen}, \citenamefont {Andersson}, \citenamefont {Cleland}, \citenamefont {Sellier}, \citenamefont {Takada}, \citenamefont {B{\"a}uerle}, \citenamefont {Wigger}, \citenamefont {Kuhn}, \citenamefont {Machnikowski}, \citenamefont {Wei{\ss}}, \citenamefont {Moody}, \citenamefont {{Hern{\'a}ndez-M{\'i}nguez}}, \citenamefont {Lazi{\'c}}, \citenamefont {Kuznetsov}, \citenamefont {K{\"u}{\ss}}, \citenamefont {Albrecht}, \citenamefont {Weiler}, \citenamefont {Puebla}, \citenamefont {Hwang}, \citenamefont {Otani}, \citenamefont {Balram}, \citenamefont {Chen}, \citenamefont {Lai}, \citenamefont {Li}, \citenamefont {Nash}, \citenamefont {Nysten}, \citenamefont {Bhattacharjee}, \citenamefont {Mishra}, \citenamefont {Iyer}, \citenamefont {Nemade}, \citenamefont {Khelif}, \citenamefont {Benchabane}, \citenamefont {Feng}, \citenamefont {Jin}, \citenamefont {Bartasyte}, \citenamefont {Margueron},
  \citenamefont {Marangolo}, \citenamefont {Thevenard}, \citenamefont {Rovillain}, \citenamefont {Gourdon}, \citenamefont {{Hage-Ali}}, \citenamefont {Elmazria}, \citenamefont {Schmidt}, \citenamefont {Yeo}, \citenamefont {Ambattu}, \citenamefont {Jeon}, \citenamefont {Kwak}, \citenamefont {Rufo}, \citenamefont {Yang},\ and\ \citenamefont {Huang}}]{krenner2026roadmap}%
  \BibitemOpen
  \bibfield  {author} {\bibinfo {author} {\bibfnamefont {H.~J.}\ \bibnamefont {Krenner}}, \bibinfo {author} {\bibfnamefont {P.~V.}\ \bibnamefont {Santos}}, \bibinfo {author} {\bibfnamefont {C.}~\bibnamefont {Westerhausen}}, \bibinfo {author} {\bibfnamefont {G.}~\bibnamefont {Andersson}}, \bibinfo {author} {\bibfnamefont {A.~N.}\ \bibnamefont {Cleland}}, \bibinfo {author} {\bibfnamefont {H.}~\bibnamefont {Sellier}}, \bibinfo {author} {\bibfnamefont {S.}~\bibnamefont {Takada}}, \bibinfo {author} {\bibfnamefont {C.}~\bibnamefont {B{\"a}uerle}}, \bibinfo {author} {\bibfnamefont {D.}~\bibnamefont {Wigger}}, \bibinfo {author} {\bibfnamefont {T.}~\bibnamefont {Kuhn}}, \bibinfo {author} {\bibfnamefont {P.}~\bibnamefont {Machnikowski}}, \bibinfo {author} {\bibfnamefont {M.}~\bibnamefont {Wei{\ss}}}, \bibinfo {author} {\bibfnamefont {G.}~\bibnamefont {Moody}}, \bibinfo {author} {\bibfnamefont {A.}~\bibnamefont {{Hern{\'a}ndez-M{\'i}nguez}}}, \bibinfo {author} {\bibfnamefont {S.}~\bibnamefont {Lazi{\'c}}}, \bibinfo
  {author} {\bibfnamefont {A.~S.}\ \bibnamefont {Kuznetsov}}, \bibinfo {author} {\bibfnamefont {M.}~\bibnamefont {K{\"u}{\ss}}}, \bibinfo {author} {\bibfnamefont {M.}~\bibnamefont {Albrecht}}, \bibinfo {author} {\bibfnamefont {M.}~\bibnamefont {Weiler}}, \bibinfo {author} {\bibfnamefont {J.}~\bibnamefont {Puebla}}, \bibinfo {author} {\bibfnamefont {Y.}~\bibnamefont {Hwang}}, \bibinfo {author} {\bibfnamefont {Y.}~\bibnamefont {Otani}}, \bibinfo {author} {\bibfnamefont {K.~C.}\ \bibnamefont {Balram}}, \bibinfo {author} {\bibfnamefont {I.-T.}\ \bibnamefont {Chen}}, \bibinfo {author} {\bibfnamefont {K.}~\bibnamefont {Lai}}, \bibinfo {author} {\bibfnamefont {M.}~\bibnamefont {Li}}, \bibinfo {author} {\bibfnamefont {G.~R.}\ \bibnamefont {Nash}}, \bibinfo {author} {\bibfnamefont {E.~D.~S.}\ \bibnamefont {Nysten}}, \bibinfo {author} {\bibfnamefont {P.}~\bibnamefont {Bhattacharjee}}, \bibinfo {author} {\bibfnamefont {H.}~\bibnamefont {Mishra}}, \bibinfo {author} {\bibfnamefont {P.~K.}\ \bibnamefont {Iyer}}, \bibinfo
  {author} {\bibfnamefont {H.~B.}\ \bibnamefont {Nemade}}, \bibinfo {author} {\bibfnamefont {A.}~\bibnamefont {Khelif}}, \bibinfo {author} {\bibfnamefont {S.}~\bibnamefont {Benchabane}}, \bibinfo {author} {\bibfnamefont {G.}~\bibnamefont {Feng}}, \bibinfo {author} {\bibfnamefont {Y.}~\bibnamefont {Jin}}, \bibinfo {author} {\bibfnamefont {A.}~\bibnamefont {Bartasyte}}, \bibinfo {author} {\bibfnamefont {S.}~\bibnamefont {Margueron}}, \bibinfo {author} {\bibfnamefont {M.}~\bibnamefont {Marangolo}}, \bibinfo {author} {\bibfnamefont {L.}~\bibnamefont {Thevenard}}, \bibinfo {author} {\bibfnamefont {P.}~\bibnamefont {Rovillain}}, \bibinfo {author} {\bibfnamefont {C.}~\bibnamefont {Gourdon}}, \bibinfo {author} {\bibfnamefont {S.}~\bibnamefont {{Hage-Ali}}}, \bibinfo {author} {\bibfnamefont {O.}~\bibnamefont {Elmazria}}, \bibinfo {author} {\bibfnamefont {H.}~\bibnamefont {Schmidt}}, \bibinfo {author} {\bibfnamefont {L.~Y.}\ \bibnamefont {Yeo}}, \bibinfo {author} {\bibfnamefont {L.~A.}\ \bibnamefont {Ambattu}},
  \bibinfo {author} {\bibfnamefont {J.~S.}\ \bibnamefont {Jeon}}, \bibinfo {author} {\bibfnamefont {D.}~\bibnamefont {Kwak}}, \bibinfo {author} {\bibfnamefont {J.}~\bibnamefont {Rufo}}, \bibinfo {author} {\bibfnamefont {S.}~\bibnamefont {Yang}},\ and\ \bibinfo {author} {\bibfnamefont {T.~J.}\ \bibnamefont {Huang}},\ }\href {https://doi.org/10.1088/1361-6463/ae258d} {\bibfield  {journal} {\bibinfo  {journal} {J. Phys. D: Appl. Phys.}\ }\textbf {\bibinfo {volume} {59}},\ \bibinfo {pages} {093001} (\bibinfo {year} {2026})}\BibitemShut {NoStop}%
\bibitem [{\citenamefont {Senellart}\ \emph {et~al.}(2017)\citenamefont {Senellart}, \citenamefont {Solomon},\ and\ \citenamefont {White}}]{Senellart2017}%
  \BibitemOpen
  \bibfield  {author} {\bibinfo {author} {\bibfnamefont {P.}~\bibnamefont {Senellart}}, \bibinfo {author} {\bibfnamefont {G.}~\bibnamefont {Solomon}},\ and\ \bibinfo {author} {\bibfnamefont {A.}~\bibnamefont {White}},\ }\href {https://doi.org/10.1038/nnano.2017.218} {\bibfield  {journal} {\bibinfo  {journal} {Nature Nanotechnology}\ }\textbf {\bibinfo {volume} {12}},\ \bibinfo {pages} {1026} (\bibinfo {year} {2017})}\BibitemShut {NoStop}%
\bibitem [{\citenamefont {Lo~Piparo}\ \emph {et~al.}(2019)\citenamefont {Lo~Piparo}, \citenamefont {Munro},\ and\ \citenamefont {Nemoto}}]{Piparo2019}%
  \BibitemOpen
  \bibfield  {author} {\bibinfo {author} {\bibfnamefont {N.}~\bibnamefont {Lo~Piparo}}, \bibinfo {author} {\bibfnamefont {W.~J.}\ \bibnamefont {Munro}},\ and\ \bibinfo {author} {\bibfnamefont {K.}~\bibnamefont {Nemoto}},\ }\href {https://doi.org/10.1103/PhysRevA.99.022337} {\bibfield  {journal} {\bibinfo  {journal} {Phys. Rev. A}\ }\textbf {\bibinfo {volume} {99}},\ \bibinfo {pages} {022337} (\bibinfo {year} {2019})}\BibitemShut {NoStop}%
\bibitem [{\citenamefont {Lu}\ \emph {et~al.}(2023)\citenamefont {Lu}, \citenamefont {Liscidini}, \citenamefont {Gaeta}, \citenamefont {Weiner},\ and\ \citenamefont {Lukens}}]{Lu2023}%
  \BibitemOpen
  \bibfield  {author} {\bibinfo {author} {\bibfnamefont {H.-H.}\ \bibnamefont {Lu}}, \bibinfo {author} {\bibfnamefont {M.}~\bibnamefont {Liscidini}}, \bibinfo {author} {\bibfnamefont {A.~L.}\ \bibnamefont {Gaeta}}, \bibinfo {author} {\bibfnamefont {A.~M.}\ \bibnamefont {Weiner}},\ and\ \bibinfo {author} {\bibfnamefont {J.~M.}\ \bibnamefont {Lukens}},\ }\href {https://doi.org/10.1364/OPTICA.506096} {\bibfield  {journal} {\bibinfo  {journal} {Optica}\ }\textbf {\bibinfo {volume} {10}},\ \bibinfo {pages} {1655} (\bibinfo {year} {2023})}\BibitemShut {NoStop}%
\bibitem [{\citenamefont {Pan}\ \emph {et~al.}(2012)\citenamefont {Pan}, \citenamefont {Chen}, \citenamefont {Lu}, \citenamefont {Weinfurter}, \citenamefont {Zeilinger},\ and\ \citenamefont {\ifmmode~\dot{Z}\else \.{Z}\fi{}ukowski}}]{Pan2012}%
  \BibitemOpen
  \bibfield  {author} {\bibinfo {author} {\bibfnamefont {J.-W.}\ \bibnamefont {Pan}}, \bibinfo {author} {\bibfnamefont {Z.-B.}\ \bibnamefont {Chen}}, \bibinfo {author} {\bibfnamefont {C.-Y.}\ \bibnamefont {Lu}}, \bibinfo {author} {\bibfnamefont {H.}~\bibnamefont {Weinfurter}}, \bibinfo {author} {\bibfnamefont {A.}~\bibnamefont {Zeilinger}},\ and\ \bibinfo {author} {\bibfnamefont {M.}~\bibnamefont {\ifmmode~\dot{Z}\else \.{Z}\fi{}ukowski}},\ }\href {https://doi.org/10.1103/RevModPhys.84.777} {\bibfield  {journal} {\bibinfo  {journal} {Rev. Mod. Phys.}\ }\textbf {\bibinfo {volume} {84}},\ \bibinfo {pages} {777} (\bibinfo {year} {2012})}\BibitemShut {NoStop}%
\bibitem [{\citenamefont {Stannigel}\ \emph {et~al.}(2010)\citenamefont {Stannigel}, \citenamefont {Rabl}, \citenamefont {S\o{}rensen}, \citenamefont {Zoller},\ and\ \citenamefont {Lukin}}]{Stannigel2010}%
  \BibitemOpen
  \bibfield  {author} {\bibinfo {author} {\bibfnamefont {K.}~\bibnamefont {Stannigel}}, \bibinfo {author} {\bibfnamefont {P.}~\bibnamefont {Rabl}}, \bibinfo {author} {\bibfnamefont {A.~S.}\ \bibnamefont {S\o{}rensen}}, \bibinfo {author} {\bibfnamefont {P.}~\bibnamefont {Zoller}},\ and\ \bibinfo {author} {\bibfnamefont {M.~D.}\ \bibnamefont {Lukin}},\ }\href {https://doi.org/10.1103/PhysRevLett.105.220501} {\bibfield  {journal} {\bibinfo  {journal} {Phys. Rev. Lett.}\ }\textbf {\bibinfo {volume} {105}},\ \bibinfo {pages} {220501} (\bibinfo {year} {2010})}\BibitemShut {NoStop}%
\bibitem [{\citenamefont {Groll}\ \emph {et~al.}(2026)\citenamefont {Groll}, \citenamefont {Wigger}, \citenamefont {Weiß}, \citenamefont {Yuan}, \citenamefont {Kuznetsov}, \citenamefont {Hernández-Mínguez}, \citenamefont {Krenner}, \citenamefont {Kuhn},\ and\ \citenamefont {Machnikowski}}]{Groll_2026}%
  \BibitemOpen
  \bibfield  {author} {\bibinfo {author} {\bibfnamefont {D.}~\bibnamefont {Groll}}, \bibinfo {author} {\bibfnamefont {D.}~\bibnamefont {Wigger}}, \bibinfo {author} {\bibfnamefont {M.}~\bibnamefont {Weiß}}, \bibinfo {author} {\bibfnamefont {M.}~\bibnamefont {Yuan}}, \bibinfo {author} {\bibfnamefont {A.}~\bibnamefont {Kuznetsov}}, \bibinfo {author} {\bibfnamefont {A.}~\bibnamefont {Hernández-Mínguez}}, \bibinfo {author} {\bibfnamefont {H.~J.}\ \bibnamefont {Krenner}}, \bibinfo {author} {\bibfnamefont {T.}~\bibnamefont {Kuhn}},\ and\ \bibinfo {author} {\bibfnamefont {P.}~\bibnamefont {Machnikowski}},\ }\href {https://doi.org/10.1088/2515-7647/ae4f43} {\bibfield  {journal} {\bibinfo  {journal} {Journal of Physics: Photonics}\ }\textbf {\bibinfo {volume} {8}},\ \bibinfo {pages} {012008} (\bibinfo {year} {2026})}\BibitemShut {NoStop}%
\bibitem [{\citenamefont {Groll}\ \emph {et~al.}(2025)\citenamefont {Groll}, \citenamefont {Paschen}, \citenamefont {Machnikowski}, \citenamefont {Hess}, \citenamefont {Wigger},\ and\ \citenamefont {Kuhn}}]{Groll2025}%
  \BibitemOpen
  \bibfield  {author} {\bibinfo {author} {\bibfnamefont {D.}~\bibnamefont {Groll}}, \bibinfo {author} {\bibfnamefont {F.}~\bibnamefont {Paschen}}, \bibinfo {author} {\bibfnamefont {P.}~\bibnamefont {Machnikowski}}, \bibinfo {author} {\bibfnamefont {O.}~\bibnamefont {Hess}}, \bibinfo {author} {\bibfnamefont {D.}~\bibnamefont {Wigger}},\ and\ \bibinfo {author} {\bibfnamefont {T.}~\bibnamefont {Kuhn}},\ }\href {https://doi.org/https://doi.org/10.1002/qute.202300153} {\bibfield  {journal} {\bibinfo  {journal} {Advanced Quantum Technologies}\ }\textbf {\bibinfo {volume} {8}},\ \bibinfo {pages} {2300153} (\bibinfo {year} {2025})}\BibitemShut {NoStop}%
\bibitem [{\citenamefont {Tran}\ \emph {et~al.}(2016)\citenamefont {Tran}, \citenamefont {Bray}, \citenamefont {Ford}, \citenamefont {Toth},\ and\ \citenamefont {Aharonovich}}]{Tran2016}%
  \BibitemOpen
  \bibfield  {author} {\bibinfo {author} {\bibfnamefont {T.~T.}\ \bibnamefont {Tran}}, \bibinfo {author} {\bibfnamefont {K.}~\bibnamefont {Bray}}, \bibinfo {author} {\bibfnamefont {M.~J.}\ \bibnamefont {Ford}}, \bibinfo {author} {\bibfnamefont {M.}~\bibnamefont {Toth}},\ and\ \bibinfo {author} {\bibfnamefont {I.}~\bibnamefont {Aharonovich}},\ }\href {https://doi.org/10.1038/nnano.2015.242} {\bibfield  {journal} {\bibinfo  {journal} {Nature Nanotechnology}\ }\textbf {\bibinfo {volume} {11}},\ \bibinfo {pages} {37} (\bibinfo {year} {2016})}\BibitemShut {NoStop}%
\bibitem [{\citenamefont {Mart\'{\i}nez}\ \emph {et~al.}(2016)\citenamefont {Mart\'{\i}nez}, \citenamefont {Pelini}, \citenamefont {Waselowski}, \citenamefont {Maze}, \citenamefont {Gil}, \citenamefont {Cassabois},\ and\ \citenamefont {Jacques}}]{Martinez2016}%
  \BibitemOpen
  \bibfield  {author} {\bibinfo {author} {\bibfnamefont {L.~J.}\ \bibnamefont {Mart\'{\i}nez}}, \bibinfo {author} {\bibfnamefont {T.}~\bibnamefont {Pelini}}, \bibinfo {author} {\bibfnamefont {V.}~\bibnamefont {Waselowski}}, \bibinfo {author} {\bibfnamefont {J.~R.}\ \bibnamefont {Maze}}, \bibinfo {author} {\bibfnamefont {B.}~\bibnamefont {Gil}}, \bibinfo {author} {\bibfnamefont {G.}~\bibnamefont {Cassabois}},\ and\ \bibinfo {author} {\bibfnamefont {V.}~\bibnamefont {Jacques}},\ }\href {https://doi.org/10.1103/PhysRevB.94.121405} {\bibfield  {journal} {\bibinfo  {journal} {Phys. Rev. B}\ }\textbf {\bibinfo {volume} {94}},\ \bibinfo {pages} {121405} (\bibinfo {year} {2016})}\BibitemShut {NoStop}%
\bibitem [{\citenamefont {Jungwirth}\ \emph {et~al.}(2016)\citenamefont {Jungwirth}, \citenamefont {Calderon}, \citenamefont {Ji}, \citenamefont {Spencer}, \citenamefont {Flatt{\'e}},\ and\ \citenamefont {Fuchs}}]{Jungwirth2016}%
  \BibitemOpen
  \bibfield  {author} {\bibinfo {author} {\bibfnamefont {N.~R.}\ \bibnamefont {Jungwirth}}, \bibinfo {author} {\bibfnamefont {B.}~\bibnamefont {Calderon}}, \bibinfo {author} {\bibfnamefont {Y.}~\bibnamefont {Ji}}, \bibinfo {author} {\bibfnamefont {M.~G.}\ \bibnamefont {Spencer}}, \bibinfo {author} {\bibfnamefont {M.~E.}\ \bibnamefont {Flatt{\'e}}},\ and\ \bibinfo {author} {\bibfnamefont {G.~D.}\ \bibnamefont {Fuchs}},\ }\href {https://doi.org/10.1021/acs.nanolett.6b01987} {\bibfield  {journal} {\bibinfo  {journal} {Nano Letters}\ }\textbf {\bibinfo {volume} {16}},\ \bibinfo {pages} {6052} (\bibinfo {year} {2016})}\BibitemShut {NoStop}%
\bibitem [{\citenamefont {Kianinia}\ \emph {et~al.}(2017)\citenamefont {Kianinia}, \citenamefont {Regan}, \citenamefont {Tawfik}, \citenamefont {Tran}, \citenamefont {Ford}, \citenamefont {Aharonovich},\ and\ \citenamefont {Toth}}]{Kianinia2017}%
  \BibitemOpen
  \bibfield  {author} {\bibinfo {author} {\bibfnamefont {M.}~\bibnamefont {Kianinia}}, \bibinfo {author} {\bibfnamefont {B.}~\bibnamefont {Regan}}, \bibinfo {author} {\bibfnamefont {S.~A.}\ \bibnamefont {Tawfik}}, \bibinfo {author} {\bibfnamefont {T.~T.}\ \bibnamefont {Tran}}, \bibinfo {author} {\bibfnamefont {M.~J.}\ \bibnamefont {Ford}}, \bibinfo {author} {\bibfnamefont {I.}~\bibnamefont {Aharonovich}},\ and\ \bibinfo {author} {\bibfnamefont {M.}~\bibnamefont {Toth}},\ }\href {https://doi.org/10.1021/acsphotonics.7b00086} {\bibfield  {journal} {\bibinfo  {journal} {ACS Photonics}\ }\textbf {\bibinfo {volume} {4}},\ \bibinfo {pages} {768} (\bibinfo {year} {2017})}\BibitemShut {NoStop}%
\bibitem [{\citenamefont {Raussendorf}\ and\ \citenamefont {Briegel}(2001)}]{raussendorf2001}%
  \BibitemOpen
  \bibfield  {author} {\bibinfo {author} {\bibfnamefont {R.}~\bibnamefont {Raussendorf}}\ and\ \bibinfo {author} {\bibfnamefont {H.~J.}\ \bibnamefont {Briegel}},\ }\href {https://doi.org/10.1103/PhysRevLett.86.5188} {\bibfield  {journal} {\bibinfo  {journal} {Phys. Rev. Lett.}\ }\textbf {\bibinfo {volume} {86}},\ \bibinfo {pages} {5188} (\bibinfo {year} {2001})}\BibitemShut {NoStop}%
\bibitem [{\citenamefont {Browne}\ and\ \citenamefont {Rudolph}(2005)}]{browne2005}%
  \BibitemOpen
  \bibfield  {author} {\bibinfo {author} {\bibfnamefont {D.~E.}\ \bibnamefont {Browne}}\ and\ \bibinfo {author} {\bibfnamefont {T.}~\bibnamefont {Rudolph}},\ }\href {https://doi.org/10.1103/PhysRevLett.95.010501} {\bibfield  {journal} {\bibinfo  {journal} {Phys. Rev. Lett.}\ }\textbf {\bibinfo {volume} {95}},\ \bibinfo {pages} {010501} (\bibinfo {year} {2005})}\BibitemShut {NoStop}%
\bibitem [{\citenamefont {Ding}\ \emph {et~al.}(2016)\citenamefont {Ding}, \citenamefont {He}, \citenamefont {Duan}, \citenamefont {Gregersen}, \citenamefont {Chen}, \citenamefont {Unsleber}, \citenamefont {Maier}, \citenamefont {Schneider}, \citenamefont {Kamp}, \citenamefont {H{\"o}fling}, \citenamefont {Lu},\ and\ \citenamefont {Pan}}]{ding2016}%
  \BibitemOpen
  \bibfield  {author} {\bibinfo {author} {\bibfnamefont {X.}~\bibnamefont {Ding}}, \bibinfo {author} {\bibfnamefont {Y.}~\bibnamefont {He}}, \bibinfo {author} {\bibfnamefont {Z.-C.}\ \bibnamefont {Duan}}, \bibinfo {author} {\bibfnamefont {N.}~\bibnamefont {Gregersen}}, \bibinfo {author} {\bibfnamefont {M.-C.}\ \bibnamefont {Chen}}, \bibinfo {author} {\bibfnamefont {S.}~\bibnamefont {Unsleber}}, \bibinfo {author} {\bibfnamefont {S.}~\bibnamefont {Maier}}, \bibinfo {author} {\bibfnamefont {C.}~\bibnamefont {Schneider}}, \bibinfo {author} {\bibfnamefont {M.}~\bibnamefont {Kamp}}, \bibinfo {author} {\bibfnamefont {S.}~\bibnamefont {H{\"o}fling}}, \bibinfo {author} {\bibfnamefont {C.-Y.}\ \bibnamefont {Lu}},\ and\ \bibinfo {author} {\bibfnamefont {J.-W.}\ \bibnamefont {Pan}},\ }\href {https://doi.org/10.1103/PhysRevLett.116.020401} {\bibfield  {journal} {\bibinfo  {journal} {Phys. Rev. Lett.}\ }\textbf {\bibinfo {volume} {116}},\ \bibinfo {pages} {020401} (\bibinfo {year} {2016})}\BibitemShut {NoStop}%
\bibitem [{\citenamefont {Wang}\ \emph {et~al.}(2016)\citenamefont {Wang}, \citenamefont {Duan}, \citenamefont {Li}, \citenamefont {Chen}, \citenamefont {Li}, \citenamefont {He}, \citenamefont {Chen}, \citenamefont {He}, \citenamefont {Ding}, \citenamefont {Peng}, \citenamefont {Schneider}, \citenamefont {Kamp}, \citenamefont {H{\"o}fling}, \citenamefont {Lu},\ and\ \citenamefont {Pan}}]{wang2016}%
  \BibitemOpen
  \bibfield  {author} {\bibinfo {author} {\bibfnamefont {H.}~\bibnamefont {Wang}}, \bibinfo {author} {\bibfnamefont {Z.-C.}\ \bibnamefont {Duan}}, \bibinfo {author} {\bibfnamefont {Y.-H.}\ \bibnamefont {Li}}, \bibinfo {author} {\bibfnamefont {S.}~\bibnamefont {Chen}}, \bibinfo {author} {\bibfnamefont {J.-P.}\ \bibnamefont {Li}}, \bibinfo {author} {\bibfnamefont {Y.-M.}\ \bibnamefont {He}}, \bibinfo {author} {\bibfnamefont {M.-C.}\ \bibnamefont {Chen}}, \bibinfo {author} {\bibfnamefont {Y.}~\bibnamefont {He}}, \bibinfo {author} {\bibfnamefont {X.}~\bibnamefont {Ding}}, \bibinfo {author} {\bibfnamefont {C.-Z.}\ \bibnamefont {Peng}}, \bibinfo {author} {\bibfnamefont {C.}~\bibnamefont {Schneider}}, \bibinfo {author} {\bibfnamefont {M.}~\bibnamefont {Kamp}}, \bibinfo {author} {\bibfnamefont {S.}~\bibnamefont {H{\"o}fling}}, \bibinfo {author} {\bibfnamefont {C.-Y.}\ \bibnamefont {Lu}},\ and\ \bibinfo {author} {\bibfnamefont {J.-W.}\ \bibnamefont {Pan}},\ }\href {https://doi.org/10.1103/PhysRevLett.116.213601}
  {\bibfield  {journal} {\bibinfo  {journal} {Phys. Rev. Lett.}\ }\textbf {\bibinfo {volume} {116}},\ \bibinfo {pages} {213601} (\bibinfo {year} {2016})}\BibitemShut {NoStop}%
\bibitem [{\citenamefont {Kuhlmann}\ \emph {et~al.}(2015)\citenamefont {Kuhlmann}, \citenamefont {Prechtel}, \citenamefont {Houel}, \citenamefont {Ludwig}, \citenamefont {Reuter}, \citenamefont {Wieck},\ and\ \citenamefont {Warburton}}]{kuhlmann2015}%
  \BibitemOpen
  \bibfield  {author} {\bibinfo {author} {\bibfnamefont {A.~V.}\ \bibnamefont {Kuhlmann}}, \bibinfo {author} {\bibfnamefont {J.~H.}\ \bibnamefont {Prechtel}}, \bibinfo {author} {\bibfnamefont {J.}~\bibnamefont {Houel}}, \bibinfo {author} {\bibfnamefont {A.}~\bibnamefont {Ludwig}}, \bibinfo {author} {\bibfnamefont {D.}~\bibnamefont {Reuter}}, \bibinfo {author} {\bibfnamefont {A.~D.}\ \bibnamefont {Wieck}},\ and\ \bibinfo {author} {\bibfnamefont {R.~J.}\ \bibnamefont {Warburton}},\ }\href {https://doi.org/10.1038/ncomms9204} {\bibfield  {journal} {\bibinfo  {journal} {Nat. Commun.}\ }\textbf {\bibinfo {volume} {6}},\ \bibinfo {pages} {8204} (\bibinfo {year} {2015})}\BibitemShut {NoStop}%
\bibitem [{\citenamefont {Somaschi}\ \emph {et~al.}(2016)\citenamefont {Somaschi}, \citenamefont {Giesz}, \citenamefont {De~Santis}, \citenamefont {Loredo}, \citenamefont {Almeida}, \citenamefont {Hornecker}, \citenamefont {Portalupi}, \citenamefont {Grange}, \citenamefont {Ant{\'o}n}, \citenamefont {Demory}, \citenamefont {G{\'o}mez}, \citenamefont {Sagnes}, \citenamefont {Lanzillotti-Kimura}, \citenamefont {Lema{\^\i}tre}, \citenamefont {Auffeves}, \citenamefont {White}, \citenamefont {Lanco},\ and\ \citenamefont {Senellart}}]{somaschi2016}%
  \BibitemOpen
  \bibfield  {author} {\bibinfo {author} {\bibfnamefont {N.}~\bibnamefont {Somaschi}}, \bibinfo {author} {\bibfnamefont {V.}~\bibnamefont {Giesz}}, \bibinfo {author} {\bibfnamefont {L.}~\bibnamefont {De~Santis}}, \bibinfo {author} {\bibfnamefont {J.~C.}\ \bibnamefont {Loredo}}, \bibinfo {author} {\bibfnamefont {M.~P.}\ \bibnamefont {Almeida}}, \bibinfo {author} {\bibfnamefont {G.}~\bibnamefont {Hornecker}}, \bibinfo {author} {\bibfnamefont {S.~L.}\ \bibnamefont {Portalupi}}, \bibinfo {author} {\bibfnamefont {T.}~\bibnamefont {Grange}}, \bibinfo {author} {\bibfnamefont {C.}~\bibnamefont {Ant{\'o}n}}, \bibinfo {author} {\bibfnamefont {J.}~\bibnamefont {Demory}}, \bibinfo {author} {\bibfnamefont {C.}~\bibnamefont {G{\'o}mez}}, \bibinfo {author} {\bibfnamefont {I.}~\bibnamefont {Sagnes}}, \bibinfo {author} {\bibfnamefont {N.~D.}\ \bibnamefont {Lanzillotti-Kimura}}, \bibinfo {author} {\bibfnamefont {A.}~\bibnamefont {Lema{\^\i}tre}}, \bibinfo {author} {\bibfnamefont {A.}~\bibnamefont {Auffeves}}, \bibinfo {author}
  {\bibfnamefont {A.~G.}\ \bibnamefont {White}}, \bibinfo {author} {\bibfnamefont {L.}~\bibnamefont {Lanco}},\ and\ \bibinfo {author} {\bibfnamefont {P.}~\bibnamefont {Senellart}},\ }\href {https://doi.org/10.1038/nphoton.2016.23} {\bibfield  {journal} {\bibinfo  {journal} {Nat. Photonics}\ }\textbf {\bibinfo {volume} {10}},\ \bibinfo {pages} {340} (\bibinfo {year} {2016})}\BibitemShut {NoStop}%
\bibitem [{\citenamefont {Tomm}\ \emph {et~al.}(2021)\citenamefont {Tomm}, \citenamefont {Javadi}, \citenamefont {Antoniadis}, \citenamefont {Najer}, \citenamefont {L{\"o}bl}, \citenamefont {Korsch}, \citenamefont {Schott}, \citenamefont {Valentin}, \citenamefont {Wieck}, \citenamefont {Ludwig},\ and\ \citenamefont {Warburton}}]{tomm2021}%
  \BibitemOpen
  \bibfield  {author} {\bibinfo {author} {\bibfnamefont {N.}~\bibnamefont {Tomm}}, \bibinfo {author} {\bibfnamefont {A.}~\bibnamefont {Javadi}}, \bibinfo {author} {\bibfnamefont {N.~O.}\ \bibnamefont {Antoniadis}}, \bibinfo {author} {\bibfnamefont {D.}~\bibnamefont {Najer}}, \bibinfo {author} {\bibfnamefont {M.~C.}\ \bibnamefont {L{\"o}bl}}, \bibinfo {author} {\bibfnamefont {A.~R.}\ \bibnamefont {Korsch}}, \bibinfo {author} {\bibfnamefont {R.}~\bibnamefont {Schott}}, \bibinfo {author} {\bibfnamefont {S.~R.}\ \bibnamefont {Valentin}}, \bibinfo {author} {\bibfnamefont {A.~D.}\ \bibnamefont {Wieck}}, \bibinfo {author} {\bibfnamefont {A.}~\bibnamefont {Ludwig}},\ and\ \bibinfo {author} {\bibfnamefont {R.~J.}\ \bibnamefont {Warburton}},\ }\href {https://doi.org/10.1038/s41565-020-00831-x} {\bibfield  {journal} {\bibinfo  {journal} {Nat. Nanotechnol.}\ }\textbf {\bibinfo {volume} {16}},\ \bibinfo {pages} {399} (\bibinfo {year} {2021})}\BibitemShut {NoStop}%
\bibitem [{\citenamefont {Wei}\ \emph {et~al.}(2014{\natexlab{a}})\citenamefont {Wei}, \citenamefont {He}, \citenamefont {He}, \citenamefont {Lu}, \citenamefont {Pan}, \citenamefont {Schneider}, \citenamefont {Kamp}, \citenamefont {H\"ofling}, \citenamefont {McCutcheon},\ and\ \citenamefont {Nazir}}]{wei2014}%
  \BibitemOpen
  \bibfield  {author} {\bibinfo {author} {\bibfnamefont {Y.-J.}\ \bibnamefont {Wei}}, \bibinfo {author} {\bibfnamefont {Y.}~\bibnamefont {He}}, \bibinfo {author} {\bibfnamefont {Y.-M.}\ \bibnamefont {He}}, \bibinfo {author} {\bibfnamefont {C.-Y.}\ \bibnamefont {Lu}}, \bibinfo {author} {\bibfnamefont {J.-W.}\ \bibnamefont {Pan}}, \bibinfo {author} {\bibfnamefont {C.}~\bibnamefont {Schneider}}, \bibinfo {author} {\bibfnamefont {M.}~\bibnamefont {Kamp}}, \bibinfo {author} {\bibfnamefont {S.}~\bibnamefont {H\"ofling}}, \bibinfo {author} {\bibfnamefont {D.~P.~S.}\ \bibnamefont {McCutcheon}},\ and\ \bibinfo {author} {\bibfnamefont {A.}~\bibnamefont {Nazir}},\ }\href {https://doi.org/10.1103/PhysRevLett.113.097401} {\bibfield  {journal} {\bibinfo  {journal} {Phys. Rev. Lett.}\ }\textbf {\bibinfo {volume} {113}},\ \bibinfo {pages} {097401} (\bibinfo {year} {2014}{\natexlab{a}})}\BibitemShut {NoStop}%
\bibitem [{\citenamefont {Wei}\ \emph {et~al.}(2014{\natexlab{b}})\citenamefont {Wei}, \citenamefont {He}, \citenamefont {Chen}, \citenamefont {Hu}, \citenamefont {He}, \citenamefont {Wu}, \citenamefont {Schneider}, \citenamefont {Kamp}, \citenamefont {H{\"o}fling}, \citenamefont {Lu},\ and\ \citenamefont {Pan}}]{wei2014a}%
  \BibitemOpen
  \bibfield  {author} {\bibinfo {author} {\bibfnamefont {Y.-J.}\ \bibnamefont {Wei}}, \bibinfo {author} {\bibfnamefont {Y.-M.}\ \bibnamefont {He}}, \bibinfo {author} {\bibfnamefont {M.-C.}\ \bibnamefont {Chen}}, \bibinfo {author} {\bibfnamefont {Y.-N.}\ \bibnamefont {Hu}}, \bibinfo {author} {\bibfnamefont {Y.}~\bibnamefont {He}}, \bibinfo {author} {\bibfnamefont {D.}~\bibnamefont {Wu}}, \bibinfo {author} {\bibfnamefont {C.}~\bibnamefont {Schneider}}, \bibinfo {author} {\bibfnamefont {M.}~\bibnamefont {Kamp}}, \bibinfo {author} {\bibfnamefont {S.}~\bibnamefont {H{\"o}fling}}, \bibinfo {author} {\bibfnamefont {C.-Y.}\ \bibnamefont {Lu}},\ and\ \bibinfo {author} {\bibfnamefont {J.-W.}\ \bibnamefont {Pan}},\ }\href {https://doi.org/10.1021/nl503081n} {\bibfield  {journal} {\bibinfo  {journal} {Nano Lett.}\ }\textbf {\bibinfo {volume} {14}},\ \bibinfo {pages} {6515} (\bibinfo {year} {2014}{\natexlab{b}})}\BibitemShut {NoStop}%
\bibitem [{\citenamefont {Wang}\ \emph {et~al.}(2019)\citenamefont {Wang}, \citenamefont {He}, \citenamefont {Chung}, \citenamefont {Hu}, \citenamefont {Yu}, \citenamefont {Chen}, \citenamefont {Ding}, \citenamefont {Chen}, \citenamefont {Qin}, \citenamefont {Yang}, \citenamefont {Liu}, \citenamefont {Duan}, \citenamefont {Li}, \citenamefont {Gerhardt}, \citenamefont {Winkler}, \citenamefont {Jurkat}, \citenamefont {Wang}, \citenamefont {Gregersen}, \citenamefont {Huo}, \citenamefont {Dai}, \citenamefont {Yu}, \citenamefont {H{\"o}fling}, \citenamefont {Lu},\ and\ \citenamefont {Pan}}]{wang2019}%
  \BibitemOpen
  \bibfield  {author} {\bibinfo {author} {\bibfnamefont {H.}~\bibnamefont {Wang}}, \bibinfo {author} {\bibfnamefont {Y.-M.}\ \bibnamefont {He}}, \bibinfo {author} {\bibfnamefont {T.-H.}\ \bibnamefont {Chung}}, \bibinfo {author} {\bibfnamefont {H.}~\bibnamefont {Hu}}, \bibinfo {author} {\bibfnamefont {Y.}~\bibnamefont {Yu}}, \bibinfo {author} {\bibfnamefont {S.}~\bibnamefont {Chen}}, \bibinfo {author} {\bibfnamefont {X.}~\bibnamefont {Ding}}, \bibinfo {author} {\bibfnamefont {M.-C.}\ \bibnamefont {Chen}}, \bibinfo {author} {\bibfnamefont {J.}~\bibnamefont {Qin}}, \bibinfo {author} {\bibfnamefont {X.}~\bibnamefont {Yang}}, \bibinfo {author} {\bibfnamefont {R.-Z.}\ \bibnamefont {Liu}}, \bibinfo {author} {\bibfnamefont {Z.-C.}\ \bibnamefont {Duan}}, \bibinfo {author} {\bibfnamefont {J.-P.}\ \bibnamefont {Li}}, \bibinfo {author} {\bibfnamefont {S.}~\bibnamefont {Gerhardt}}, \bibinfo {author} {\bibfnamefont {K.}~\bibnamefont {Winkler}}, \bibinfo {author} {\bibfnamefont {J.}~\bibnamefont {Jurkat}}, \bibinfo {author}
  {\bibfnamefont {L.-J.}\ \bibnamefont {Wang}}, \bibinfo {author} {\bibfnamefont {N.}~\bibnamefont {Gregersen}}, \bibinfo {author} {\bibfnamefont {Y.-H.}\ \bibnamefont {Huo}}, \bibinfo {author} {\bibfnamefont {Q.}~\bibnamefont {Dai}}, \bibinfo {author} {\bibfnamefont {S.}~\bibnamefont {Yu}}, \bibinfo {author} {\bibfnamefont {S.}~\bibnamefont {H{\"o}fling}}, \bibinfo {author} {\bibfnamefont {C.-Y.}\ \bibnamefont {Lu}},\ and\ \bibinfo {author} {\bibfnamefont {J.-W.}\ \bibnamefont {Pan}},\ }\href {https://doi.org/10.1038/s41566-019-0494-3} {\bibfield  {journal} {\bibinfo  {journal} {Nat. Photonics}\ }\textbf {\bibinfo {volume} {13}},\ \bibinfo {pages} {770} (\bibinfo {year} {2019})}\BibitemShut {NoStop}%
\bibitem [{\citenamefont {Besombes}\ \emph {et~al.}(2001)\citenamefont {Besombes}, \citenamefont {Kheng}, \citenamefont {Marsal},\ and\ \citenamefont {Mariette}}]{Besombes2001}%
  \BibitemOpen
  \bibfield  {author} {\bibinfo {author} {\bibfnamefont {L.}~\bibnamefont {Besombes}}, \bibinfo {author} {\bibfnamefont {K.}~\bibnamefont {Kheng}}, \bibinfo {author} {\bibfnamefont {L.}~\bibnamefont {Marsal}},\ and\ \bibinfo {author} {\bibfnamefont {H.}~\bibnamefont {Mariette}},\ }\href {https://doi.org/10.1103/PhysRevB.63.155307} {\bibfield  {journal} {\bibinfo  {journal} {Phys. Rev. B}\ }\textbf {\bibinfo {volume} {63}},\ \bibinfo {pages} {155307} (\bibinfo {year} {2001})}\BibitemShut {NoStop}%
\bibitem [{\citenamefont {Favero}\ \emph {et~al.}(2003)\citenamefont {Favero}, \citenamefont {Cassabois}, \citenamefont {Ferreira}, \citenamefont {Darson}, \citenamefont {Voisin}, \citenamefont {Tignon}, \citenamefont {Delalande}, \citenamefont {Bastard}, \citenamefont {Roussignol},\ and\ \citenamefont {G\'erard}}]{Favero2003}%
  \BibitemOpen
  \bibfield  {author} {\bibinfo {author} {\bibfnamefont {I.}~\bibnamefont {Favero}}, \bibinfo {author} {\bibfnamefont {G.}~\bibnamefont {Cassabois}}, \bibinfo {author} {\bibfnamefont {R.}~\bibnamefont {Ferreira}}, \bibinfo {author} {\bibfnamefont {D.}~\bibnamefont {Darson}}, \bibinfo {author} {\bibfnamefont {C.}~\bibnamefont {Voisin}}, \bibinfo {author} {\bibfnamefont {J.}~\bibnamefont {Tignon}}, \bibinfo {author} {\bibfnamefont {C.}~\bibnamefont {Delalande}}, \bibinfo {author} {\bibfnamefont {G.}~\bibnamefont {Bastard}}, \bibinfo {author} {\bibfnamefont {P.}~\bibnamefont {Roussignol}},\ and\ \bibinfo {author} {\bibfnamefont {J.~M.}\ \bibnamefont {G\'erard}},\ }\href {https://doi.org/10.1103/PhysRevB.68.233301} {\bibfield  {journal} {\bibinfo  {journal} {Phys. Rev. B}\ }\textbf {\bibinfo {volume} {68}},\ \bibinfo {pages} {233301} (\bibinfo {year} {2003})}\BibitemShut {NoStop}%
\bibitem [{\citenamefont {Vagov}\ \emph {et~al.}(2004)\citenamefont {Vagov}, \citenamefont {Axt}, \citenamefont {Kuhn}, \citenamefont {Langbein}, \citenamefont {Borri},\ and\ \citenamefont {Woggon}}]{Vagov2004}%
  \BibitemOpen
  \bibfield  {author} {\bibinfo {author} {\bibfnamefont {A.}~\bibnamefont {Vagov}}, \bibinfo {author} {\bibfnamefont {V.~M.}\ \bibnamefont {Axt}}, \bibinfo {author} {\bibfnamefont {T.}~\bibnamefont {Kuhn}}, \bibinfo {author} {\bibfnamefont {W.}~\bibnamefont {Langbein}}, \bibinfo {author} {\bibfnamefont {P.}~\bibnamefont {Borri}},\ and\ \bibinfo {author} {\bibfnamefont {U.}~\bibnamefont {Woggon}},\ }\href {https://doi.org/10.1103/PhysRevB.70.201305} {\bibfield  {journal} {\bibinfo  {journal} {Phys. Rev. B}\ }\textbf {\bibinfo {volume} {70}},\ \bibinfo {pages} {201305(R)} (\bibinfo {year} {2004})}\BibitemShut {NoStop}%
\bibitem [{\citenamefont {Krummheuer}\ \emph {et~al.}(2002)\citenamefont {Krummheuer}, \citenamefont {Axt},\ and\ \citenamefont {Kuhn}}]{Krummheuer2002}%
  \BibitemOpen
  \bibfield  {author} {\bibinfo {author} {\bibfnamefont {B.}~\bibnamefont {Krummheuer}}, \bibinfo {author} {\bibfnamefont {V.~M.}\ \bibnamefont {Axt}},\ and\ \bibinfo {author} {\bibfnamefont {T.}~\bibnamefont {Kuhn}},\ }\href {https://doi.org/10.1103/PhysRevB.65.195313} {\bibfield  {journal} {\bibinfo  {journal} {Phys. Rev. B}\ }\textbf {\bibinfo {volume} {65}},\ \bibinfo {pages} {195313} (\bibinfo {year} {2002})}\BibitemShut {NoStop}%
\bibitem [{\citenamefont {Huang}\ and\ \citenamefont {Rhys}(1950)}]{Huang1950}%
  \BibitemOpen
  \bibfield  {author} {\bibinfo {author} {\bibfnamefont {K.}~\bibnamefont {Huang}}\ and\ \bibinfo {author} {\bibfnamefont {A.}~\bibnamefont {Rhys}},\ }\href@noop {} {\bibfield  {journal} {\bibinfo  {journal} {Proc. R. Soc. Lond.}\ }\textbf {\bibinfo {volume} {204}},\ \bibinfo {pages} {406} (\bibinfo {year} {1950})}\BibitemShut {NoStop}%
\bibitem [{\citenamefont {Wigger}\ \emph {et~al.}(2019)\citenamefont {Wigger}, \citenamefont {Schmidt}, \citenamefont {Del Pozo-Zamudio}, \citenamefont {Preuß}, \citenamefont {Tonndorf}, \citenamefont {Schneider}, \citenamefont {Steeger}, \citenamefont {Kern}, \citenamefont {Khodaei}, \citenamefont {Sperling}, \citenamefont {de~Vasconcellos}, \citenamefont {Bratschitsch},\ and\ \citenamefont {Kuhn}}]{Wigger2019}%
  \BibitemOpen
  \bibfield  {author} {\bibinfo {author} {\bibfnamefont {D.}~\bibnamefont {Wigger}}, \bibinfo {author} {\bibfnamefont {R.}~\bibnamefont {Schmidt}}, \bibinfo {author} {\bibfnamefont {O.}~\bibnamefont {Del Pozo-Zamudio}}, \bibinfo {author} {\bibfnamefont {J.~A.}\ \bibnamefont {Preuß}}, \bibinfo {author} {\bibfnamefont {P.}~\bibnamefont {Tonndorf}}, \bibinfo {author} {\bibfnamefont {R.}~\bibnamefont {Schneider}}, \bibinfo {author} {\bibfnamefont {P.}~\bibnamefont {Steeger}}, \bibinfo {author} {\bibfnamefont {J.}~\bibnamefont {Kern}}, \bibinfo {author} {\bibfnamefont {Y.}~\bibnamefont {Khodaei}}, \bibinfo {author} {\bibfnamefont {J.}~\bibnamefont {Sperling}}, \bibinfo {author} {\bibfnamefont {S.~M.}\ \bibnamefont {de~Vasconcellos}}, \bibinfo {author} {\bibfnamefont {R.}~\bibnamefont {Bratschitsch}},\ and\ \bibinfo {author} {\bibfnamefont {T.}~\bibnamefont {Kuhn}},\ }\href {https://doi.org/10.1088/2053-1583/ab1188} {\bibfield  {journal} {\bibinfo  {journal} {2D Materials}\ }\textbf {\bibinfo {volume} {6}},\
  \bibinfo {pages} {035006} (\bibinfo {year} {2019})}\BibitemShut {NoStop}%
\bibitem [{\citenamefont {Khatri}\ \emph {et~al.}(2019)\citenamefont {Khatri}, \citenamefont {Luxmoore},\ and\ \citenamefont {Ramsay}}]{Khatri2019}%
  \BibitemOpen
  \bibfield  {author} {\bibinfo {author} {\bibfnamefont {P.}~\bibnamefont {Khatri}}, \bibinfo {author} {\bibfnamefont {I.~J.}\ \bibnamefont {Luxmoore}},\ and\ \bibinfo {author} {\bibfnamefont {A.~J.}\ \bibnamefont {Ramsay}},\ }\href {https://doi.org/10.1103/PhysRevB.100.125305} {\bibfield  {journal} {\bibinfo  {journal} {Phys. Rev. B}\ }\textbf {\bibinfo {volume} {100}},\ \bibinfo {pages} {125305} (\bibinfo {year} {2019})}\BibitemShut {NoStop}%
\bibitem [{\citenamefont {Preuss}\ \emph {et~al.}(2022)\citenamefont {Preuss}, \citenamefont {Groll}, \citenamefont {Schmidt}, \citenamefont {Hahn}, \citenamefont {Machnikowski}, \citenamefont {Bratschitsch}, \citenamefont {Kuhn}, \citenamefont {de~Vasconcellos},\ and\ \citenamefont {Wigger}}]{Preuss2022}%
  \BibitemOpen
  \bibfield  {author} {\bibinfo {author} {\bibfnamefont {J.~A.}\ \bibnamefont {Preuss}}, \bibinfo {author} {\bibfnamefont {D.}~\bibnamefont {Groll}}, \bibinfo {author} {\bibfnamefont {R.}~\bibnamefont {Schmidt}}, \bibinfo {author} {\bibfnamefont {T.}~\bibnamefont {Hahn}}, \bibinfo {author} {\bibfnamefont {P.}~\bibnamefont {Machnikowski}}, \bibinfo {author} {\bibfnamefont {R.}~\bibnamefont {Bratschitsch}}, \bibinfo {author} {\bibfnamefont {T.}~\bibnamefont {Kuhn}}, \bibinfo {author} {\bibfnamefont {S.~M.}\ \bibnamefont {de~Vasconcellos}},\ and\ \bibinfo {author} {\bibfnamefont {D.}~\bibnamefont {Wigger}},\ }\href {https://doi.org/10.1364/OPTICA.448124} {\bibfield  {journal} {\bibinfo  {journal} {Optica}\ }\textbf {\bibinfo {volume} {9}},\ \bibinfo {pages} {522} (\bibinfo {year} {2022})}\BibitemShut {NoStop}%
\bibitem [{\citenamefont {Muljarov}\ and\ \citenamefont {Zimmermann}(2004)}]{muljarov2004}%
  \BibitemOpen
  \bibfield  {author} {\bibinfo {author} {\bibfnamefont {E.~A.}\ \bibnamefont {Muljarov}}\ and\ \bibinfo {author} {\bibfnamefont {R.}~\bibnamefont {Zimmermann}},\ }\href {https://doi.org/10.1103/PhysRevLett.93.237401} {\bibfield  {journal} {\bibinfo  {journal} {Phys. Rev. Lett.}\ }\textbf {\bibinfo {volume} {93}},\ \bibinfo {pages} {237401} (\bibinfo {year} {2004})}\BibitemShut {NoStop}%
\bibitem [{\citenamefont {Machnikowski}(2006{\natexlab{a}})}]{machnikowski2006prl}%
  \BibitemOpen
  \bibfield  {author} {\bibinfo {author} {\bibfnamefont {P.}~\bibnamefont {Machnikowski}},\ }\href {https://doi.org/10.1103/PhysRevLett.96.140405} {\bibfield  {journal} {\bibinfo  {journal} {Phys. Rev. Lett.}\ }\textbf {\bibinfo {volume} {96}},\ \bibinfo {pages} {140405} (\bibinfo {year} {2006}{\natexlab{a}})}\BibitemShut {NoStop}%
\bibitem [{\citenamefont {Machnikowski}(2006{\natexlab{b}})}]{machnikowski2006pss}%
  \BibitemOpen
  \bibfield  {author} {\bibinfo {author} {\bibfnamefont {P.}~\bibnamefont {Machnikowski}},\ }\bibfield  {journal} {\bibinfo  {journal} {Physica Status Solidi (B) Basic Research}\ }\textbf {\bibinfo {volume} {243}},\ \href {https://doi.org/10.1002/pssb.200668035} {10.1002/pssb.200668035} (\bibinfo {year} {2006}{\natexlab{b}})\BibitemShut {NoStop}%
\bibitem [{\citenamefont {Kabuss}\ \emph {et~al.}(2010)\citenamefont {Kabuss}, \citenamefont {Werner}, \citenamefont {Hoffmann}, \citenamefont {Hildebrandt}, \citenamefont {Knorr},\ and\ \citenamefont {Richter}}]{Kabuss2010}%
  \BibitemOpen
  \bibfield  {author} {\bibinfo {author} {\bibfnamefont {J.}~\bibnamefont {Kabuss}}, \bibinfo {author} {\bibfnamefont {S.}~\bibnamefont {Werner}}, \bibinfo {author} {\bibfnamefont {A.}~\bibnamefont {Hoffmann}}, \bibinfo {author} {\bibfnamefont {P.}~\bibnamefont {Hildebrandt}}, \bibinfo {author} {\bibfnamefont {A.}~\bibnamefont {Knorr}},\ and\ \bibinfo {author} {\bibfnamefont {M.}~\bibnamefont {Richter}},\ }\href {https://doi.org/10.1103/PhysRevB.81.075314} {\bibfield  {journal} {\bibinfo  {journal} {Phys. Rev. B}\ }\textbf {\bibinfo {volume} {81}},\ \bibinfo {pages} {075314} (\bibinfo {year} {2010})}\BibitemShut {NoStop}%
\bibitem [{\citenamefont {Kabuss}\ and\ \citenamefont {Richter}(2011)}]{Kabuss2011b}%
  \BibitemOpen
  \bibfield  {author} {\bibinfo {author} {\bibfnamefont {J.}~\bibnamefont {Kabuss}}\ and\ \bibinfo {author} {\bibfnamefont {M.}~\bibnamefont {Richter}},\ }\href {https://doi.org/https://doi.org/10.1016/j.photonics.2011.06.001} {\bibfield  {journal} {\bibinfo  {journal} {Photonics Nanostructures - Fundam. Appl.}\ }\textbf {\bibinfo {volume} {9}},\ \bibinfo {pages} {296} (\bibinfo {year} {2011})}\BibitemShut {NoStop}%
\bibitem [{\citenamefont {McCutcheon}\ and\ \citenamefont {Nazir}(2013)}]{McCutcheon2013}%
  \BibitemOpen
  \bibfield  {author} {\bibinfo {author} {\bibfnamefont {D.~P.~S.}\ \bibnamefont {McCutcheon}}\ and\ \bibinfo {author} {\bibfnamefont {A.}~\bibnamefont {Nazir}},\ }\href {https://doi.org/10.1103/PhysRevLett.110.217401} {\bibfield  {journal} {\bibinfo  {journal} {Phys. Rev. Lett.}\ }\textbf {\bibinfo {volume} {110}},\ \bibinfo {pages} {217401} (\bibinfo {year} {2013})}\BibitemShut {NoStop}%
\bibitem [{\citenamefont {Ahn}\ \emph {et~al.}(2005)\citenamefont {Ahn}, \citenamefont {F\"orstner},\ and\ \citenamefont {Knorr}}]{Ahn2005}%
  \BibitemOpen
  \bibfield  {author} {\bibinfo {author} {\bibfnamefont {K.~J.}\ \bibnamefont {Ahn}}, \bibinfo {author} {\bibfnamefont {J.}~\bibnamefont {F\"orstner}},\ and\ \bibinfo {author} {\bibfnamefont {A.}~\bibnamefont {Knorr}},\ }\href {https://doi.org/10.1103/PhysRevB.71.153309} {\bibfield  {journal} {\bibinfo  {journal} {Phys. Rev. B}\ }\textbf {\bibinfo {volume} {71}},\ \bibinfo {pages} {153309} (\bibinfo {year} {2005})}\BibitemShut {NoStop}%
\bibitem [{\citenamefont {Bogaczewicz}\ and\ \citenamefont {Machnikowski}(2023)}]{Bogaczewicz2023}%
  \BibitemOpen
  \bibfield  {author} {\bibinfo {author} {\bibfnamefont {R.~A.}\ \bibnamefont {Bogaczewicz}}\ and\ \bibinfo {author} {\bibfnamefont {P.}~\bibnamefont {Machnikowski}},\ }\href {https://doi.org/10.1088/1367-2630/acfb2f} {\bibfield  {journal} {\bibinfo  {journal} {New Journal of Physics}\ }\textbf {\bibinfo {volume} {25}},\ \bibinfo {pages} {093057} (\bibinfo {year} {2023})}\BibitemShut {NoStop}%
\bibitem [{\citenamefont {Bogaczewicz}\ and\ \citenamefont {Machnikowski}(2025)}]{Bogaczewicz2025}%
  \BibitemOpen
  \bibfield  {author} {\bibinfo {author} {\bibfnamefont {R.~A.}\ \bibnamefont {Bogaczewicz}}\ and\ \bibinfo {author} {\bibfnamefont {P.}~\bibnamefont {Machnikowski}},\ }\href {https://doi.org/10.1364/OL.539414} {\bibfield  {journal} {\bibinfo  {journal} {Opt. Lett.}\ }\textbf {\bibinfo {volume} {50}},\ \bibinfo {pages} {888} (\bibinfo {year} {2025})}\BibitemShut {NoStop}%
\bibitem [{\citenamefont {Fano}(1961)}]{Fano1961}%
  \BibitemOpen
  \bibfield  {author} {\bibinfo {author} {\bibfnamefont {U.}~\bibnamefont {Fano}},\ }\href {https://doi.org/10.1103/PhysRev.124.1866} {\bibfield  {journal} {\bibinfo  {journal} {Phys. Rev.}\ }\textbf {\bibinfo {volume} {124}},\ \bibinfo {pages} {1866} (\bibinfo {year} {1961})}\BibitemShut {NoStop}%
\bibitem [{\citenamefont {Riffe}(2011)}]{Riffe2011}%
  \BibitemOpen
  \bibfield  {author} {\bibinfo {author} {\bibfnamefont {D.~M.}\ \bibnamefont {Riffe}},\ }\href {https://doi.org/10.1103/PhysRevB.84.064308} {\bibfield  {journal} {\bibinfo  {journal} {Phys. Rev. B}\ }\textbf {\bibinfo {volume} {84}},\ \bibinfo {pages} {064308} (\bibinfo {year} {2011})}\BibitemShut {NoStop}%
\bibitem [{\citenamefont {Lee}\ \emph {et~al.}(2006)\citenamefont {Lee}, \citenamefont {Inoue},\ and\ \citenamefont {Hase}}]{leeUltrafastFanoResonance2006}%
  \BibitemOpen
  \bibfield  {author} {\bibinfo {author} {\bibfnamefont {J.~D.}\ \bibnamefont {Lee}}, \bibinfo {author} {\bibfnamefont {J.}~\bibnamefont {Inoue}},\ and\ \bibinfo {author} {\bibfnamefont {M.}~\bibnamefont {Hase}},\ }\href {https://doi.org/10.1103/PhysRevLett.97.157405} {\bibfield  {journal} {\bibinfo  {journal} {Phys. Rev. Lett.}\ }\textbf {\bibinfo {volume} {97}},\ \bibinfo {pages} {157405} (\bibinfo {year} {2006})}\BibitemShut {NoStop}%
\bibitem [{\citenamefont {Misochko}\ and\ \citenamefont {Lebedev}(2015)}]{misochkoFanoInterferenceExcitation2015}%
  \BibitemOpen
  \bibfield  {author} {\bibinfo {author} {\bibfnamefont {O.~V.}\ \bibnamefont {Misochko}}\ and\ \bibinfo {author} {\bibfnamefont {M.~V.}\ \bibnamefont {Lebedev}},\ }\href {https://doi.org/10.1134/S1063776115020168} {\bibfield  {journal} {\bibinfo  {journal} {J. Exp. Theor. Phys.}\ }\textbf {\bibinfo {volume} {120}},\ \bibinfo {pages} {651} (\bibinfo {year} {2015})}\BibitemShut {NoStop}%
\bibitem [{\citenamefont {Yoshino}\ \emph {et~al.}(2015)\citenamefont {Yoshino}, \citenamefont {Oohata},\ and\ \citenamefont {Mizoguchi}}]{yoshinoDynamicalFanoLikeInterference2015}%
  \BibitemOpen
  \bibfield  {author} {\bibinfo {author} {\bibfnamefont {S.}~\bibnamefont {Yoshino}}, \bibinfo {author} {\bibfnamefont {G.}~\bibnamefont {Oohata}},\ and\ \bibinfo {author} {\bibfnamefont {K.}~\bibnamefont {Mizoguchi}},\ }\href {https://doi.org/10.1103/PhysRevLett.115.157402} {\bibfield  {journal} {\bibinfo  {journal} {Phys. Rev. Lett.}\ }\textbf {\bibinfo {volume} {115}},\ \bibinfo {pages} {157402} (\bibinfo {year} {2015})}\BibitemShut {NoStop}%
\bibitem [{\citenamefont {Watanabe}\ \emph {et~al.}(2017)\citenamefont {Watanabe}, \citenamefont {Hino}, \citenamefont {Hase},\ and\ \citenamefont {Maeshima}}]{watanabePolaronicQuasiparticlePicture2017}%
  \BibitemOpen
  \bibfield  {author} {\bibinfo {author} {\bibfnamefont {Y.}~\bibnamefont {Watanabe}}, \bibinfo {author} {\bibfnamefont {K.-i.}\ \bibnamefont {Hino}}, \bibinfo {author} {\bibfnamefont {M.}~\bibnamefont {Hase}},\ and\ \bibinfo {author} {\bibfnamefont {N.}~\bibnamefont {Maeshima}},\ }\href {https://doi.org/10.1103/PhysRevB.95.014301} {\bibfield  {journal} {\bibinfo  {journal} {Phys. Rev. B}\ }\textbf {\bibinfo {volume} {95}},\ \bibinfo {pages} {014301} (\bibinfo {year} {2017})}\BibitemShut {NoStop}%
\bibitem [{\citenamefont {Vinod}\ \emph {et~al.}(2018)\citenamefont {Vinod}, \citenamefont {Raghavan},\ and\ \citenamefont {Sivasubramanian}}]{vinodFanoResonanceCoherent2018}%
  \BibitemOpen
  \bibfield  {author} {\bibinfo {author} {\bibfnamefont {M.}~\bibnamefont {Vinod}}, \bibinfo {author} {\bibfnamefont {G.}~\bibnamefont {Raghavan}},\ and\ \bibinfo {author} {\bibfnamefont {V.}~\bibnamefont {Sivasubramanian}},\ }\href {https://doi.org/10.1038/s41598-018-35866-7} {\bibfield  {journal} {\bibinfo  {journal} {Sci Rep}\ }\textbf {\bibinfo {volume} {8}},\ \bibinfo {pages} {17706} (\bibinfo {year} {2018})}\BibitemShut {NoStop}%
\bibitem [{\citenamefont {Watanabe}\ \emph {et~al.}(2019)\citenamefont {Watanabe}, \citenamefont {Hino}, \citenamefont {Maeshima}, \citenamefont {Petek},\ and\ \citenamefont {Hase}}]{watanabeUltrafastAsymmetricRosenZenerlike2019}%
  \BibitemOpen
  \bibfield  {author} {\bibinfo {author} {\bibfnamefont {Y.}~\bibnamefont {Watanabe}}, \bibinfo {author} {\bibfnamefont {K.-i.}\ \bibnamefont {Hino}}, \bibinfo {author} {\bibfnamefont {N.}~\bibnamefont {Maeshima}}, \bibinfo {author} {\bibfnamefont {H.}~\bibnamefont {Petek}},\ and\ \bibinfo {author} {\bibfnamefont {M.}~\bibnamefont {Hase}},\ }\href {https://doi.org/10.1103/PhysRevB.99.174304} {\bibfield  {journal} {\bibinfo  {journal} {Phys. Rev. B}\ }\textbf {\bibinfo {volume} {99}},\ \bibinfo {pages} {174304} (\bibinfo {year} {2019})}\BibitemShut {NoStop}%
\bibitem [{\citenamefont {Vasileiadis}\ \emph {et~al.}(2020)\citenamefont {Vasileiadis}, \citenamefont {Zhang}, \citenamefont {Wang}, \citenamefont {Bonn}, \citenamefont {Fytas},\ and\ \citenamefont {Graczykowski}}]{vasileiadisFrequencydomainStudyNonthermal2020}%
  \BibitemOpen
  \bibfield  {author} {\bibinfo {author} {\bibfnamefont {T.}~\bibnamefont {Vasileiadis}}, \bibinfo {author} {\bibfnamefont {H.}~\bibnamefont {Zhang}}, \bibinfo {author} {\bibfnamefont {H.}~\bibnamefont {Wang}}, \bibinfo {author} {\bibfnamefont {M.}~\bibnamefont {Bonn}}, \bibinfo {author} {\bibfnamefont {G.}~\bibnamefont {Fytas}},\ and\ \bibinfo {author} {\bibfnamefont {B.}~\bibnamefont {Graczykowski}},\ }\href {https://doi.org/10.1126/sciadv.abd4540} {\bibfield  {journal} {\bibinfo  {journal} {Sci. Adv.}\ }\textbf {\bibinfo {volume} {6}},\ \bibinfo {pages} {eabd4540} (\bibinfo {year} {2020})}\BibitemShut {NoStop}%
\bibitem [{\citenamefont {Kitzman}\ \emph {et~al.}(2023)\citenamefont {Kitzman}, \citenamefont {Lane}, \citenamefont {Undershute}, \citenamefont {Beysengulov}, \citenamefont {Mikolas}, \citenamefont {Murch},\ and\ \citenamefont {Pollanen}}]{kitzmanQuantumAcousticFano2023}%
  \BibitemOpen
  \bibfield  {author} {\bibinfo {author} {\bibfnamefont {J.~M.}\ \bibnamefont {Kitzman}}, \bibinfo {author} {\bibfnamefont {J.~R.}\ \bibnamefont {Lane}}, \bibinfo {author} {\bibfnamefont {C.}~\bibnamefont {Undershute}}, \bibinfo {author} {\bibfnamefont {N.~R.}\ \bibnamefont {Beysengulov}}, \bibinfo {author} {\bibfnamefont {C.~A.}\ \bibnamefont {Mikolas}}, \bibinfo {author} {\bibfnamefont {K.~W.}\ \bibnamefont {Murch}},\ and\ \bibinfo {author} {\bibfnamefont {J.}~\bibnamefont {Pollanen}},\ }\href {https://doi.org/10.1103/PhysRevA.108.L010601} {\bibfield  {journal} {\bibinfo  {journal} {Phys. Rev. A}\ }\textbf {\bibinfo {volume} {108}},\ \bibinfo {pages} {L010601} (\bibinfo {year} {2023})}\BibitemShut {NoStop}%
\bibitem [{\citenamefont {Weiss}(1998)}]{Weiss1998}%
  \BibitemOpen
  \bibfield  {author} {\bibinfo {author} {\bibfnamefont {U.}~\bibnamefont {Weiss}},\ }\href@noop {} {\emph {\bibinfo {title} {Quantum Dissipative Systems.}}},\ \bibinfo {edition} {2nd}\ ed.\ (\bibinfo  {publisher} {World Scientific Publishing Company},\ \bibinfo {address} {Singapore},\ \bibinfo {year} {1998})\BibitemShut {NoStop}%
\bibitem [{\citenamefont {Reiter}\ \emph {et~al.}(2019)\citenamefont {Reiter}, \citenamefont {Kuhn},\ and\ \citenamefont {Axt}}]{Reiter2019}%
  \BibitemOpen
  \bibfield  {author} {\bibinfo {author} {\bibfnamefont {D.~E.}\ \bibnamefont {Reiter}}, \bibinfo {author} {\bibfnamefont {T.}~\bibnamefont {Kuhn}},\ and\ \bibinfo {author} {\bibfnamefont {V.~M.}\ \bibnamefont {Axt}},\ }\href {https://doi.org/10.1080/23746149.2019.1655478} {\bibfield  {journal} {\bibinfo  {journal} {Advances in Physics: X}\ }\textbf {\bibinfo {volume} {4}},\ \bibinfo {pages} {1655478} (\bibinfo {year} {2019})}\BibitemShut {NoStop}%
\bibitem [{\citenamefont {L\"uker}\ \emph {et~al.}(2017)\citenamefont {L\"uker}, \citenamefont {Kuhn},\ and\ \citenamefont {Reiter}}]{Lueker2017}%
  \BibitemOpen
  \bibfield  {author} {\bibinfo {author} {\bibfnamefont {S.}~\bibnamefont {L\"uker}}, \bibinfo {author} {\bibfnamefont {T.}~\bibnamefont {Kuhn}},\ and\ \bibinfo {author} {\bibfnamefont {D.~E.}\ \bibnamefont {Reiter}},\ }\href {https://doi.org/10.1103/PhysRevB.96.245306} {\bibfield  {journal} {\bibinfo  {journal} {Phys. Rev. B}\ }\textbf {\bibinfo {volume} {96}},\ \bibinfo {pages} {245306} (\bibinfo {year} {2017})}\BibitemShut {NoStop}%
\bibitem [{\citenamefont {Mahan}(2000)}]{Mahan2000}%
  \BibitemOpen
  \bibfield  {author} {\bibinfo {author} {\bibfnamefont {G.~D.}\ \bibnamefont {Mahan}},\ }\href@noop {} {\emph {\bibinfo {title} {Many Particle Physics, Third Edition}}}\ (\bibinfo  {publisher} {Plenum},\ \bibinfo {address} {New York},\ \bibinfo {year} {2000})\BibitemShut {NoStop}%
\bibitem [{\citenamefont {Jayich}\ \emph {et~al.}(2012)\citenamefont {Jayich}, \citenamefont {Sankey}, \citenamefont {B{\o}rkje}, \citenamefont {Lee}, \citenamefont {Yang}, \citenamefont {Underwood}, \citenamefont {Childress}, \citenamefont {Petrenko}, \citenamefont {Girvin},\ and\ \citenamefont {Harris}}]{Jayich2012}%
  \BibitemOpen
  \bibfield  {author} {\bibinfo {author} {\bibfnamefont {A.~M.}\ \bibnamefont {Jayich}}, \bibinfo {author} {\bibfnamefont {J.~C.}\ \bibnamefont {Sankey}}, \bibinfo {author} {\bibfnamefont {K.}~\bibnamefont {B{\o}rkje}}, \bibinfo {author} {\bibfnamefont {D.}~\bibnamefont {Lee}}, \bibinfo {author} {\bibfnamefont {C.}~\bibnamefont {Yang}}, \bibinfo {author} {\bibfnamefont {M.}~\bibnamefont {Underwood}}, \bibinfo {author} {\bibfnamefont {L.}~\bibnamefont {Childress}}, \bibinfo {author} {\bibfnamefont {A.}~\bibnamefont {Petrenko}}, \bibinfo {author} {\bibfnamefont {S.~M.}\ \bibnamefont {Girvin}},\ and\ \bibinfo {author} {\bibfnamefont {J.~G.~E.}\ \bibnamefont {Harris}},\ }\href {https://doi.org/10.1088/1367-2630/14/11/115018} {\bibfield  {journal} {\bibinfo  {journal} {New J. Phys.}\ }\textbf {\bibinfo {volume} {14}},\ \bibinfo {pages} {115018} (\bibinfo {year} {2012})}\BibitemShut {NoStop}%
\bibitem [{\citenamefont {Timberlake}\ \emph {et~al.}(2019)\citenamefont {Timberlake}, \citenamefont {Toro{\v s}}, \citenamefont {Hempston}, \citenamefont {Winstone}, \citenamefont {Rashid},\ and\ \citenamefont {Ulbricht}}]{Timberlake2019}%
  \BibitemOpen
  \bibfield  {author} {\bibinfo {author} {\bibfnamefont {C.}~\bibnamefont {Timberlake}}, \bibinfo {author} {\bibfnamefont {M.}~\bibnamefont {Toro{\v s}}}, \bibinfo {author} {\bibfnamefont {D.}~\bibnamefont {Hempston}}, \bibinfo {author} {\bibfnamefont {G.}~\bibnamefont {Winstone}}, \bibinfo {author} {\bibfnamefont {M.}~\bibnamefont {Rashid}},\ and\ \bibinfo {author} {\bibfnamefont {H.}~\bibnamefont {Ulbricht}},\ }\href {https://doi.org/10.1063/1.5081045} {\bibfield  {journal} {\bibinfo  {journal} {Applied Physics Letters}\ }\textbf {\bibinfo {volume} {114}},\ \bibinfo {pages} {023104} (\bibinfo {year} {2019})}\BibitemShut {NoStop}%
\bibitem [{\citenamefont {Tribelsky}\ \emph {et~al.}(2008)\citenamefont {Tribelsky}, \citenamefont {Flach}, \citenamefont {Miroshnichenko}, \citenamefont {Gorbach},\ and\ \citenamefont {Kivshar}}]{Tribelsky2008}%
  \BibitemOpen
  \bibfield  {author} {\bibinfo {author} {\bibfnamefont {M.~I.}\ \bibnamefont {Tribelsky}}, \bibinfo {author} {\bibfnamefont {S.}~\bibnamefont {Flach}}, \bibinfo {author} {\bibfnamefont {A.~E.}\ \bibnamefont {Miroshnichenko}}, \bibinfo {author} {\bibfnamefont {A.~V.}\ \bibnamefont {Gorbach}},\ and\ \bibinfo {author} {\bibfnamefont {Y.~S.}\ \bibnamefont {Kivshar}},\ }\href {https://doi.org/10.1103/PhysRevLett.100.043903} {\bibfield  {journal} {\bibinfo  {journal} {Phys. Rev. Lett.}\ }\textbf {\bibinfo {volume} {100}},\ \bibinfo {pages} {043903} (\bibinfo {year} {2008})}\BibitemShut {NoStop}%
\bibitem [{\citenamefont {Liu}\ \emph {et~al.}(2009)\citenamefont {Liu}, \citenamefont {Langguth}, \citenamefont {Weiss}, \citenamefont {K{\"a}stel}, \citenamefont {Fleischhauer}, \citenamefont {Pfau},\ and\ \citenamefont {Giessen}}]{Liu2009}%
  \BibitemOpen
  \bibfield  {author} {\bibinfo {author} {\bibfnamefont {N.}~\bibnamefont {Liu}}, \bibinfo {author} {\bibfnamefont {L.}~\bibnamefont {Langguth}}, \bibinfo {author} {\bibfnamefont {T.}~\bibnamefont {Weiss}}, \bibinfo {author} {\bibfnamefont {J.}~\bibnamefont {K{\"a}stel}}, \bibinfo {author} {\bibfnamefont {M.}~\bibnamefont {Fleischhauer}}, \bibinfo {author} {\bibfnamefont {T.}~\bibnamefont {Pfau}},\ and\ \bibinfo {author} {\bibfnamefont {H.}~\bibnamefont {Giessen}},\ }\href {https://doi.org/10.1038/nmat2495} {\bibfield  {journal} {\bibinfo  {journal} {Nature Materials}\ }\textbf {\bibinfo {volume} {8}},\ \bibinfo {pages} {758} (\bibinfo {year} {2009})}\BibitemShut {NoStop}%
\bibitem [{\citenamefont {Hao}\ \emph {et~al.}(2008)\citenamefont {Hao}, \citenamefont {Sonnefraud}, \citenamefont {Dorpe}, \citenamefont {Maier}, \citenamefont {Halas},\ and\ \citenamefont {Nordlander}}]{Hao2008}%
  \BibitemOpen
  \bibfield  {author} {\bibinfo {author} {\bibfnamefont {F.}~\bibnamefont {Hao}}, \bibinfo {author} {\bibfnamefont {Y.}~\bibnamefont {Sonnefraud}}, \bibinfo {author} {\bibfnamefont {P.~V.}\ \bibnamefont {Dorpe}}, \bibinfo {author} {\bibfnamefont {S.~A.}\ \bibnamefont {Maier}}, \bibinfo {author} {\bibfnamefont {N.~J.}\ \bibnamefont {Halas}},\ and\ \bibinfo {author} {\bibfnamefont {P.}~\bibnamefont {Nordlander}},\ }\href {https://doi.org/10.1021/nl802509r} {\bibfield  {journal} {\bibinfo  {journal} {Nano Letters}\ }\textbf {\bibinfo {volume} {8}},\ \bibinfo {pages} {3983} (\bibinfo {year} {2008})}\BibitemShut {NoStop}%
\bibitem [{\citenamefont {Verellen}\ \emph {et~al.}(2009)\citenamefont {Verellen}, \citenamefont {Sonnefraud}, \citenamefont {Sobhani}, \citenamefont {Hao}, \citenamefont {Moshchalkov}, \citenamefont {Dorpe}, \citenamefont {Nordlander},\ and\ \citenamefont {Maier}}]{Verellen2009}%
  \BibitemOpen
  \bibfield  {author} {\bibinfo {author} {\bibfnamefont {N.}~\bibnamefont {Verellen}}, \bibinfo {author} {\bibfnamefont {Y.}~\bibnamefont {Sonnefraud}}, \bibinfo {author} {\bibfnamefont {H.}~\bibnamefont {Sobhani}}, \bibinfo {author} {\bibfnamefont {F.}~\bibnamefont {Hao}}, \bibinfo {author} {\bibfnamefont {V.~V.}\ \bibnamefont {Moshchalkov}}, \bibinfo {author} {\bibfnamefont {P.~V.}\ \bibnamefont {Dorpe}}, \bibinfo {author} {\bibfnamefont {P.}~\bibnamefont {Nordlander}},\ and\ \bibinfo {author} {\bibfnamefont {S.~A.}\ \bibnamefont {Maier}},\ }\href {https://doi.org/10.1021/nl9001876} {\bibfield  {journal} {\bibinfo  {journal} {Nano Letters}\ }\textbf {\bibinfo {volume} {9}},\ \bibinfo {pages} {1663} (\bibinfo {year} {2009})}\BibitemShut {NoStop}%
\bibitem [{\citenamefont {Li}\ \emph {et~al.}(2011)\citenamefont {Li}, \citenamefont {Xiao}, \citenamefont {Zou}, \citenamefont {Liu}, \citenamefont {Jiang}, \citenamefont {Chen}, \citenamefont {Li},\ and\ \citenamefont {Gong}}]{Li2011}%
  \BibitemOpen
  \bibfield  {author} {\bibinfo {author} {\bibfnamefont {B.-B.}\ \bibnamefont {Li}}, \bibinfo {author} {\bibfnamefont {Y.-F.}\ \bibnamefont {Xiao}}, \bibinfo {author} {\bibfnamefont {C.-L.}\ \bibnamefont {Zou}}, \bibinfo {author} {\bibfnamefont {Y.-C.}\ \bibnamefont {Liu}}, \bibinfo {author} {\bibfnamefont {X.-F.}\ \bibnamefont {Jiang}}, \bibinfo {author} {\bibfnamefont {Y.-L.}\ \bibnamefont {Chen}}, \bibinfo {author} {\bibfnamefont {Y.}~\bibnamefont {Li}},\ and\ \bibinfo {author} {\bibfnamefont {Q.}~\bibnamefont {Gong}},\ }\href {https://doi.org/10.1063/1.3541884} {\bibfield  {journal} {\bibinfo  {journal} {Applied Physics Letters}\ }\textbf {\bibinfo {volume} {98}},\ \bibinfo {pages} {021116} (\bibinfo {year} {2011})}\BibitemShut {NoStop}%
\bibitem [{\citenamefont {Lee}\ and\ \citenamefont {Poon}(2004)}]{Lee2004}%
  \BibitemOpen
  \bibfield  {author} {\bibinfo {author} {\bibfnamefont {H.-T.}\ \bibnamefont {Lee}}\ and\ \bibinfo {author} {\bibfnamefont {A.~W.}\ \bibnamefont {Poon}},\ }\href {https://doi.org/10.1364/OL.29.000005} {\bibfield  {journal} {\bibinfo  {journal} {Opt. Lett.}\ }\textbf {\bibinfo {volume} {29}},\ \bibinfo {pages} {5} (\bibinfo {year} {2004})}\BibitemShut {NoStop}%
\bibitem [{\citenamefont {Iizawa}\ \emph {et~al.}(2021)\citenamefont {Iizawa}, \citenamefont {Kosugi}, \citenamefont {Koike},\ and\ \citenamefont {Azuma}}]{Iizawa2021}%
  \BibitemOpen
  \bibfield  {author} {\bibinfo {author} {\bibfnamefont {M.}~\bibnamefont {Iizawa}}, \bibinfo {author} {\bibfnamefont {S.}~\bibnamefont {Kosugi}}, \bibinfo {author} {\bibfnamefont {F.}~\bibnamefont {Koike}},\ and\ \bibinfo {author} {\bibfnamefont {Y.}~\bibnamefont {Azuma}},\ }\href {https://doi.org/10.1088/1402-4896/abe580} {\bibfield  {journal} {\bibinfo  {journal} {Phys. Scr.}\ }\textbf {\bibinfo {volume} {96}},\ \bibinfo {pages} {055401} (\bibinfo {year} {2021})}\BibitemShut {NoStop}%
\bibitem [{\citenamefont {Garrido~Alzar}\ \emph {et~al.}(2002)\citenamefont {Garrido~Alzar}, \citenamefont {Martinez},\ and\ \citenamefont {Nussenzveig}}]{GarridoAlzar2002}%
  \BibitemOpen
  \bibfield  {author} {\bibinfo {author} {\bibfnamefont {C.~L.}\ \bibnamefont {Garrido~Alzar}}, \bibinfo {author} {\bibfnamefont {M.~A.~G.}\ \bibnamefont {Martinez}},\ and\ \bibinfo {author} {\bibfnamefont {P.}~\bibnamefont {Nussenzveig}},\ }\href {https://doi.org/10.1119/1.1412644} {\bibfield  {journal} {\bibinfo  {journal} {American Journal of Physics}\ }\textbf {\bibinfo {volume} {70}},\ \bibinfo {pages} {37} (\bibinfo {year} {2002})}\BibitemShut {NoStop}%
\bibitem [{\citenamefont {Joe}\ \emph {et~al.}(2006)\citenamefont {Joe}, \citenamefont {Satanin},\ and\ \citenamefont {Kim}}]{Joe2006}%
  \BibitemOpen
  \bibfield  {author} {\bibinfo {author} {\bibfnamefont {Y.~S.}\ \bibnamefont {Joe}}, \bibinfo {author} {\bibfnamefont {A.~M.}\ \bibnamefont {Satanin}},\ and\ \bibinfo {author} {\bibfnamefont {C.~S.}\ \bibnamefont {Kim}},\ }\href {https://doi.org/10.1088/0031-8949/74/2/020} {\bibfield  {journal} {\bibinfo  {journal} {Physica Scripta}\ }\textbf {\bibinfo {volume} {74}},\ \bibinfo {pages} {259} (\bibinfo {year} {2006})}\BibitemShut {NoStop}%
\bibitem [{\citenamefont {Satpathy}\ \emph {et~al.}(2012)\citenamefont {Satpathy}, \citenamefont {Roy},\ and\ \citenamefont {Mohapatra}}]{Satpathy2012}%
  \BibitemOpen
  \bibfield  {author} {\bibinfo {author} {\bibfnamefont {S.}~\bibnamefont {Satpathy}}, \bibinfo {author} {\bibfnamefont {A.}~\bibnamefont {Roy}},\ and\ \bibinfo {author} {\bibfnamefont {A.}~\bibnamefont {Mohapatra}},\ }\href {https://doi.org/10.1088/0143-0807/33/4/863} {\bibfield  {journal} {\bibinfo  {journal} {European Journal of Physics}\ }\textbf {\bibinfo {volume} {33}},\ \bibinfo {pages} {863} (\bibinfo {year} {2012})}\BibitemShut {NoStop}%
\bibitem [{\citenamefont {Stassi}\ \emph {et~al.}(2017)\citenamefont {Stassi}, \citenamefont {Chiad{\`o}}, \citenamefont {Calafiore}, \citenamefont {Palmara}, \citenamefont {Cabrini},\ and\ \citenamefont {Ricciardi}}]{Stassi2017}%
  \BibitemOpen
  \bibfield  {author} {\bibinfo {author} {\bibfnamefont {S.}~\bibnamefont {Stassi}}, \bibinfo {author} {\bibfnamefont {A.}~\bibnamefont {Chiad{\`o}}}, \bibinfo {author} {\bibfnamefont {G.}~\bibnamefont {Calafiore}}, \bibinfo {author} {\bibfnamefont {G.}~\bibnamefont {Palmara}}, \bibinfo {author} {\bibfnamefont {S.}~\bibnamefont {Cabrini}},\ and\ \bibinfo {author} {\bibfnamefont {C.}~\bibnamefont {Ricciardi}},\ }\href {https://doi.org/10.1038/s41598-017-01147-y} {\bibfield  {journal} {\bibinfo  {journal} {Scientific Reports}\ }\textbf {\bibinfo {volume} {7}},\ \bibinfo {pages} {1065} (\bibinfo {year} {2017})}\BibitemShut {NoStop}%
\end{thebibliography}%

\end{document}